\documentclass[fleqn,usenatbib]{mnras}
\usepackage{mathptmx}
\usepackage[T1]{fontenc}
\usepackage{ae,aecompl}

\usepackage{graphicx}	%
\usepackage{amsmath}	%
\usepackage{amssymb}	%
\usepackage[]{siunitx}
\usepackage[dvipsnames]{xcolor}
\usepackage[export]{adjustbox}
\usepackage{makecell}
\usepackage{booktabs}
\usepackage{enumitem}

\usepackage{fixme}
\fxsetup{
  status=draft,
  author=CC,
  layout=inline,nomargin,nofootnote,
  theme=colorsig,
}
\usepackage{makecell}

\FXRegisterAuthor{cc}{acc}{CC}
\FXRegisterAuthor{mr}{amr}{MR}	%

\newcommand{\hide}[1]{}
\newcommand\hiide[1]{}
\newcommand{\annot}[1]{}

\newcommand{\A}{\textit{A-hr}}
\newcommand{\B}{\textit{B}}
\newcommand{\ramses}{\textsc{ramses-rt}}
\newcommand{\rsph}{\ensuremath{r_{\rm sph}}}
\newcommand{\rcyl}{\ensuremath{r_{\rm cyl}}}

\newcommand{\HII}{H\,{\sc ii}}
\newcommand{\msun}{\ensuremath{{\rm M}_{\odot}}}
\newcommand{\pcc}{\ensuremath{{\rm cm}^{-3}}}	%

\newcommand{\ie}{\textit{i.e.}}
\newcommand{\eg}{\textit{e.g.}}

\title[Massive Prestellar Cores]{Massive Prestellar Cores in Radiation-magneto-turbulent Simulations of Molecular Clouds}
\author[C.-C. He and M. Ricotti]{
  Chong-Chong He$^{1}$\thanks{E-mail: che1234@umd.edu} and Massimo Ricotti$^1$\\
  $^{1}$Department of Astronomy, University of Maryland, College Park, MD, 20742, US\\
}
\date{Accepted XXX. Received YYY; in original form ZZZ}
\pubyear{2022}

\begin{document}
\label{firstpage}
\pagerange{\pageref{firstpage}--\pageref{lastpage}}
\maketitle

\begin{abstract}
We simulate the formation and collapse of prestellar cores at few-AU resolution in a set of radiation-magneto-hydrodynamic simulations of giant molecular clouds (GMCs) using the grid-based code RAMSES-RT. We adopt, for the first time to our best knowledge, realistic initial/boundary conditions by zooming-in onto individual massive prestellar cores within the GMC. 
We identify two distinct modes of fragmentation: ``quasi-spherical'' and ``filamentary''. In both modes the fragments eventually become embedded in a quasi-steady accretion disk or toroid with radii $\sim 500-5000$~AU and opening angles $H/R \sim 0.5-1$. The disks/toroids are Toomre stable but the accreted pre-existing fragments are found orbiting the outer disk, appearing as disk fragmentation. Each core converts nearly 100 percent of the gas mass into a few massive stars forming near the disk center. Large and massive disks around high-mass stars are supported by magnetic pressure in the outer disk, at radii $>200-1000$~AU, and turbulent pressure in the inner disk. The most massive core accretes several times more mass than its initial mass, forming a (proto)star cluster of 8 massive stars enshrouded by a toroid, suggesting a competitive accretion scenario for ultra-high-mass star formation. We also find that the \HII{} regions produced by a single massive star remain trapped in the dense circumstellar disks for a few hundred kiloyears, while the dynamic motions of massive stars in wide binaries or multiple systems displace the stars from the densest parts of the disk, allowing UV radiation to escape producing steady or pulsating bipolar HII regions.

\end{abstract}


\section{Introduction}
\label{sec:intro}

Whether high-mass (HM) stars form from the monolithic collapse of massive prestellar cores -- supported by turbulence, and/or magnetic fields rather than thermal pressure -- known as the Turbulent Core (TC) scenario \citep{McKee2003,Tan2014}, or via accretion inflows from larger scales, known as the Competitive Accretion (CA) scenario \citep{Bonnell2001, Padoan2020}, still remains an open question. 
Observations \citep{Fuller2005, vanderTak2019} of accreting HM young stellar objects (YSOs) suggest that HM stars form similarly to their low-mass counterparts via infall from a surrounding envelope and from anisotropic accretion flow from an accretion disk. However, the physical processes involved are not well understood partially due to the lack of high-resolution observations of structures below $\sim 1000$~AU owing to the large distances of the sources, high dust extinction, high multiplicity, and complexity of the environment typical of high-mass star formation. The shorter timescale of formation and rarity of the objects result in a low probability of finding a Class O massive (proto)star or massive starless core.

There is growing evidence from ALMA observations that accretion disks around massive proto-stars undergo fragmentation and produce companion stars \citep[{\it e.g.},][]{Johnston2015,Guzman2020,Williams2021}. \cite{Ilee2018} reported the observation of a fragmented Keplerian disk around an O-type protostar, with a fragment in the outskirts of the disk at $\sim 2000$ AU from the primary. \cite{Johnston2020} observed spiral arms and instability in a disk of radius $\sim 1000$~AU around an O-type star.

The CA model postulates that low-mass protostellar seeds accrete unbound gas within the clump from large scales in a hierarchical structure. To test the idea of the CA scenario from a theoretical perspective, we need to simulate the formation of prestellar cores from the collapse of turbulent giant molecular clouds (GMCs), which is the site of star formation.
Several numerical studies have investigated the formation of star clusters from GMCs \citep{Jones2018,Lee2018,Kim2018,Bate2019,Fukushima2020a,Grudic2021,Kim2021}. 
In \cite*{He2019,He2020} we have conducted a series of simulations of the collapse of isolated turbulent GMCs using \ramses{}. In these works we have run a large grid of GMC simulations and pushed the parameters of the GMC mass and density to include very massive ($\sim 10^5 \ \msun$) and extremely dense ($\sim 10^4~{\rm cm}^{-3}$) clouds, resolving the formation of individual stars with masses $M \gtrsim 1 M_\odot$, significantly improving the resolution with respect to previous works (see a summary in Table 2 of \citealt{Lee2018}).
The initial mass functions (IMFs) of the stars forming in these simulations have not only characteristic power-law slopes very close to \cite{Kroupa2002} at the high-mass end, but also the correct normalization to a mass-normalized Kroupa IMF if we assume that each sink particle converts $\sim 40\%$ of its mass to a single star and the remaining mass forms several smaller mass stars.  This scaling is also inferred from the mapping between the observed core mass function (CMF) and stellar IMF that preserves the slope and normalization of the IMF.  Hence, we hypothesize that the unresolved sinks in the simulation form stars with high efficiency but fragment into lower-mass stars.

Motivated by these previous results, in this work we aim at testing our assumption on the fragmentation of sink particles in order to understand the mass function, star multiplicity, and kinematics which is important to eventually understand the long-term evolution of the star cluster and a possible role of high-z compact star clusters in forming and growing intermediate mass black holes (IMBHs) seeds. The methodology we use is to perform higher-resolution ``zoom'' simulations of the fragmenting protostellar cores, while simultaneously following the collapse of the GMC in which the cores are located.

Relatively thin disks of cool gas extending for tens to hundreds of AU are observed around almost all low-mass, young stars \citep{Williams2011}. They are an inevitable consequence of angular momentum conservation during the formation of a protostar through gravitational collapse. 
There is growing evidence from ALMA observations that accretion disks around massive proto-stars undergo fragmentation and produce companion stars \citep[{\it e.g.},][]{Johnston2015,Guzman2020,Williams2021}.
The kinematics of the dense gas in prestellar cores has been studied observationally through molecular line emission and it is shown to exhibit velocity gradients consistent with rotation \citep{Goodman1993}. 
Recent high dynamic range observations of the rotational motions of a sample of Class 0 protostars \citep{Gaudel2020} reveal systematic dispersion of the directions of velocity gradients at envelope scales > 1600 AU, challenging the presence of a rotation-dominated motion of the envelopes.
Furthermore, observations of T-Tauri disks suggest that their specific angular momentum is larger, by about one order of magnitude, than that observed in low-mass protostellar cores at scales of a few thousand AU \citep{Simon2000, Kurtovic2018}.
These observations, if correct, suggest that simple models of disk formation from the collapse of an isolated  spherical prestellar core that conserves angular momentum is missing some important physics.

Disks are observed to not only possess substructures in the ($r$, $\theta$) plane, but also show clear signs of substructures in the vertical ($z$) direction \citep{Muzerolle2009, Espaillat2011}.
Warped geometries or misalignments (``broken'' disks) have been inferred kinematically with resolved spectral line data \citep{Rosenfeld2012, Casassus2015} and scatter light shadows at larger $r$ \citep{Marino2015}.

A considerable amount of research has studied disks around nearby solar-type stars. However, the number of disk studied around more distant, massive stars (type A and earlier) is comparatively small. This is because massive protostellar cores that may form massive stars, multiple systems or even a mini-cluster of stars are fewer and short lived, and hence are less likely to be found nearby.

Recent advances in radio/mm and optical/IR interferometers have enabled important progress in the field of disks around intermediate-mass (IM) and HM YSOs.
These observations of embedded IM protostars (A to late-B spectral type) \citep{Zapata2007, Sanchez-Monge2010, vanKempen2012, Takahashi2012} have revealed circumstellar disks with typical radii of a few hundreds of AU. These disks are geometrically thick with a scale height that is more than 20-30 percent of their radius. The disks have masses of a few solar masses and could be in Keplerian rotation.
Evidence for circumstellar disks has been reported \citep{Cesaroni2005, Patel2005, Kraus2010, deWit2011, Ginsburg2018, Law2022} for HM (proto)stars (early-B to late-O type) that correspond to zero-age main sequence stars of about 25-30 \msun{} with typical radii of a few thousands of AU, although radii smaller than 1000~AU have been estimated in some samples. These geometrically thick structures have scale heights of >30-40 percent of their radii and masses that range from a few \msun{} to a few tens of \msun{}. They are gravitationally stable as suggested by Toomre's stability parameter $Q>1$. In short, the basic properties of the disks around HM (proto)stars appear as a scaled-up version of those found for disks around low-mass and intermediate-mass protostars \citep[see][for a review]{Beltran2016}.

For stars of extremely high mass ($>30\msun$), the existence of a circumstellar disk has been elusive in observations. Simulations have shown that radiation pressure does not prevent disk accretion to from stars up to $140 \msun{}$ \citep{Krumholz2009aa, Kuiper2010}. However, no models of protostars allow there to be a hydrostatic object beyond this limit. 
Large, dense ($n \gtrsim 10^7 \ \pcc$) and massive (a few $\times 100 \ \msun$) rotating cores have been detected around early-O-type protostars. These are likely non-equilibrium structures that favor the formation of young stellar mini-clusters instead of individual massive stars \citep{Cesaroni2007,Beltran2011}.

In this work, we study the collapse of prestellar cores and the structure of protostellar disks around massive stars in realistic simulations of turbulent GMCs. 
These disks span a large range in sizes and masses. 
In this paper, we emphasize the dominant role of turbulence and magnetic field in determining the formation and support against the gravity of massive  disks within prestellar cores. 
In a companion paper, we will address in more detail the structure and evolution of the magnetic field and the problem of magnetic braking. 

The rest of this article is organized as follows. We describe our simulation method in Section~\ref{sec:1}. We present the basic results in Section~\ref{sec:2}, where we discuss the formation of turbulent massive disks. In Section~\ref{sec:3}, we discuss the properties of the core fragments. We provide discussions and the summary in Section~\ref{sec:dis}.

\section{Methods and Simulations}\label{sec:1}

\begin{table*}
\centering
\caption{\label{tab:init}
  Summary of the properties of the simulated cores, protostellar disks, and forming stars. The disks with $^*$ are the primary focus of this article and are discussed extensively. The columns are, from left to right, 
  (1) the names of the cores; 
  (2) the mass of the gas/core above a threshold density 3000 \pcc{}, $M_{\rm core}$;
  (3) the radius of the gas/core above a threshold density 3000 \pcc{}, $R_{\rm core}$;
  (4) the mass of the sink particle that forms in the lower-resolution ($l_{\rm max} = 14$) baseline run;
  (5) the maximum density, $n_{\rm sink,base}$, in the baseline simulations that the gas can reach before it is replaced with a sink particle, 
  (6) the maximum level of refinement $l_{\rm max}$ at the core center, based on top of the base grids with $l=7$ which has $2^7=128$ pixels in each dimension;
  (7) the maximum density $n_{\rm sink}$ that the gas can reach before it is replaced with a sink particle; 
  (8) the maximum resolution (minimum cell size) $\Delta x_{\rm min}$; 
  (9) the total stellar mass $M_{\rm stars, tot}$ in the ``zoom-in'' run;
  (10) the maximum mass $M_{\rm stars, max}$ of the stars in the ``zoom-in'' run;
  (11) the number of stars $N_{\rm stars}$ that form in the core;
  (12) disk radius $R_{\rm disk}$,
  (13) disk thickness $H_{\rm disk}$ 
  (14) disk aspect ratio $H_{\rm disk}$/$R_{\rm disk}$.
}
\begin{tabular}{lrrrlrrrrrrrrr}
\toprule
Core &  \makecell{ \(M_{\rm core}\) \\ (\msun{}) } & \makecell{ $R_{\rm core}$ \\ (pc)} &\makecell{\(M_{\rm \star, base}\) \\ (\msun{})} & \makecell{$n_{\rm sink,base}$ \\ (\({\rm cm}^{-3}\))} & \(l_{\rm max}\) & \makecell{ \(n_{\rm sink}\) \\ (\pcc{}) } & \makecell{ \(\Delta x_{\rm min}\) \\ (AU)} & \makecell{ \(M_{\star, \rm tot}\) \\ (\msun{}) } & \makecell{ \(M_{\star, \rm max}\) \\ (\msun{}) } & \(N_{\rm stars}\) & \makecell{\(R_{\rm disk}\) \\ (AU)} & \makecell{\(H_{\rm disk}\) \\ (AU)} & \(H_{\rm disk}/R_{\rm disk}\)\\
\hline
\textit{Ahr$^*$} & 26.9 & 0.25 & 12.0 & \(\num{1.4e7}\) & 20 & \(\num{5.7e+10}\) & 7.2 & 12.6 & 4.9 & 12 & 600 & 100-200 & 0.17-0.33\\
\textit{A  } & 26.9 & 0.25 & 12.0 & \(\num{1.4e7}\) & 18 & \(\num{ 3.6e+9}\) & 28.8 & 18.2 & 10.0 & 4 & 600 & 100-200 & 0.17-0.33\\
\textit{B$^*$} & 131 & 0.5 & 309 & \(\num{3.0e6}\) & 18 & \(\num{ 7.7e+8}\) & 60 & >601 & 75.1 & 9 & 6000 & 3000-8000 & 0.5-1.3\\
\textit{C  } &  &  & 32.6 & \(\num{1.4e7}\) & 18 & \(\num{ 3.6e+9}\) & 29.0 & 43.1 & 42.7 & 4 & 200 & 100 & 0.5\\
\textit{D  } &  &  & 8.4 & \(\num{6.5e6}\) & 18 & \(\num{ 1.7e+9}\) & 42.5 & 14.7 & 14.7 & 1 & 600 & 200 & 0.33\\
\textit{E  } &  &  & 14.5 & \(\num{6.5e6}\) & 18 & \(\num{ 1.7e+9}\) & 42.5 & 21.5 & 9.0 & 6 & 300-800 & 200 & 0.25-0.67\\
\textit{F  } &  &  & 7.3 & \(\num{3.0e6}\) & 18 & \(\num{ 7.7e+8}\) & 60 & 27 & 21.7 & 6 & 600-1000 & 300 & 0.3-0.5\\
\bottomrule
\end{tabular}
\end{table*}

We conduct ``zoom-in'' radiation-MHD simulations of collapsing molecular 
clouds resolving individual prestellar cores. We focus on the fragmentation of 
the cores and the formation of protostellar disks. We have conducted a set of 
6 simulations on a large range of sink masses selected from several parental 
GMCs. We summarize the key parameters and results in Table~\ref{tab:init}.

We perform simulations using the grid-based adaptive mesh refinement (AMR) MHD code \ramses{} \citep{Teyssier2002, Fromang2006}. Radiation transfer is modeled using a moment-based method with the M1 closure relation for the Eddington tensor \citep{Rosdahl2013}. 
The ionizing photons emitted from stars interact with neutral gas and we keep track of the ionization chemistry of hydrogen and helium, but we do not include the chemical evolution of the molecular phase. Heating from photoionization and cooling from hydrogen, helium, metals, and dust grains are implemented (see \citealt{Geen2017} for details). Cooling below 10~K is shut down to keep the temperature floor at 10 K. 
We carry out simulations starting from a subset of simulations presented in \cite{He2019} and zooming into prestellar cores to resolve their fragmentation and disk formation.
We refer to the original paper for details of the method and key results of these baseline simulations. Here we briefly summarize the AMR technique and sink particle recipe before getting into the zoom-in method.

The baseline simulations \citep{He2019} are started from idealized spherical isothermal clouds in hydrostatic equilibrium surrounded by a low-density shell, in which gravity is nearly balanced by turbulent motions ($\alpha\equiv K/|W|=0.4$). We let the GMC evolve for three free-fall times with gravity reduced by 1/2, to allow the turbulence to develop. Then, gravity is fully turned on and the cloud undergoes filamentary collapse and fragments to form sink particles in dense regions that represent singular stars or small clusters of stars, as described in \cite{Bleuler2014}. 
Adaptive mesh refinement is applied to the whole domain to make sure at any time and any location the local Jeans length, $L_J = c_s \sqrt{\pi/(G \rho)}$, is resolved by at least 10 grid points. 
The maximum refinement level $l_{\rm max}$ is set to 14, reaching a minimum grid size $\Delta x_{\rm min}$ that is $1/2^{14}$ of the box size, or around 200 - 1600~AU. 
When the density reaches the critical density, $n_{\rm sink}$, defined such that the corresponding Jeans length equals to 5$\times$ the grid size at the maximum refinement level $l_{\rm max}$, a sink particle is placed to prevent the increase of the gas density beyond $n_{\rm sink}$ at which a local Jeans length is not fully resolved.
The critical density for sink formation is set to \( n_{\rm sink} = \num{2.16e10} \ \pcc (\Delta x_{\rm min}/10~{\rm AU})^{-2} \), motivated by the criterion that a Jeans length must be resolved by no less than 5 grids. The corresponding Jeans mass is $0.0055 \ \msun{} (\Delta x_{\rm min}/10~{\rm AU})$. 
In the baseline simulations in \cite{He2019} used as initial conditions, $n_{\rm sink}$ ranges from $10^6 \ {\rm cm}^{-3}$ to $\num{6e7} \ {\rm cm}^{-3}$, while in the zoom-in simulations presented here we reach densities three orders of magnitude higher: $n_{\rm sink}$ ranges from $10^9 \ {\rm cm}^{-3}$ to $\num{5e10} \ {\rm cm}^{-3}$.
Accretion onto the sink particles is modeled using a threshold method such that $75 \%$ of the mass above $n_{\rm sink}$ is transferred to the sink particle in each time step \citep{Bleuler2014}. 
Ionizing photos are emitted from the sink particles to ionize and heat the gas, dispersing the cloud and quenching star formation. These sink particles in the baseline simulations are shown to resemble prestellar cores observed in local star-forming regions.
\begin{figure*}
  \centering
  \includegraphics[width=\textwidth]{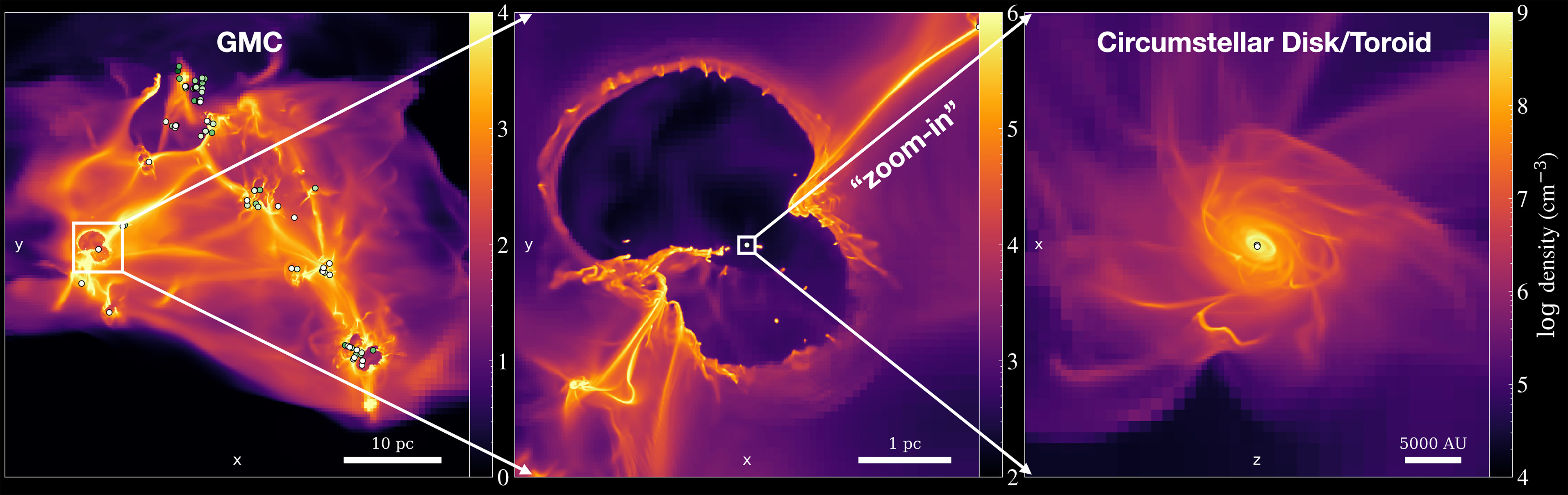}
  \caption{Outline of the simulation method. We pick a zoom region which is a box at the formation location of a sink particle. Inside the box, we allow a high level of refinement to reach high density before star formation. The whole simulation spans a dynamic range up to $2^{20}\approx 10^6$ in linear scale, or a volumetric dynamic range of 18 orders of magnitude.
  \label{fig:zoom}}
\end{figure*}

In the ``zoom-in'' simulations presented in this paper, we start a simulation from a snapshot of a baseline AMR simulation right before the formation of a sink particle (Figure~\ref{fig:zoom}). 
We define a ``zoom'' region, about 0.5 to 1 pc in size, where the sink particle is about to form and allow a higher $l_{\rm max}$ only within this region.
To reach the best possible resolution, we use a nested refinement structure where $l_{\rm max}$ increases as it gets closer to the domain center. In the simulation with the best resolution (Core \A{}, as we will introduce later), $l_{\rm max}$ at the center of the ``zoom-in'' region is set to 20, reaching a dynamic range of $2^{20} \approx 10^6$, or 18 orders of magnitude in volume. 
The corresponding critical density of sink formation $\rho_{\rm sink}$ is $\num{1.4e-13} \ {\rm g}~{\rm cm}^{-3}$, approaching the density at which the core transitions from isothermal to adiabatic \citep{Masunaga1998,Masunaga2000}.
The corresponding spatial resolution and other parameters for all simulations are listed in Table~\ref{tab:init}.

Hydrogen and helium-ionizing photons are emitted from sinks and heat the gas.
The hydrogen-ionizing luminosity of a star with mass $M$ is given by: 
\begin{equation}
S = 9.63 \times 10^{48}\ {\rm s}^{-1}\left(\frac{M}{27.28 M_\odot}\right)^{1.86}.
\end{equation}
This is the same as the fits given in \cite{Vacca1996} for high-mass stars ($\gtrsim 30 \ \msun$). However, \cite{Vacca1996}  fits are described by a broken power law. Hence, we are overestimating the ionizing radiation emitted by stars with masses smaller than $\sim 10\ \msun{}$.
The excess of ionizing photons from low/intermediate-mass stars is used to compensate for the lack of protostellar feedback. This recipe is proved effective in reproducing the IMF at the high-mass end in our simulations \citep{He2019}\footnote{In \citep{He2019} we adopted a broken power-law as in \cite{Vacca1996}, but due to a bug in the code the change in power-law slope at the low-mass end did happen at much lower masses. When we found and fixed the bug, we observed that the IMF at the high-mass end was not reproduced as well as before. We interpreted this result as the need for stronger feedback from the low-mass end, perhaps produced by protostellar outflows. We plan on implementing more realistic feedback from low-mass stars in future work. }.  
Further simulations with more realistic feedback mechanisms are left for future work. 
\begin{figure*}
  \centering
  \def\thisw{0.22}
  \begin{tabular}{cc|cc}
  \multicolumn{2}{c}{Core {\it Ahr}} & \multicolumn{2}{c}{Core {\it B}} \\
	earlier & later & earlier & later \\
 \includegraphics[width=\thisw\textwidth]{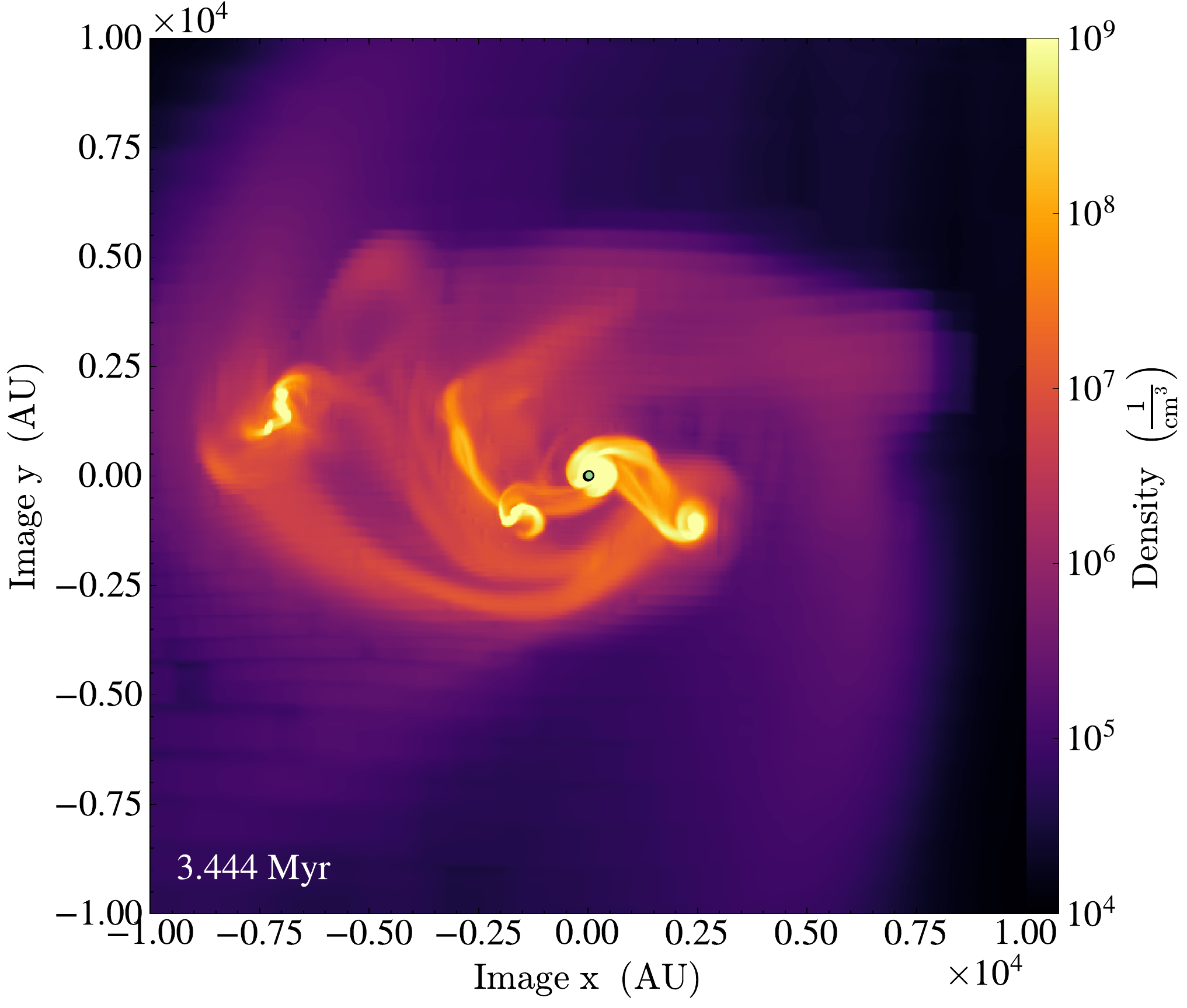} &
 \includegraphics[width=\thisw\textwidth]{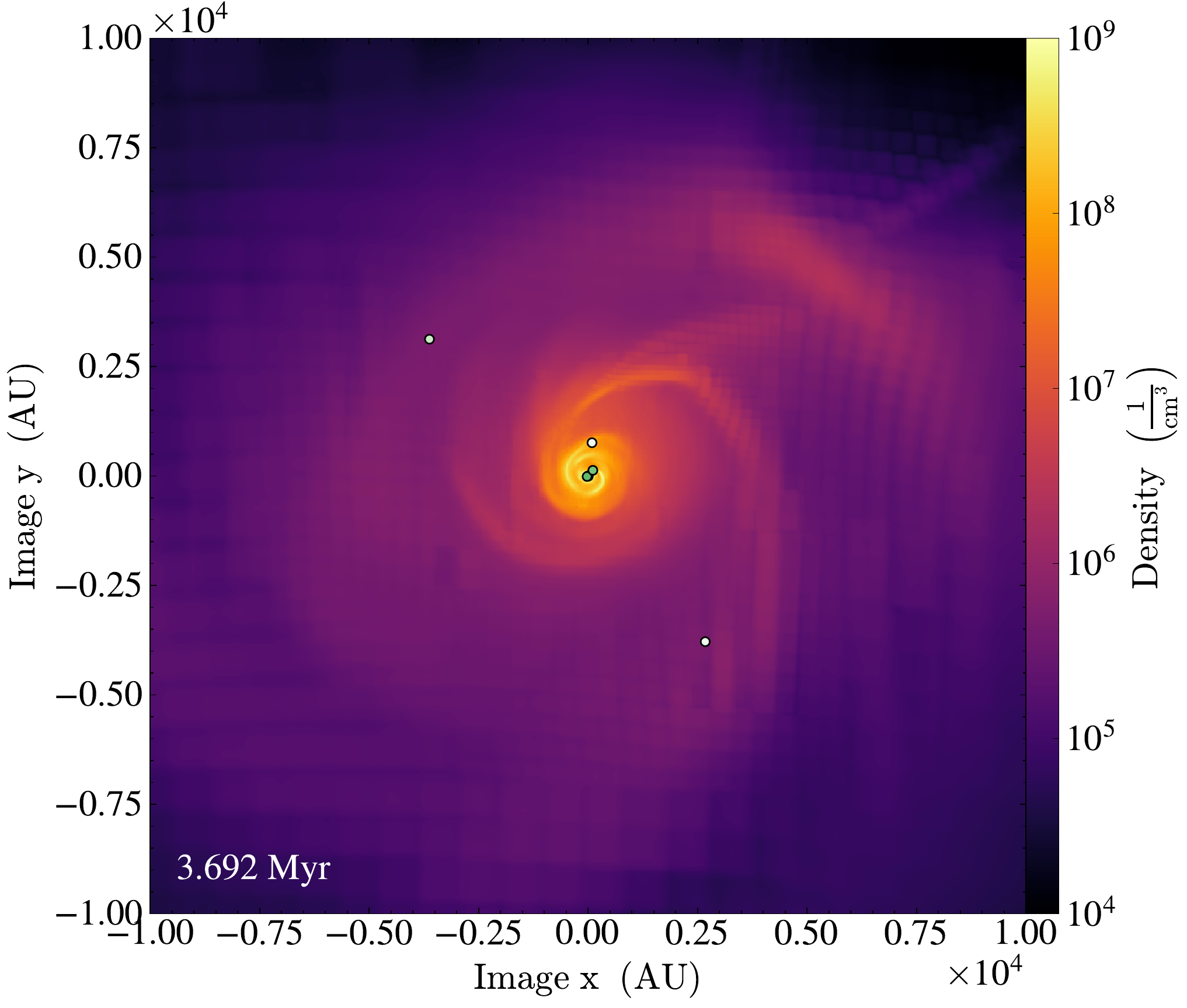}&
  \includegraphics[width=\thisw\textwidth]{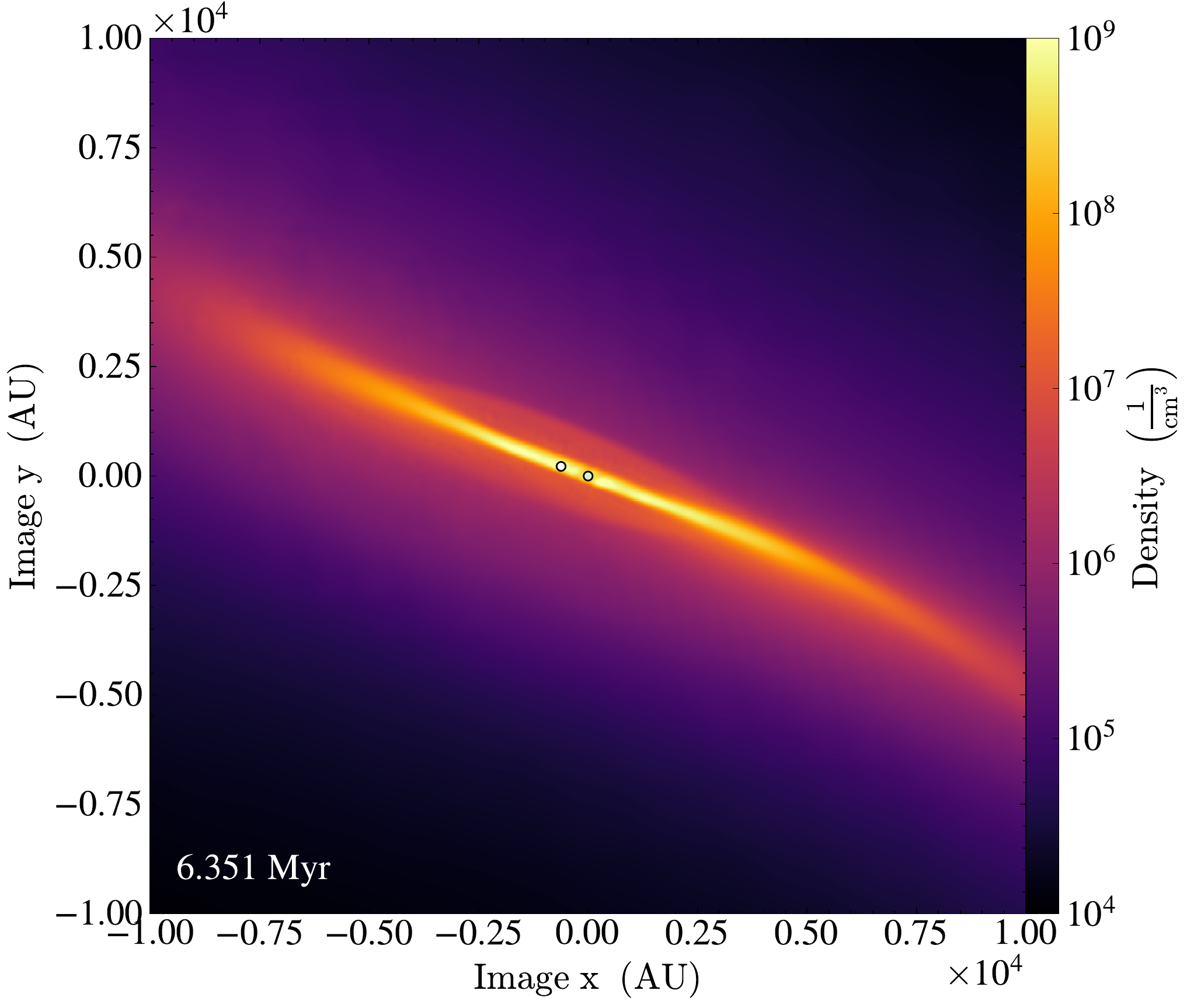}&
  \includegraphics[width=\thisw\textwidth]{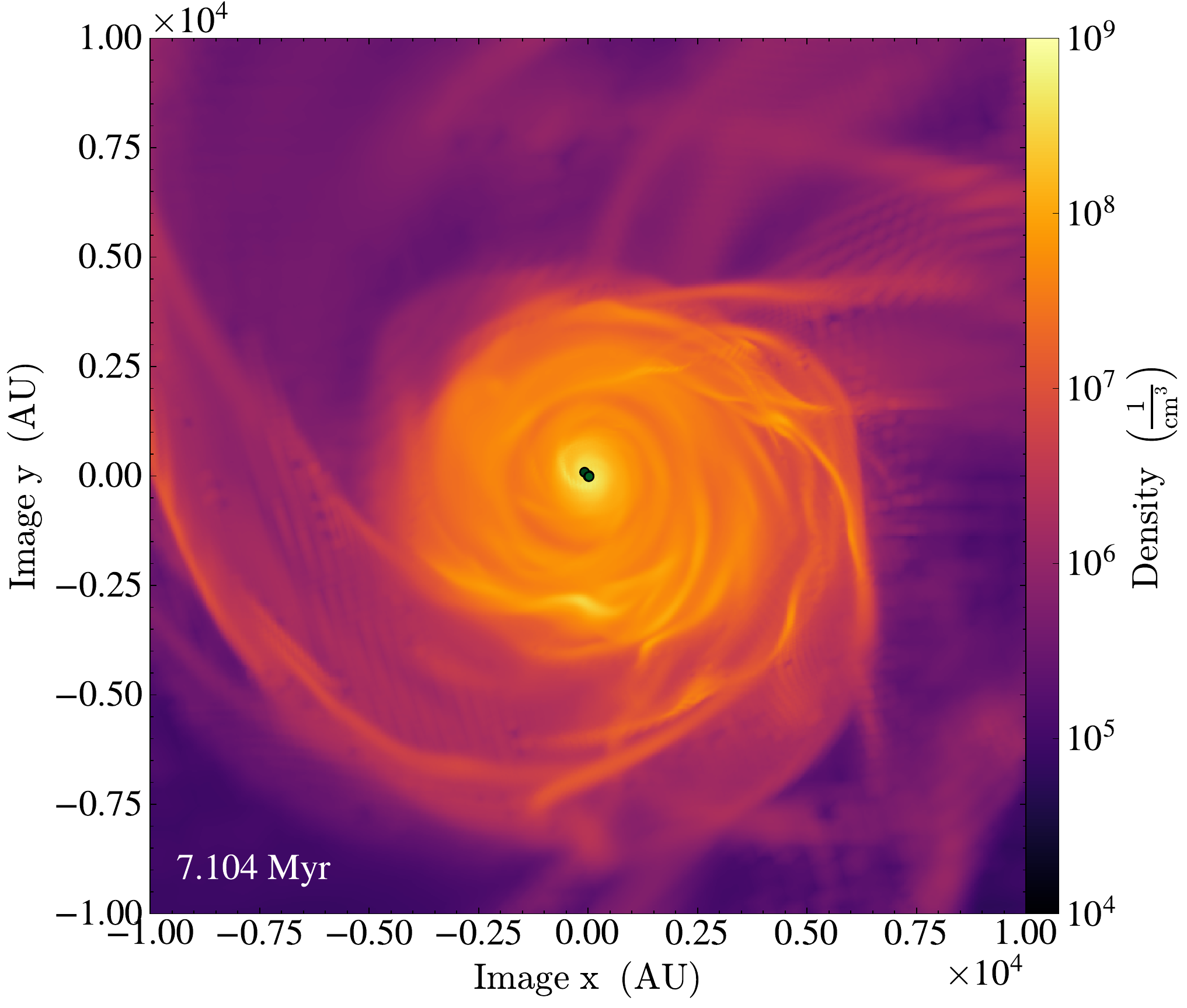}\\
  \multicolumn{2}{c}{Core {\it C}} & \multicolumn{2}{c}{Core {\it D}} \\
  \includegraphics[width=\thisw\textwidth]{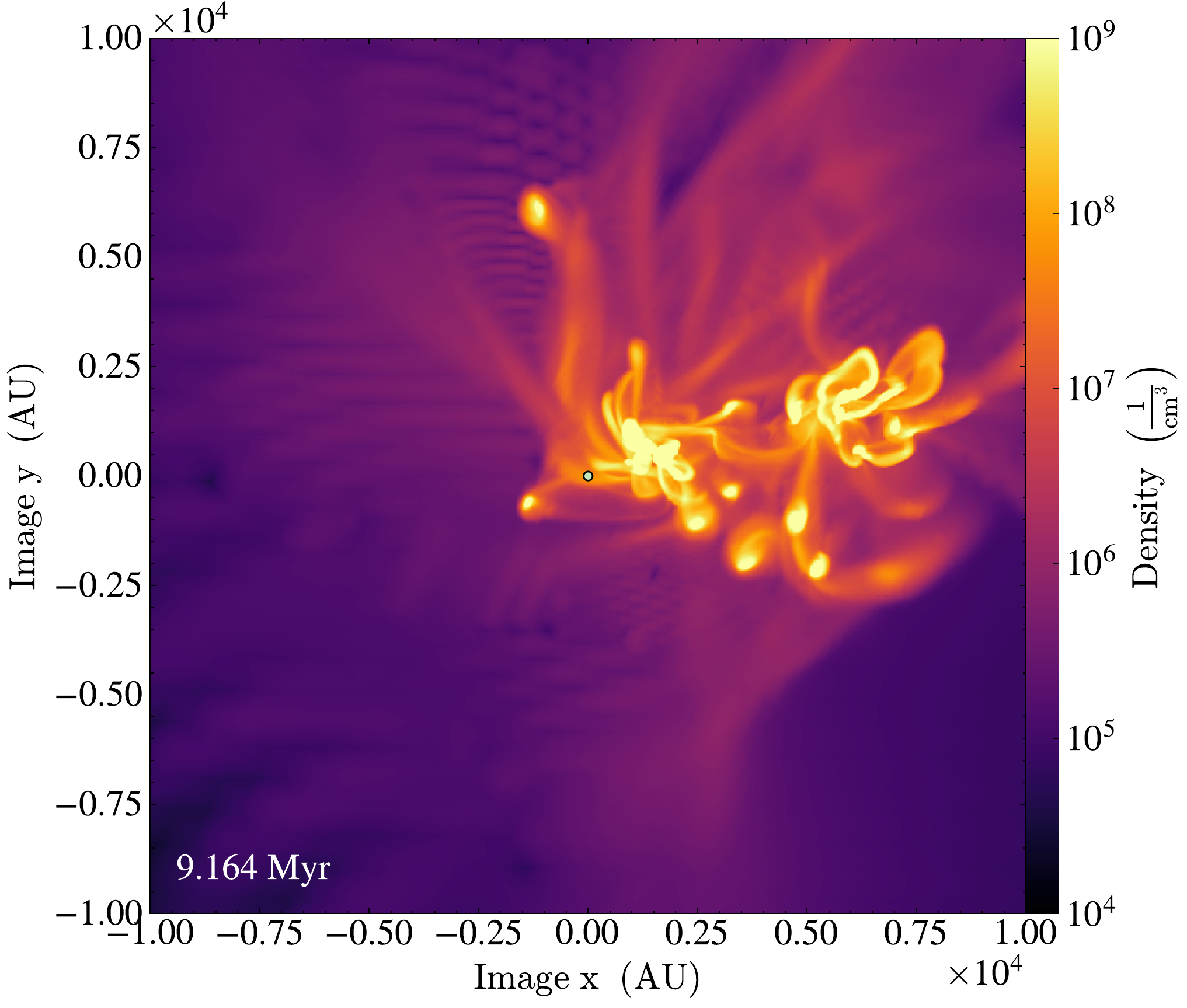}&
  \includegraphics[width=\thisw\textwidth]{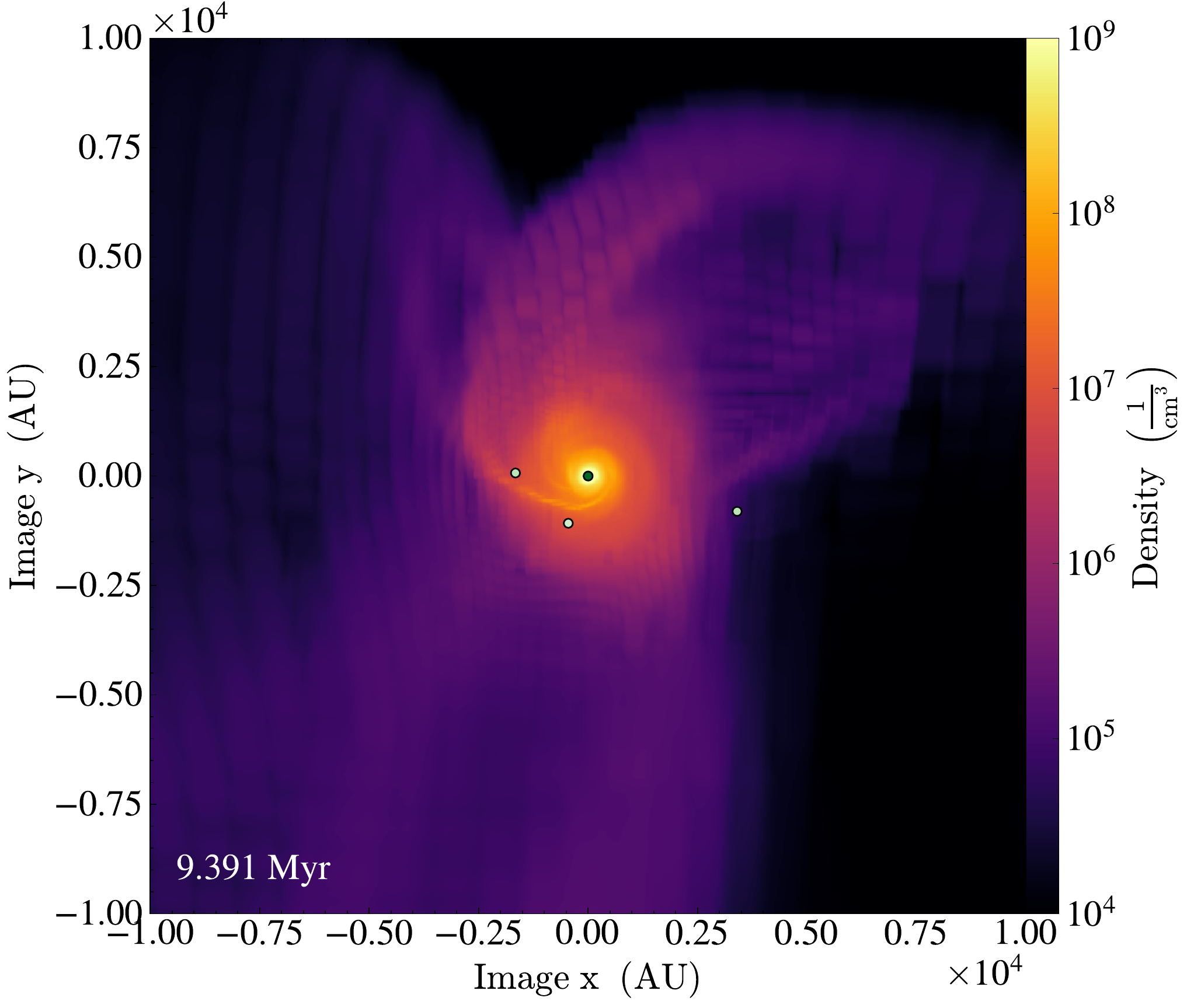}&
  \includegraphics[width=\thisw\textwidth]{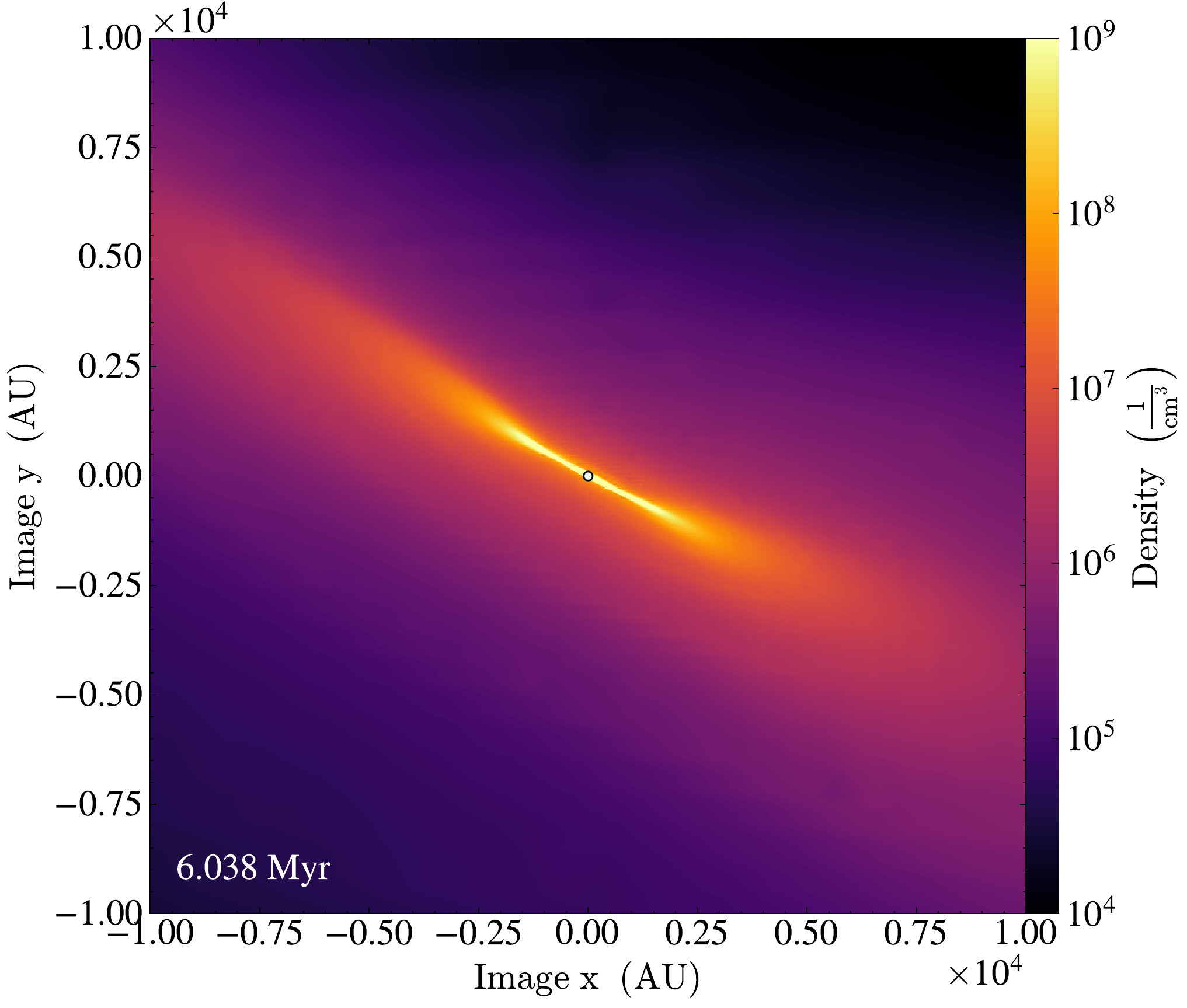}&
  \includegraphics[width=\thisw\textwidth]{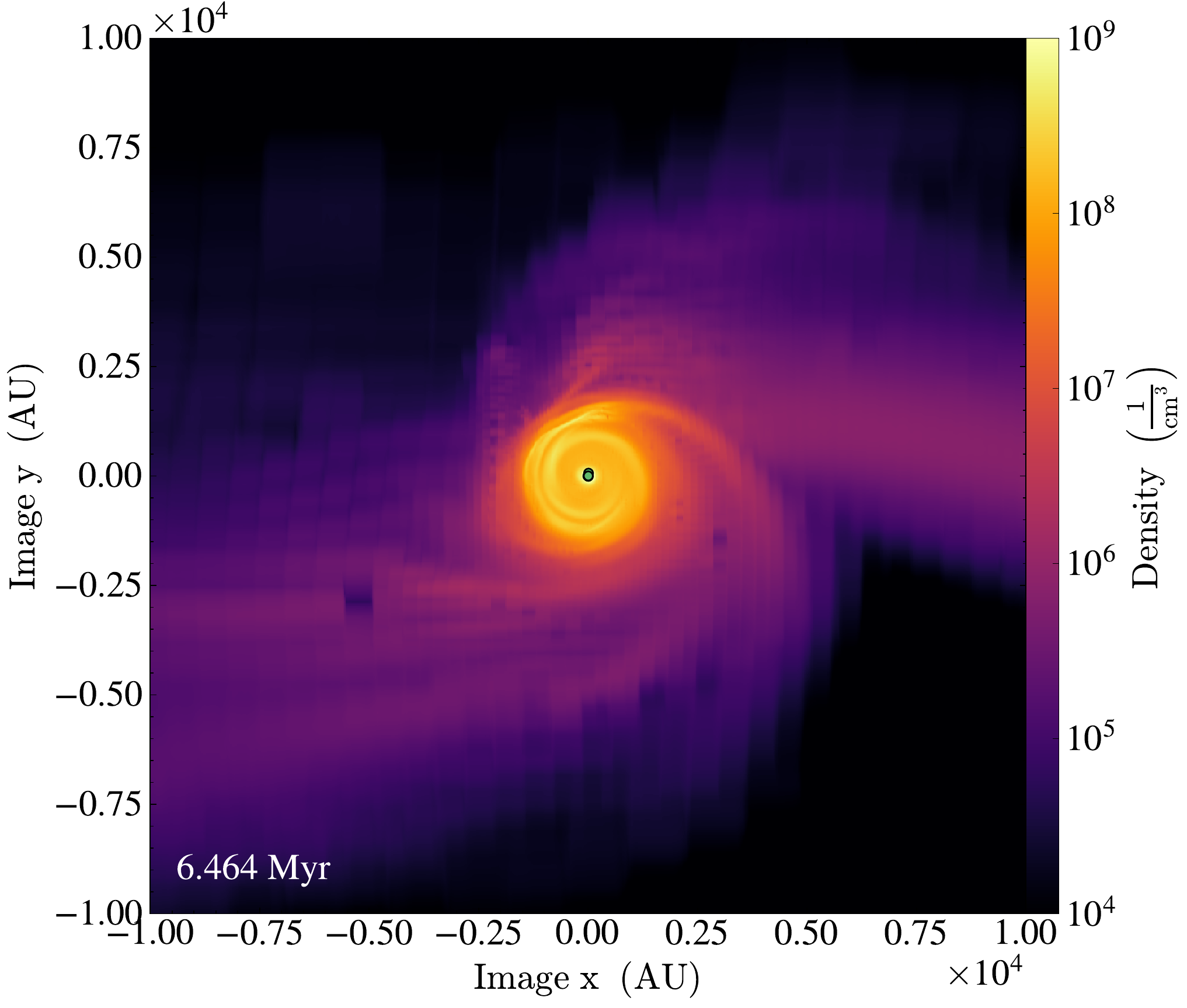}\\
  \multicolumn{2}{c}{Core {\it E}} & \multicolumn{2}{c}{Core {\it F}} \\
  \includegraphics[width=\thisw\textwidth]{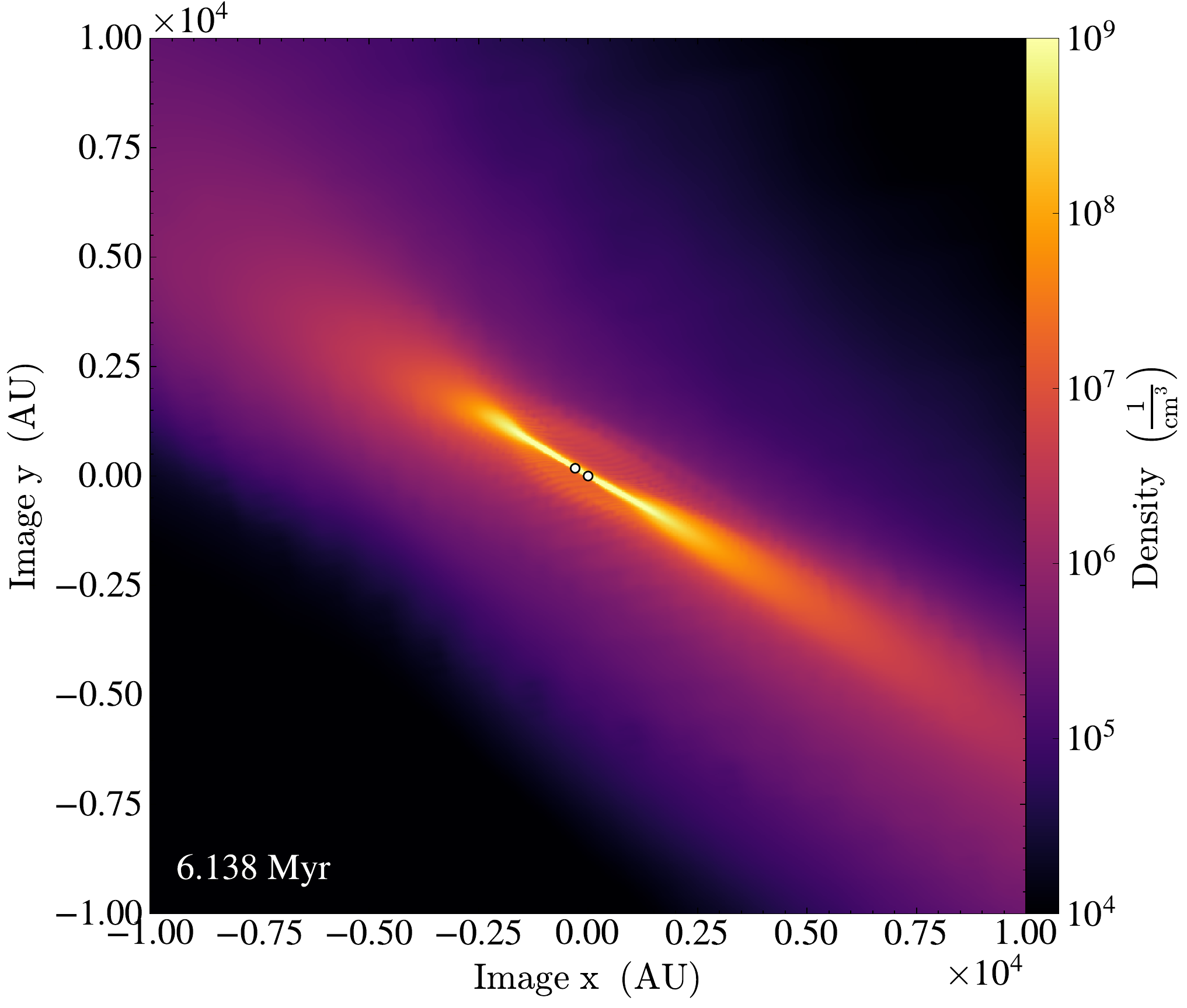}&
  \includegraphics[width=\thisw\textwidth]{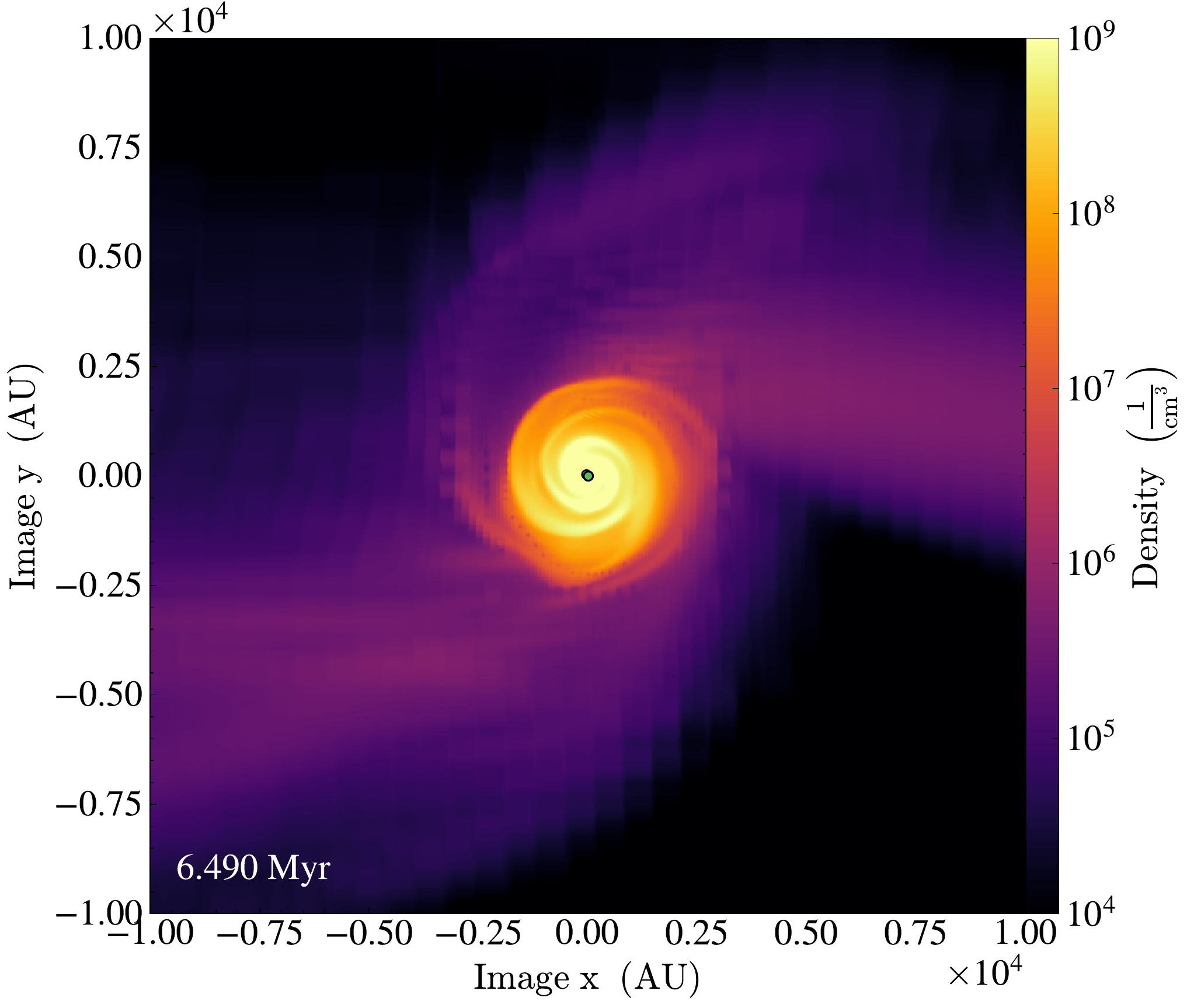}&
  \includegraphics[width=\thisw\textwidth]{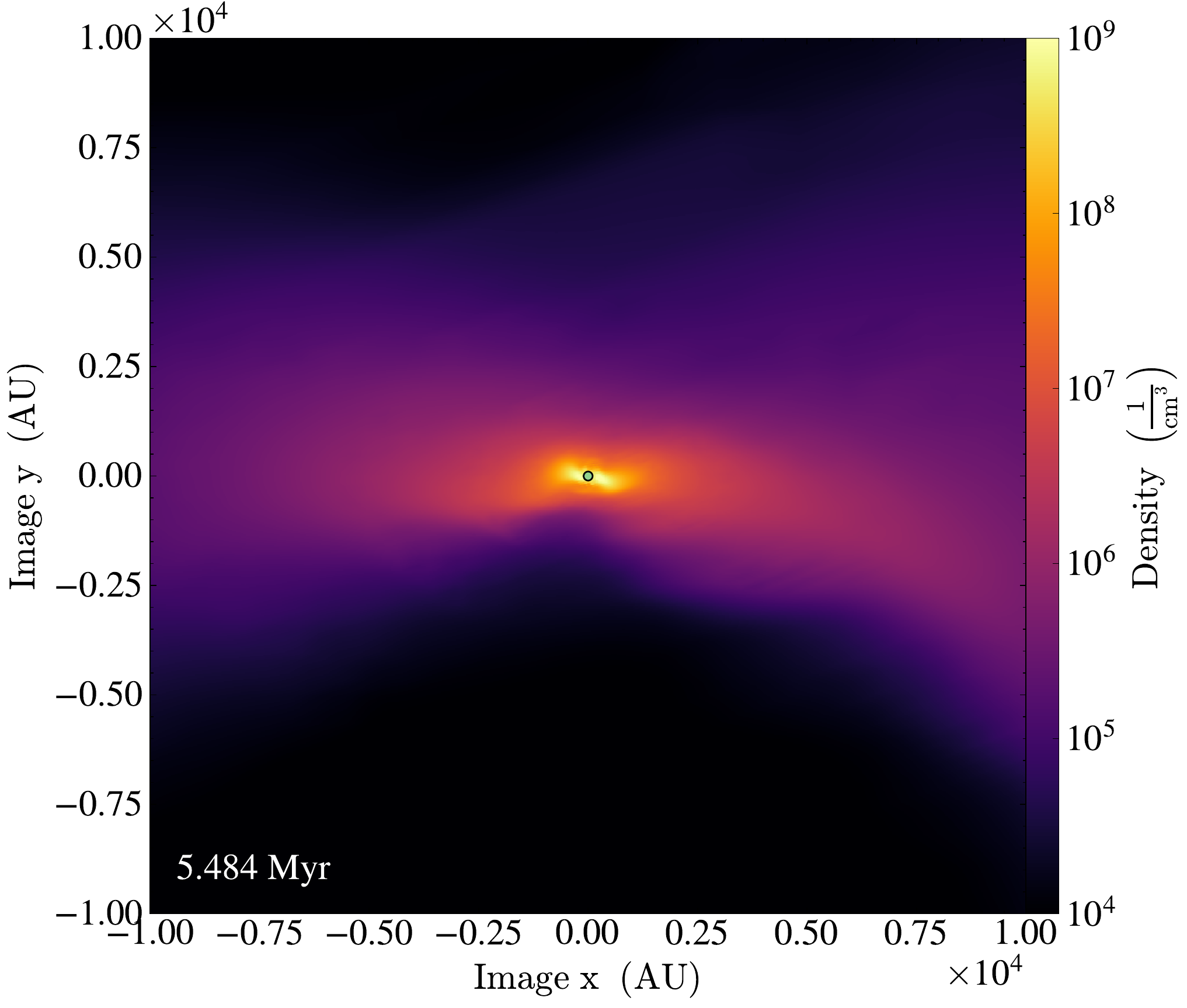}&
  \includegraphics[width=\thisw\textwidth]{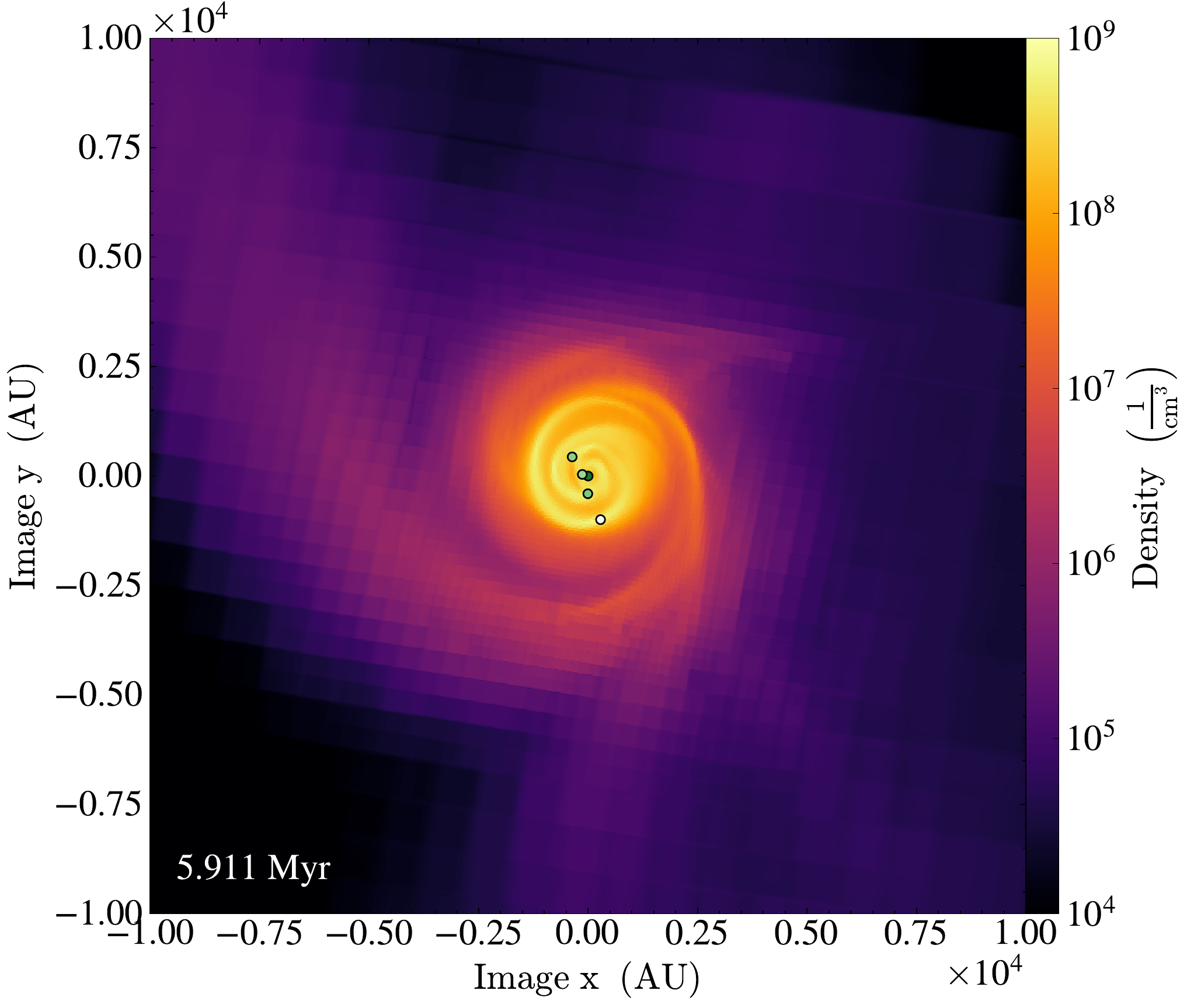} \\
	\end{tabular}
	\caption{Gallery of the collapsing prestellar cores from a set of six simulations presented in this work.
	  The colors are projected density of the gas from a direction parallel to the angular momentum of the gas.
	  The first snapshot of each core is when the first sink particle forms, and the second snapshot is when the disk comes to a quiescent phase.
	  Sink particles representing single stars are plotted on top as colored circles with darker shades indicating higher masses.
	  The cores display two distinct kinds of morphology. Cores \A{} and \textit{C} shows the spherical mode and Cores \B{}, \textit{D}, \textit{E}, and \textit{F} show the filamentary mode. 
	}
\label{fig:gallery}
\end{figure*}

We set a magnetic field with moderate strength in the $x$-direction threading the isothermal cloud in the initial conditions. 
The GMC starts from an idealized sphere with an isothermal core surrounded by a low-density shell that extends twice the radius.  
The magnetic intensity is about $10 - 25 \mu$G at a density of $10^3 \pcc{}$ and the mass-to-flux ratio is $\mu \approx 5$ averaged over the whole cloud. The value of $\mu$ in the core is higher ($\approx 8$) due to the fact that the mass is more concentrated in the core but the magnetic field is more evenly distributed. After a period of relaxation to let the turbulence develop, the initial mean density decreases slightly and $\mu$ settles at 3 -- 4, averaged over the whole cloud.
Instead of the traditional definition of the mass-to-flux ratio, $\mu \equiv M / M_\Phi$, where $M_{\Phi}$ is the magnetic critical mass, we adopt a definition that takes into account the non-homogeneity of the density and magnetic field, $\mu = \sqrt{|{\cal W}| / {\cal B}}$, where ${\cal W}$ is the gravitational binding energy and ${\cal B}$ is the magnetic energy. The two definitions are equivalent for a uniform spherical cloud with uniform magnetic field. We will explain further the derivation and significance of the $\mu$ parameter in \S~\ref{sec:2}.

The motion of the sink particles is determined by combining direct N-body integration (using the leapfrog method) between the sinks and between the sinks and the gas based on the particle mesh method. A softening length of $2 \Delta x_{\rm min}$ is set to avoid singularities.

\section{Results. I. Turbulent massive disks}
\label{sec:2}

The main result of this study is the formation of rotationally supported thick disks, characterized by supersonic turbulence and a moderately strong magnetic field.
Figure~\ref{fig:gallery} shows snapshots for a grid of simulations illustrating the evolution of the prestellar cores inside turbulent GMCs. In all six simulated prestellar cores with various initial masses and morphologies, quasi-Keplerian disks form around the central proto-star/binary.
The morphologies of the disks are very similar to those found in many previous studies \citep{Bate2003, Goodwin2004a, Hennebelle2008a}: spiral arms that could potentially transport angular momentum are prominent features. 
We summarize the key properties of the disks in Table~\ref{tab:init}.

In the classical picture, the gravitational collapse of a magnetized prestellar core occurs from an initially spherical structure that tends to flatten along magnetic field lines, leading to the formation of an oblate pseudo-disk \citep{Galli1993}. These pseudo disks are disk-like but are not supported by centrifugal force and may transition into centrifugally supported disks \citep[][]{Galli1993, Joos2012}.
The collapse of turbulent cores in our simulations spans a wide variety of morphologies that are far different from idealized spherical collapse, as shown in Figure~\ref{fig:gallery}.

\begin{figure}
  \centering
  \includegraphics[width=3in]{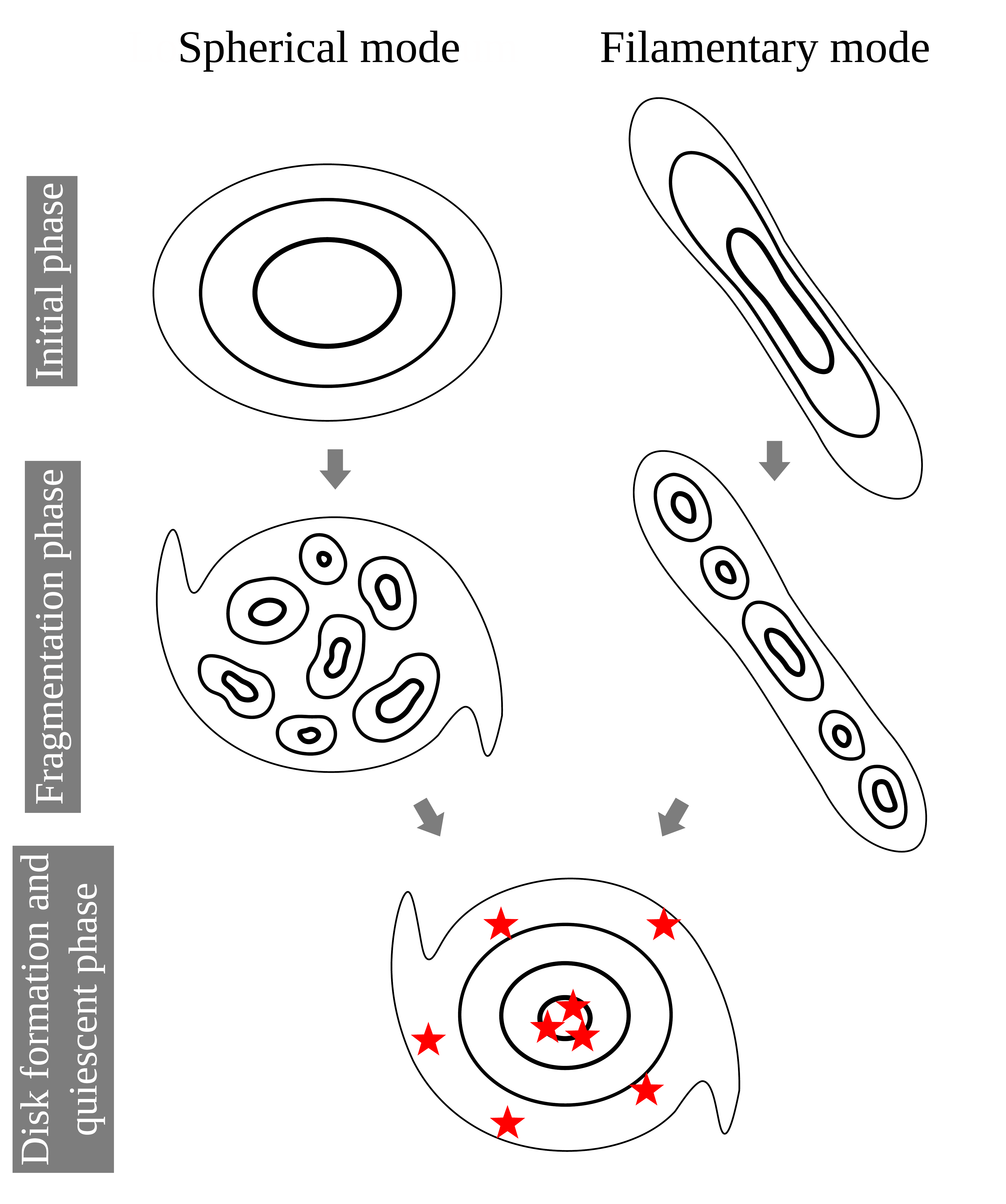}
  \caption{Schematic of the two modes of prestellar core fragmentation and disk formation. In the spherical mode, the core begins with a spherical/oblate shape and fragmentation occurs uniformly inside the core. In the filamentary mode, or mode B, the core starts from a long, thin tube and instabilities occur on the arms of the filament. In both scenarios, a centrifugal disk forms at the core center a few hundred of kiloyears after the initial phase. The disks have sizes up to several thousand AU. The central star/binary and secondary stars that form from the core fragments orbit around the disk center, entering a quiescent phase. %
    \label{fig:schematic}}
\end{figure}

We identify two main geometries of core formation and fragmentation: ``spherical'' and ``filamentary'', sketched in Figure~\ref{fig:schematic}. In the first mode, the collapse is quasi-spherical and fragmentation occurs during the collapse at random locations inside the spherical core.
Examples of this geometry are cores \A{} and {\it C} in Figure~\ref{fig:gallery}. 
In the second geometry, the core collapses into a thin elongated filament, which breaks into aligned quasi-spherical fragments. This fragmentation mode is represented by cores \B{}, {\it D}, {\it E}, and {\it F}. 
For both geometries, the collapse eventually evolves into the formation of a geometrically thick massive disk, in which the pre-existing fragments are collected. The orbiting fragments may lead to the formation of low-mass stars that are either ejected or spiral toward the center.

As the gas collapses and the density becomes higher than $\sim 10^6-10^7~\pcc$, almost inevitably conservation of angular momentum produces quasi-Keplerian protostellar disks. Almost all the gas with density above $10^7 \ \pcc$ is in disks rather than turbulent quasi-spherical cores (see Figure~\ref{fig:gallery}). This is in contrast to the hierarchical structure of the molecular cloud at larger scales and lower densities, better described as clumps composed of more compact mini-clumps \citep[see,][]{He2019}.
The cores in our zoom-in simulations have masses between $\sim 27$~M$_\odot$ and $\sim 130$~M$_\odot$, and the protostellar disks forming from their collapse are thick and supported in the vertical direction by magnetic pressure and turbulent pressure rather than thermal pressure. This is contrary to what is observed in simulations of standard lower mass protostellar disks around solar mass protostars, in which the disk scale height is determined by thermal pressure \citep[\eg,][]{AndreOliva2020}. We will discuss this result in detail in \S{}~\ref{sec:turb}.

Apart from a central star/binary that eventually grows to have a large fraction of the total mass of the core, multiple secondary stars form at the outskirt of the pseudo-disk. 
These stars form from pre-existing fragments formed uniformly inside a quasi-spherical turbulent core or from the fragmentation of a collapsing filament.
Shortly after their formation, some of these fragments spiral into the center of the disk owning to either dynamic friction or torques exerted by accretion or gravity from the asymmetric core, and some are ejected from the system. 
In the last phase of the evolution, between 1 to 12 stars form in the core. This is consistent with previous numerical studies \citep{Bate1997, Goodwin2004a, Goodwin2004}.
Nonetheless, these small N-body systems are unlikely to be observed because they evaporate into the field on timescales shorter than the lifetime of the disks. It remains to be understood whether the stars produced by the dissolution of this small hierarchical N-body system retain their original binary fraction. 

In the rest of this section, we closely examine the properties of the centrifugal disks in cores \A{} and \B{}, demonstrating that the disk's scale-height is primarily determined by magnetic support and turbulent motions.

Let us first introduce some definitions useful to describe the stability of the cloud to fragmentation and the importance of the magnetic field.

The dynamical importance of the magnetic field in a cloud of mass $M$ is often parameterized in terms of the dimensionless ratio  $\mu \equiv M / M_\Phi$, where $M_{\Phi}$ is the magnetic critical mass:
the mass at which the pressure from the magnetic energy, ${\cal B}$, balances the gravitational binding energy, ${\cal W}$, of the cloud.
For a spherical cloud of uniform density and uniform magnetic intensity, ${\cal W}=-3GM^2/(5R)$ and ${\cal B}=B^2R^3 / 6 = \Phi_B^2/(6 \pi^2 R)$, where $\Phi_B \equiv \int B_{\perp} dS = \pi R^2B$ is the magnetic flux.
By setting $|{\cal W}| = {\cal B}$, we get the magnetic critical mass
\begin{equation}\label{eq:mphi}
M_{\Phi} = \sqrt{\frac{5}{2}} \frac{\Phi_B}{3 \pi G^{1/2}}. 
\end{equation}
Then,
\begin{equation}\label{eq:eWeB}
  \frac{|{W}|}{{\cal B}} = \frac{18 \pi^2}{5} \frac{GM^2}{\Phi_B^2} = \frac{M^2}{M_{\Phi}^2} = \mu^2. 
\end{equation}
In our analysis we adopt the equivalent definition $\mu = \sqrt{|{\cal W}| / {\cal B}}$ to calculate the mass-to-flux ratio in our simulations. The advantage is that it accounts for the inhomogeneity of the density and magnetic field distribution as well as the binding energy between the central stars and the disk. For a cloud or a core that is centrally concentrated, $\mu$ calculated using this definition is slightly higher than the classical definition because the binding energy is increased by the mass concentration in the center.

Simulations \citep{Joos2012} have shown that if $\mu \lesssim 1$, the cloud does not collapse due to the support of the magnetic field. If $1 \lesssim \mu \lesssim 5$ the cloud collapses but magnetic braking prevents the formation of a disk and the gas collapses quasi-spherically. If $\mu \gtrsim 5$ a quasi-Keplerian disk can form. We will show that our in our simulations pre-stellar cores have $\mu \sim 1-5$, however, we nevertheless observe the formation of quasi-Keplerian disks.

Neglecting the effect of the magnetic forces, we define the Toomre Q parameter 
\begin{equation}
Q = \frac{\Omega \ \sigma(v_z)}{\pi G \Sigma},
\end{equation}
where we have replaced $c_s$ with the vertical turbulent velocity $\sigma(v_z)$ since, as we will show later, the support of the disk against gravity in the vertical direction is dominated by turbulent motions rather than thermal pressure.

\subsection{Evolution of a 27 solar mass core (Core \A{})}
\label{sec:coreA}

\definecolor{gr}{RGB}{0,249,0}
\definecolor{bl}{RGB}{0,174,239}

\begin{figure*}
  \centering
  \includegraphics[width=\textwidth]{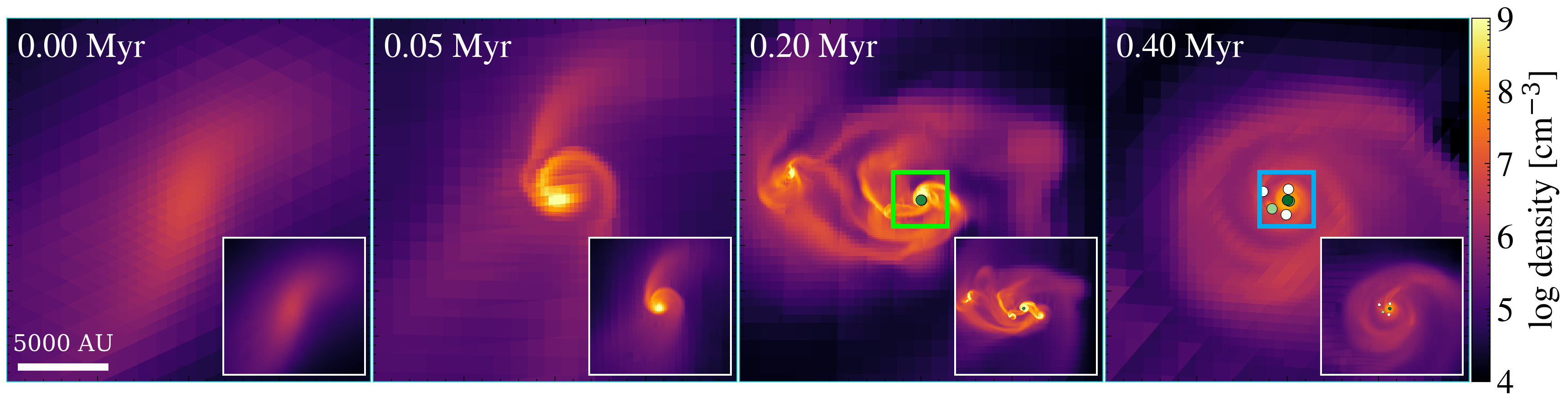}\\
  \includegraphics[width=\textwidth]{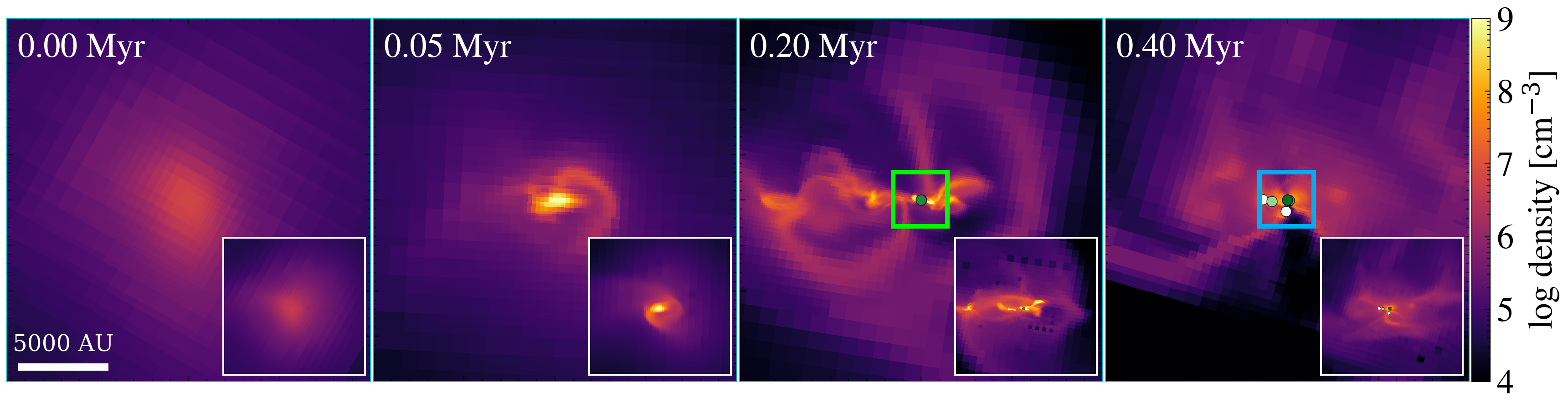}\\
  \hspace{-0.8cm}
  {\setlength{\fboxrule}{0pt}\setlength{\fboxsep}{2pt}
  \fcolorbox{white}{gr}{\includegraphics[width=1.4in]{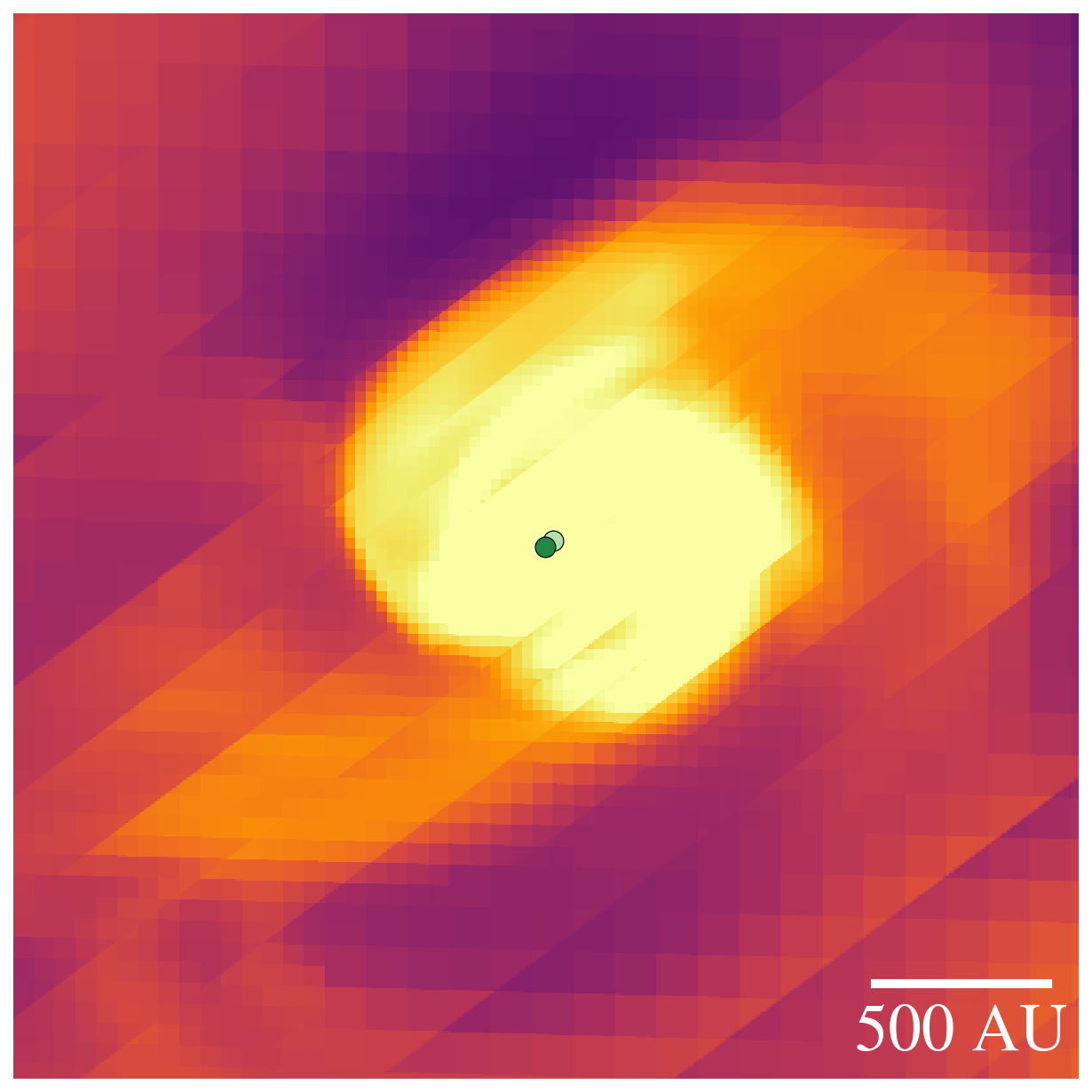}}
  \fcolorbox{white}{bl}{\includegraphics[width=1.4in]{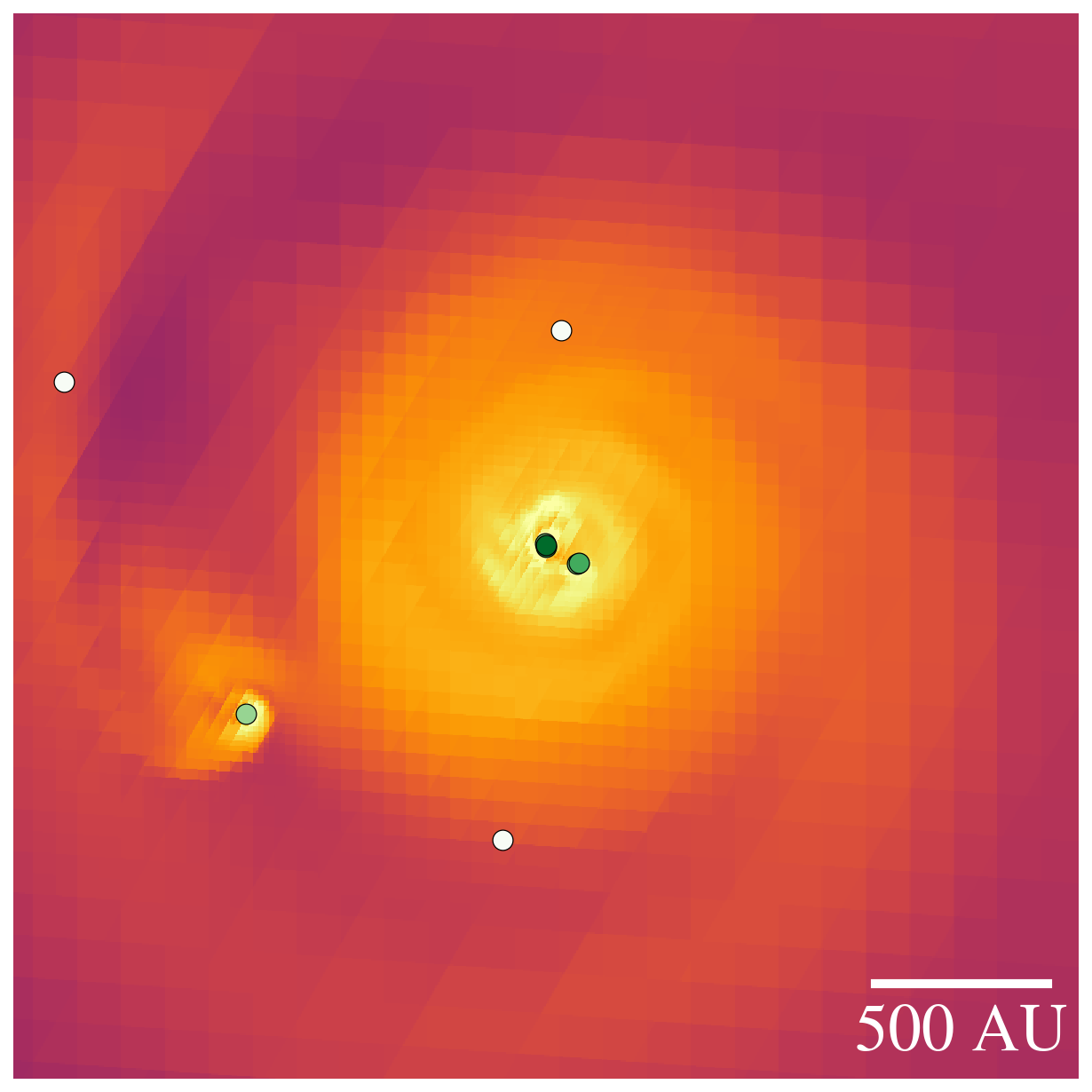}}
  \fcolorbox{white}{gr}{\includegraphics[width=1.72in]{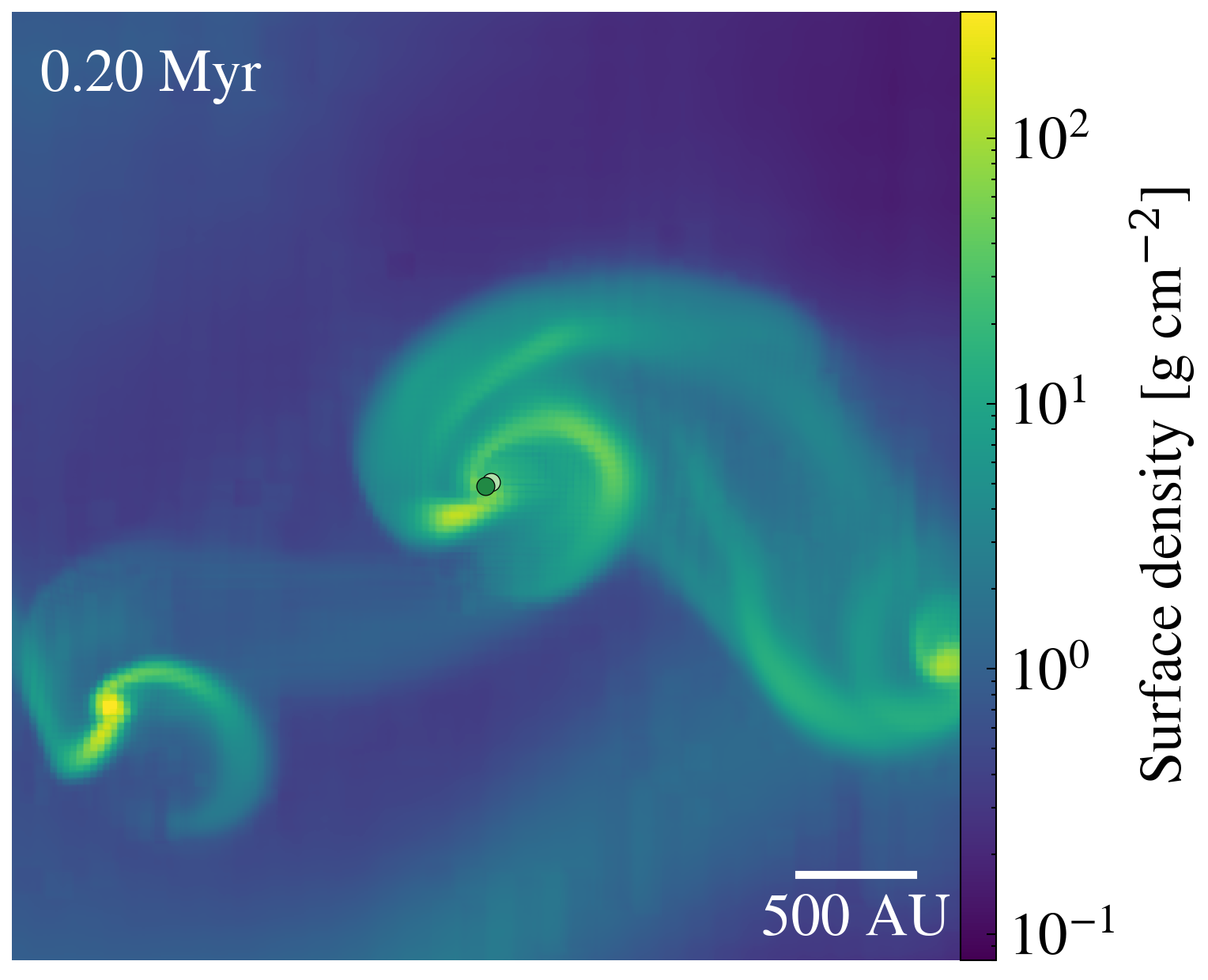}}
  \fcolorbox{white}{bl}{\includegraphics[width=1.72in]{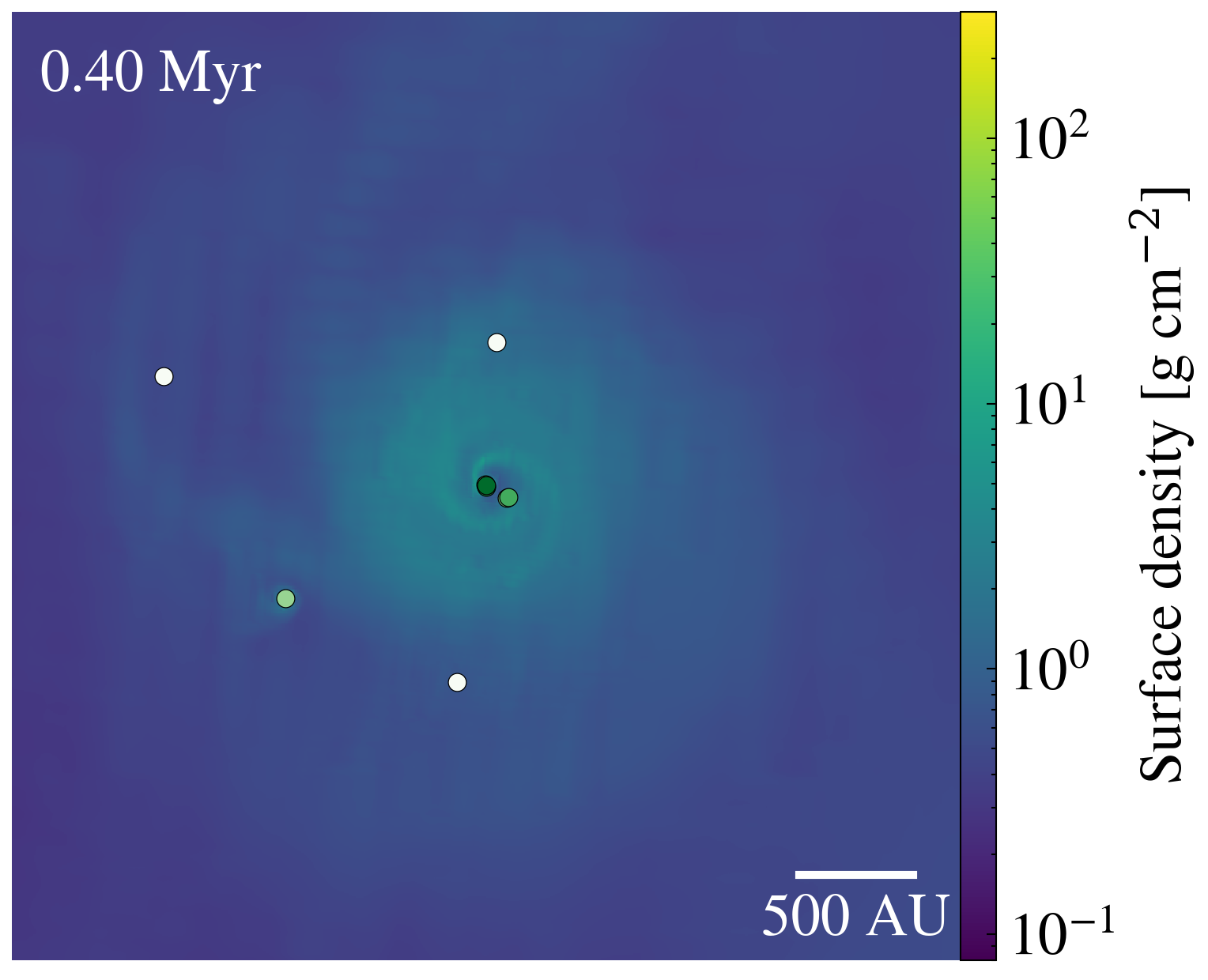}}}\\
  \hspace{-0.8cm}
  {\setlength{\fboxrule}{0pt}\setlength{\fboxsep}{2pt}
  \fcolorbox{white}{gr}{\includegraphics[width=1.4in]{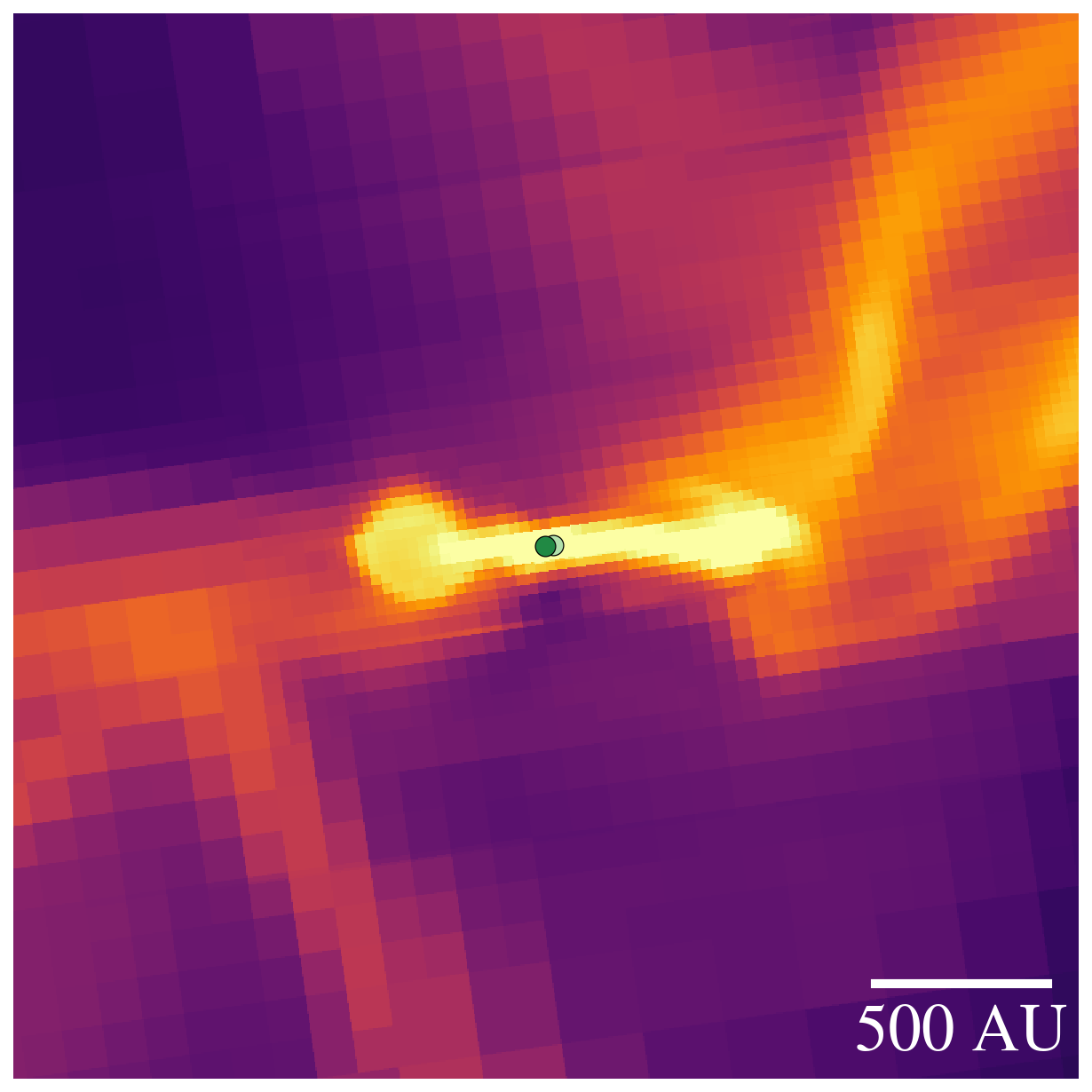}}
  \fcolorbox{white}{bl}{\includegraphics[width=1.4in]{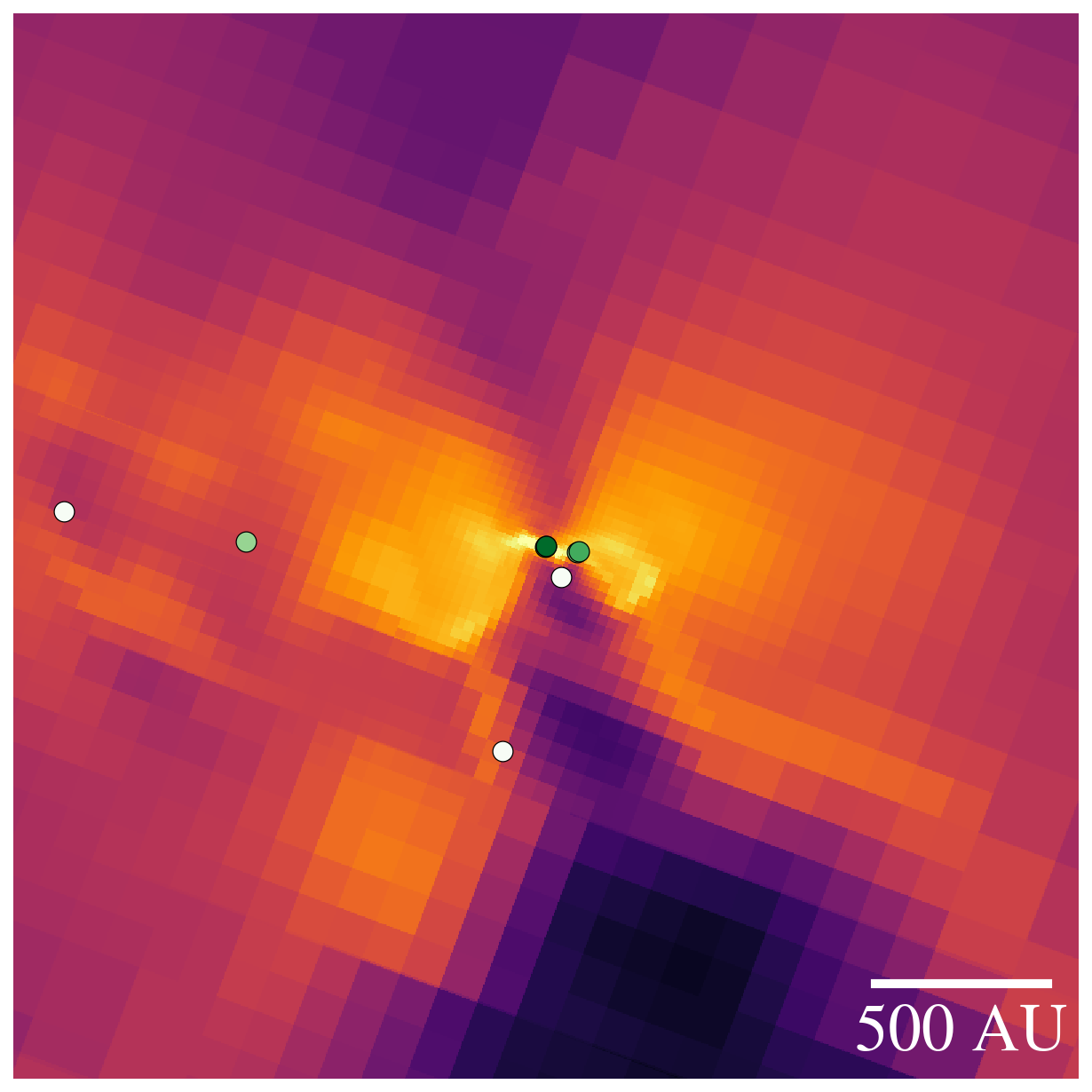}}
  \fcolorbox{white}{gr}{\includegraphics[width=1.72in]{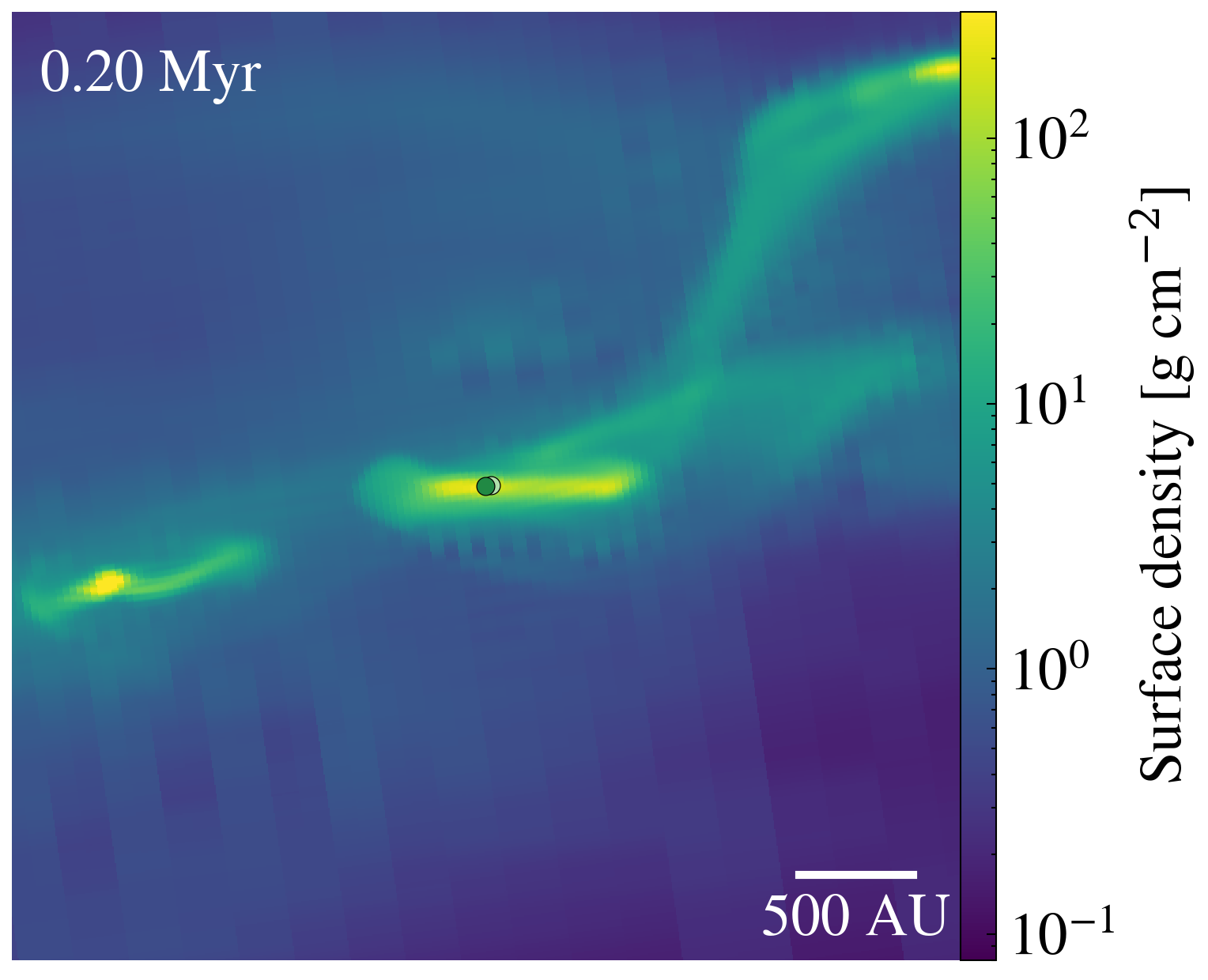}}
  \fcolorbox{white}{bl}{\includegraphics[width=1.72in]{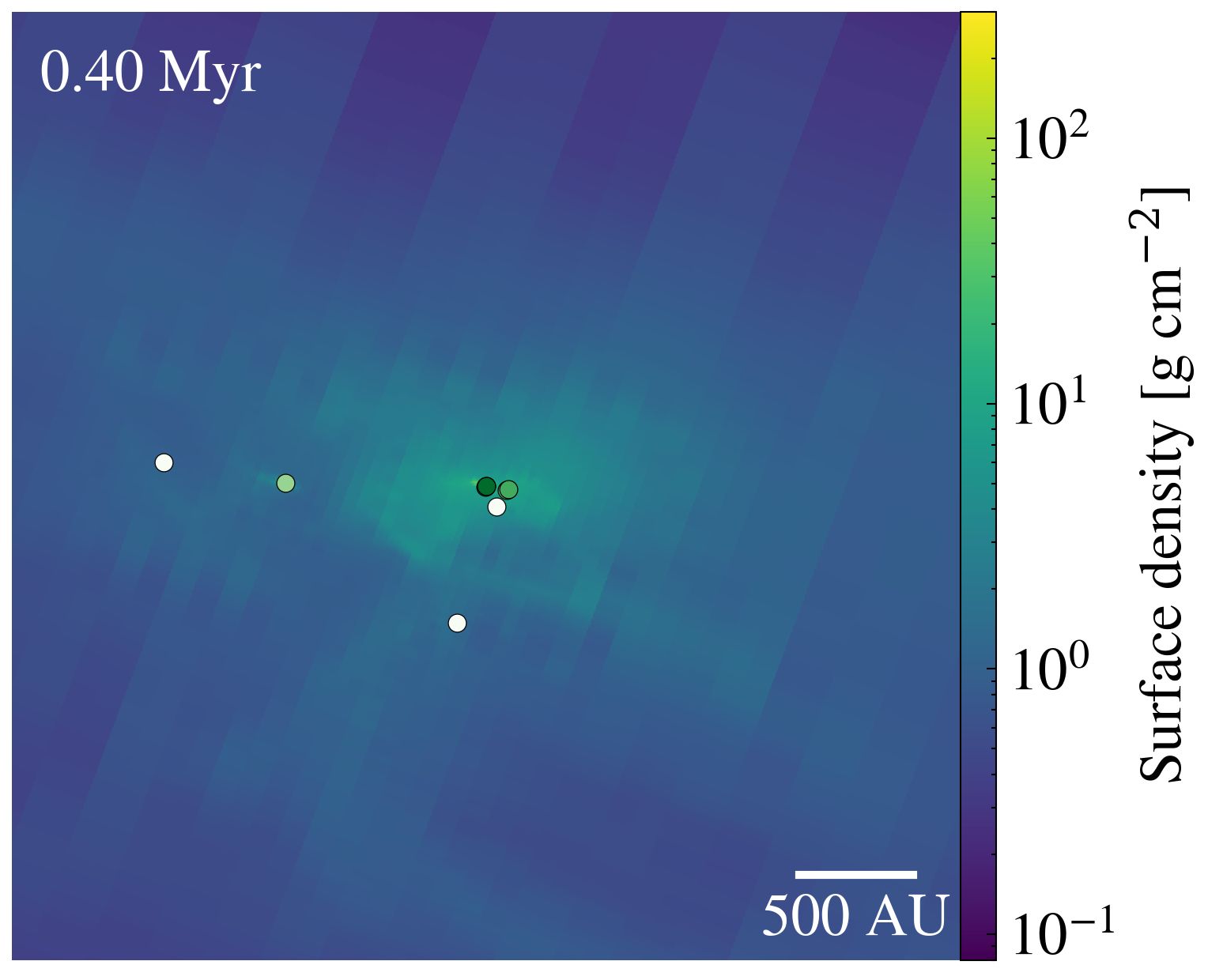}}}
  \caption{\label{fig:coreA} Density slices of the $\sim 27 \ \msun$ core (Core \A{}) at the mid-plane (first row) or cross-section (second row) at various ages. The images measure $\num{2e4}$ AU ($\sim 0.1$~pc) on a side, and the circles mark the positions of the star particles with darker shades indicating higher mass.
  Density-weighted projections of the gas density are inserted at the bottom right corners of each panel to better demonstrate the core structure.
  The first snapshot is chosen at the initial isothermal phase when the central density reaches $10^6 \ \pcc$. The second snapshot is picked when the instability occurs in the center. The third snapshot is picked when the first stars form at the center. The fourth snapshot is picked when the disk comes to a quiescent phase. 
  The last two rows displays a zoomed view of the disk which is characterized by volume density above $10^{8-9} \ \pcc$, or surface density above $\sim 10 \ {\rm g~cm^{-2}}$. 
  The core spherically collapses from the initial sphere into a disk whose spiral arms could potentially transport angular momentum outward. A mini-cluster of stars orbits around the disk center.}
\end{figure*}

In this subsection, we describe the properties of one of the smaller cores in our set of simulations: core \A{}. The general properties of this core are shown in Table~\ref{tab:init}, also showing that the resolution of this simulation is $\sim 7$~AU, the highest in our set. This core forms in the  quasi-spherical geometry of fragmentation. Each panel in the top two rows of Figure~\ref{fig:coreA} shows the gas density on a slice through the center of the core/disk and the smaller insert shows the projected gas density in a view parallel to the angular momentum of the gas (face-on, first row) and a view perpendicular to it (edge-on, second row). From left to right, each panel shows the time evolution of the core as indicated by the labels.

When the core starts collapsing it has a spheroidal shape; we define time $t=0$ when the central density of the core reaches $\sim 10^6 \ \pcc$. In Figure~\ref{fig:coreA}, the snapshots are shown at times $t=0, 50, 200$~kyr and $400$~kyr, from left to right. The last three snapshots correspond to times when: i) a disk forms and becomes Toomre unstable at its center; ii) when the first star forms; and iii) when a quasi-steady (quiescent) disk forms.

Due to the conservation of angular momentum, the spherically symmetric collapse transitions into a rotation-dominated but turbulent fragmentation. As the central density increases and the disk becomes more gravitationally (Toomre) unstable, a spiral arm forms at the center of the core.
The subsequent evolution of the core is crucially dependent on supersonic turbulence. Eddies of eddies emerge and dense sub-cores at their centers spiral inward. This is a well-known mechanism that allows rapid gas accretion and transport of angular momentum in an unstable disk. These dense blobs are the locations where later on a single-star or a multiple-star system will form.
When the first star forms at the center of the system, accretion of gas into the central star clearly appear as a protostellar disk with prominent spiral arms, similar to those seen in many previous studies \citep{Bate2003, Goodwin2004a, Hennebelle2008a}. A zoom-in view of this smaller disk at times 200 kyr and 400 kyr is shown in the last two rows of Figure~\ref{fig:coreA}. 
The first two columns show the gas density in a slice through the disk face-on (third row) and edge-on (fourth row) 
to emphasize the typical flared shape of the disk in the edge-on view. 
The last two columns show the gas surface density, emphasizing the presence of spiral arms and the presence of other smaller disks forming from the contraction of other nearby smaller fragments, in good agreement with recent observations discussed above.

\begin{figure*}
  \centering
  \includegraphics[scale=0.8]{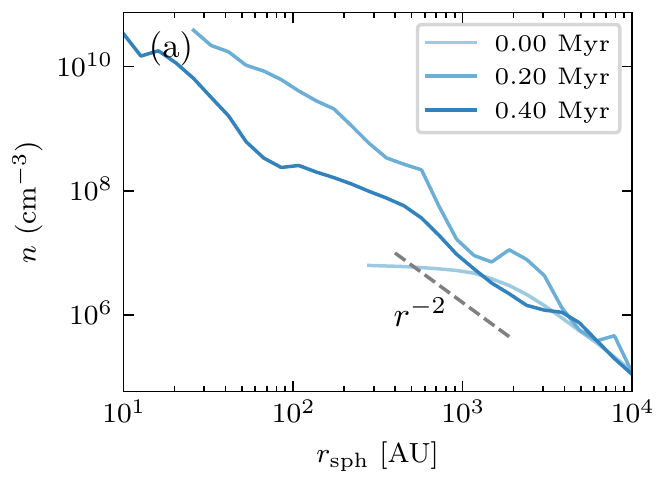}\hspace{-0.12cm}
   \includegraphics[scale=0.8]{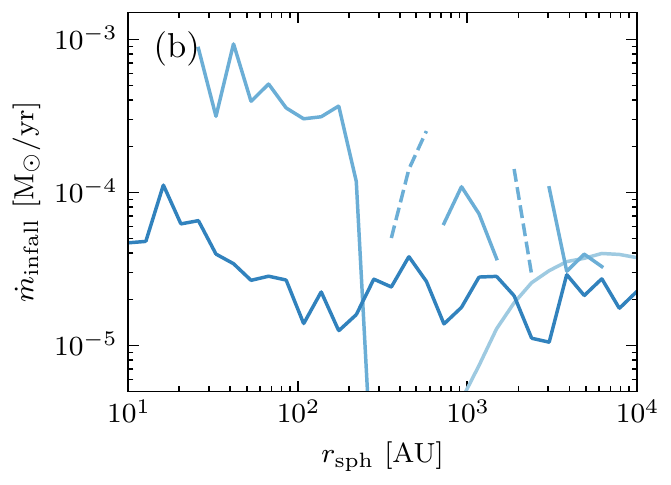}\hspace{-0.12cm}
  \includegraphics[scale=0.8]{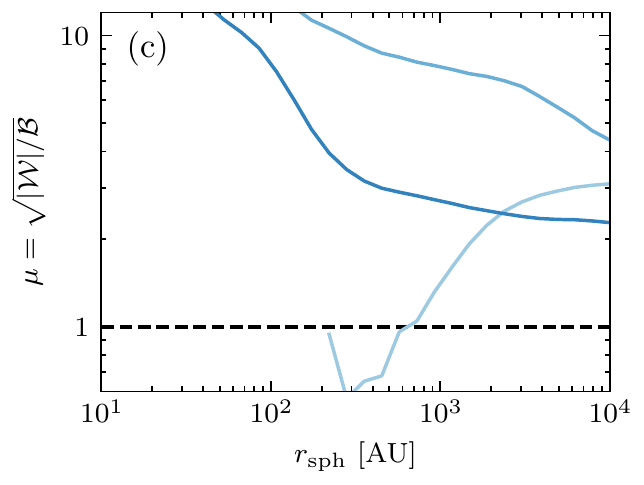}\hspace{-0.12cm}\\
  \includegraphics[scale=0.8]{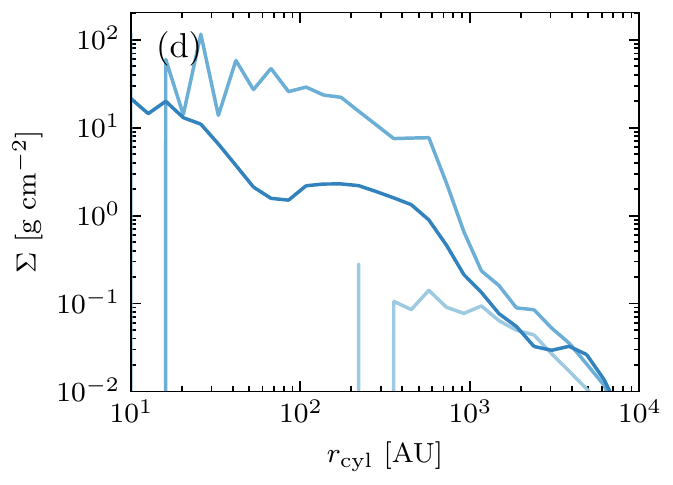}\hspace{-0.12cm}
  \includegraphics[scale=0.8]{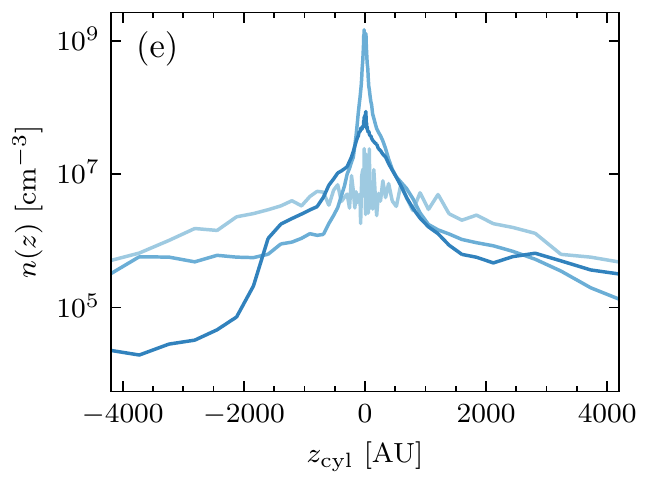}\hspace{-0.12cm}
  \includegraphics[scale=0.8]{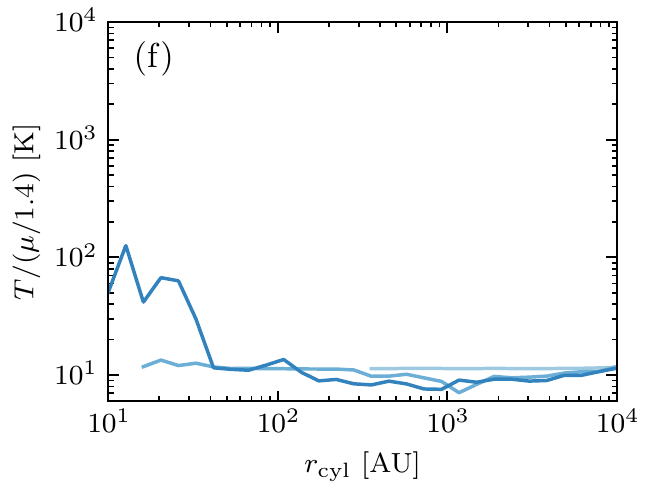}\hspace{-0.12cm}\\
  \includegraphics[scale=0.8]{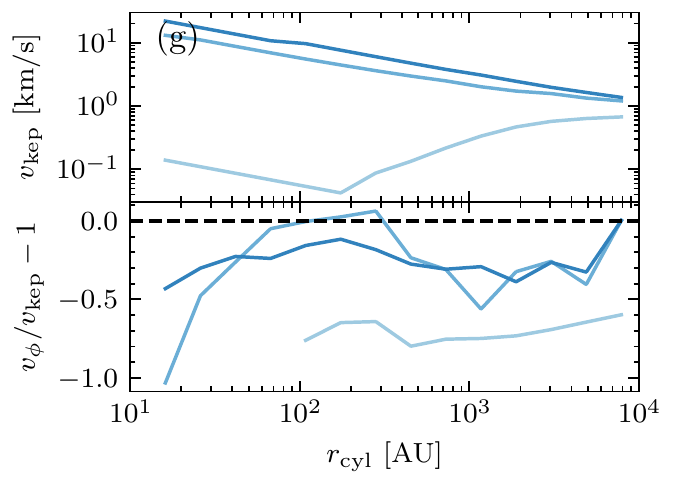}\hspace{-0.12cm}
  \includegraphics[scale=0.8]{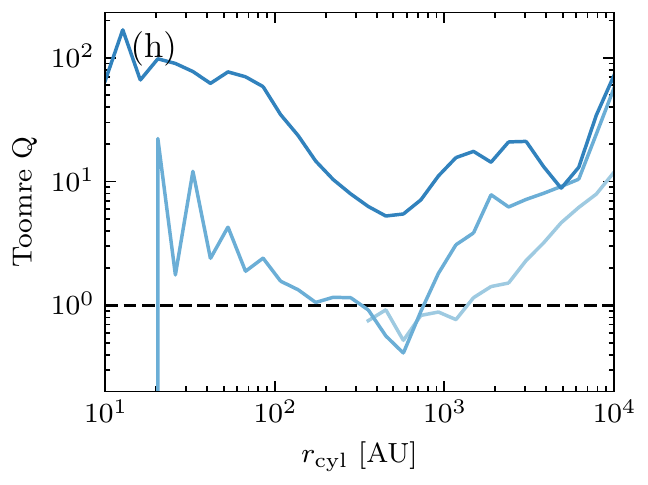}\hspace{-0.12cm}
  \includegraphics[scale=0.8]{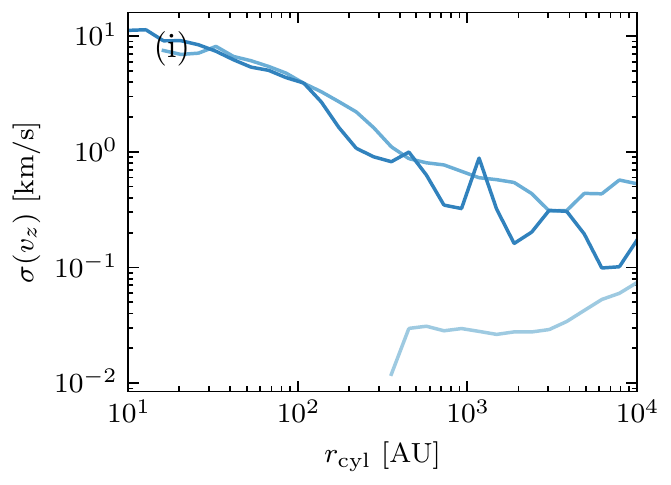}\hspace{-0.12cm}\\
 \includegraphics[scale=0.8]{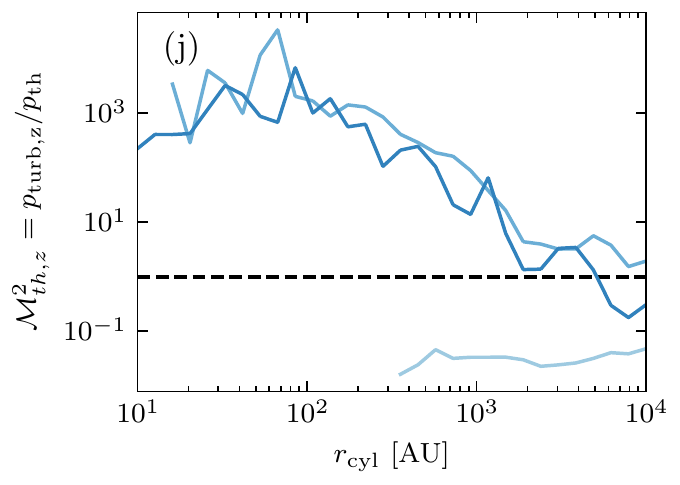}\hspace{-0.12cm}
  \includegraphics[scale=0.8]{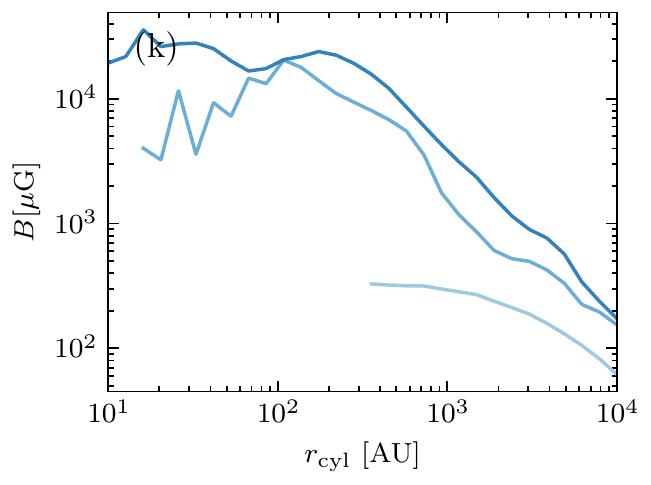}\hspace{-0.12cm}
  \includegraphics[scale=0.8]{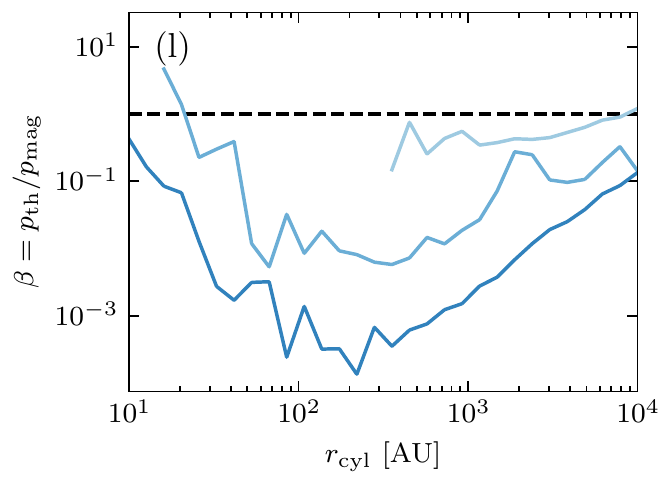}\hspace{-0.12cm}\\
  \includegraphics[scale=0.8]{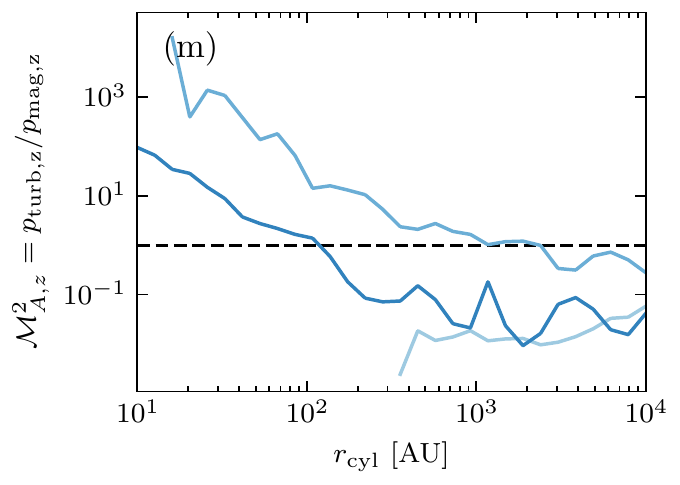}\hspace{-0.12cm}
   \includegraphics[scale=0.8]{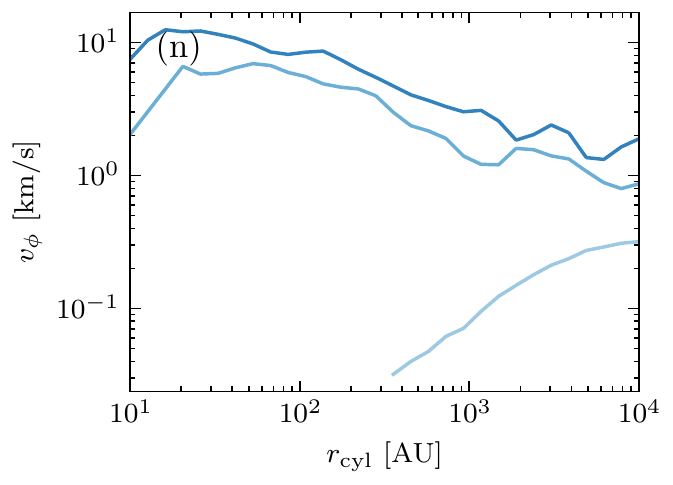}\hspace{-0.12cm}
  \includegraphics[scale=0.8]{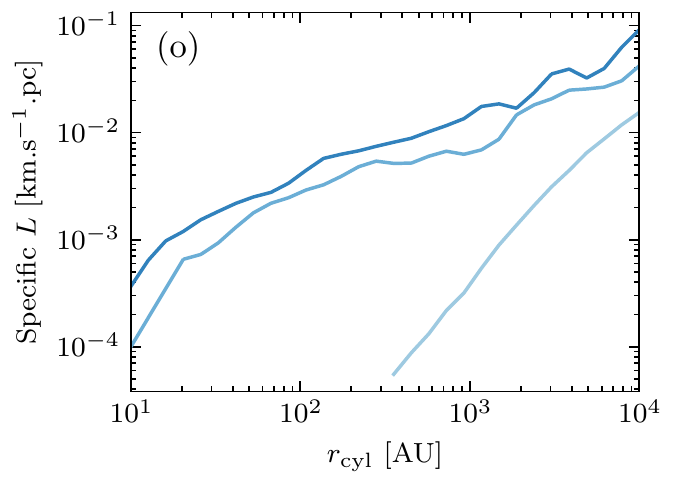}
   \caption{Radial and vertical profiles of the $27 \ \msun$ core \A{}. The $x$ axis represents the radius in a spherical coordinate $\rsph{}$, i.e. the distance to the center, or the radial distance in a cylindrical coordinate $\rcyl{}$, or the longitudinal position $z_{\rm cyl}$. The curves are averaged values of the following physical quantitates:
	(a) Number density at spherical radius \rsph{}.
	(b) Mass infall rate at \rsph{}. The dashed curve indicates a negative infall rate.
	(c) Mass-to-flux ratio at \rsph{}, defined as the square root of the ratio of gravitational binding energy to magnetic energy, $\beta = \sqrt{{\cal W} / {\cal B}}$, enclosed within \rsph{}. This is equivalent to the common mass-to-critical mass definition.
	(d) Column density at cylindrical radius \rcyl{} in a depth of 5000 AU.
	(e) Number density of a cylinder with radius 6000 AU at cylindrical height $z_{\rm cyl}$.
	(f) Temperature at \rcyl{}.
	(g) Non-Keplerianity on the disk plane defined as $(v_{\rm \phi} - v_{\rm kep})/v_{\rm kep}$.
	(h) Toomre Q at \rcyl{}.
	(i) Vertical-component velocity dispersion at \rcyl{}.
	(j) Square of $z$-component thermal Mach number, ${\cal M}_{\rm th}^2 = p_{\rm turb,z} / p_{\rm th}$ at \rcyl{}.
	(k) Magnetic field strength at \rcyl{}.
	(l) Plasma beta, $\beta = p_{\rm th}/p_{\rm mag}$, at \rcyl{}.
	(m) Square of the $z$-component Alfven Mach number, ${\cal M}_A^2 = p_{\rm turb,z} / p_{\rm mag}$, at \rcyl{}.
	(n) Disk's azimuthal velocity as a function of \rcyl{}. 
	(o) Disk's specific angular momentum as a function of \rcyl{}.
  \label{fig:pro1}}
\end{figure*}

In Figure~\ref{fig:pro1} we show a detailed quantitative characterization of the time evolution and properties of the collapsing core leading to the formation of the disk. We consider either spherically averaged profiles, most appropriate to describe the initial phases of the evolution and the outer parts of the core that maintain quasi-spherical geometry, or cylindrical coordinates, most appropriate to describe the disk structure. The origin of the coordinate system is set at the center of the disk or core. For the cylindrical coordinate system, the z-axis is set along the direction of the angular momentum of the gas. The different lines show the profiles at times corresponding to the three times shown in the legend, which also corresponds to the 1st, 3rd, and 4th columns in the top two rows of Figure~\ref{fig:coreA}, with an increasing shade of darkness indicating later times in the evolution. 

\textbf{1) Evolution and structure of the collapsing core and the disk.}
Panel (a) shows the density profile of the gas in spherical coordinates, $n(\rsph)$. At the time $t=0$ the core can be approximated by an isothermal cloud in hydrostatic equilibrium, showing a Bonnor-Ebert density profile with a central density $6 \times 10^6 \ \pcc$ and core radius $\sim 900$~AU. The envelope of the isothermal sphere extends up to 0.25 parsec ($5 \times 10^4$ AU), despite being beyond the range of the $x$ axis and not visible in the figure. The collapsing core has an enclosed mass of $\sim 27 \ \msun$ within 0.25 parsec, or above a density of $3000~\pcc{}$.
The density of the isothermal core increases self-similarly as the core collapses and reaches a density of $5 \times 10^{9} \ \pcc$ before a protostar (sink particle) forms at the center. The density profile has a power-law slope of about $-2$ in the outskirt of the core, consistent with an isothermal sphere in hydrostatic equilibrium.
The net mass infall rate (panel b) is between $10^{-5}$ and $10^{-4}$ $\msun{}/{\rm yr}$ and the total accreted mass into the center is about 17 \msun{} by the end of the simulation at $t=0.4$~Myr.

At the time $t=0$ the core is marginally magnetically supercritical as shown in panel (c): $\mu$ ranges
 from 0.6 to 3 from the inner region to the outer region. Over time, as the mass accumulates into the central stars in a compact region, the gravitational binding energy increases dramatically while the magnetic energy does not increase as rapidly, due to the decoupling of the magnetic fields from gas as a result of star formation.
 We will discuss the properties of the magnetic fields and the disk formation in a followup work. %

The remaining panels show the properties of the pseudo-disk in cylindrical coordinates.
Panel (d) shows the face-on surface density profile of the disk, which increases from $1\ {\rm g}\,{\rm cm}^{-2}$ at $r_{cyl}=10^4$~AU to $10-100\ {\rm g}\,{\rm cm}^{-2}$ at the center.
This surface density is comparable to what is observed in Class II disks around young stellar objects in Ophiuchus \citep[\eg{}][]{Andrews2009}. 
 
The density profile of the disk in the vertical (z-axis) direction $n(|z|)$ (the average of $z$ and $-z$) is shown in panel (e). The disk thickness is 400 - 500~AU at a threshold density of $10^7 \ \pcc$ at times $t > 200$~kyr. At the same mean density cutoff, the radius of the disk is roughly 1000-2000~AU, therefore the disk is rather thick with an aspect ratio $H/R \sim 1/2-1/4$. 

The average gas temperature in the disk (panel f) remains near $10$ Kelvin throughout the simulation. In the center of the disk after the formation of the first star, the gas temperature increases to $\sim 100$~K as a result of photoionization heating from massive stars. As will be discussed in \S{}~\ref{sec:HII}, the ionizing UV radiation from the stars at these early times is trapped in the thick dense disk and the disk remains cold at radii $>100$~AU in most simulations. 

\textbf{2) Keplerianity and stability of the disk.}
The disk has a quasi-Keplerian rotation, with a deviation from Keplerianity  $\beta_{\rm kep} \equiv (v_{\phi} - v_{\rm kep}) / v_{\rm kep}$ of the order 50\% (see bottom of panel g). The deviation is mainly due to the relatively large accretion rate of gas: as shown in panel (b), where and when the mass accretion rate is higher, corresponds to larger deviation from Keplerianity. In addition, both the radial infall rate and the deviation from Keplerianity is larger when the disk is more strongly gravitationally unstable as illustrated in panel (i) showing the profile of the Toomre Q parameter. The pseudo-disk starts more strongly gravitationally unstable ($Q<1$) and transitions into a stable disk after about 0.3~Myr from $t=0$.

{\bf 3) Turbulent, thermal and magnetic support of the disk.}
The disk is very turbulent: the turbulent velocity in the vertical direction is between 1~km/s and 10~km/s (panel i) and the turbulence is highly supersonic as shown by the square of the thermal Mach number in the vertical direction (or the ratio of the turbulence pressure over the thermal pressure in the z-direction) shown in panel (j). The core starts with negligible turbulence, supported by thermal pressure. The pseudo-disk phase is instead dominated by turbulence motions, with rms velocity in the z-direction of about a third of the Keplerian velocity and significantly higher than the sound speed of the gas ($c_s \sim 0.3$~km/s), reaching Mach numbers of 10-30. Hence, we expect a geometrically thick disk supported by supersonic turbulent motions, as discussed in more detail in \S~\ref{sec:turb}.

The core is magnetized with an initial magnetic strength of about 50 -- 300 $\mu$B (panel k). The magnetic field is amplified by the accretion of gas and increase of the surface density of the gas, which can be partially explained owning to magnetic flux freezing, \ie{}, $\mu \propto M/\Phi_B = \Sigma / B \sim {\rm const}$.
The turbulence also grows dramatically due to nonaxisymmetric collapse, despite the existence of relatively strong magnetic fields. 
During this process, the core transitions from a thermal and magnetic pressure-dominated phase into a turbulent pressure-dominated phase (panels j and l), with a plasma $\beta$ as low as $10^{-2}$ at a few 100~AU and thermal Mach number as high as $10-10^2$ in the inner region (<100~AU).
Finally, in the quasi-steady phase of the disk, the turbulent pressure in the inner parts of the disk still dominates over the magnetic pressure but this is reversed at radii $r_{\rm cyl} \gtrsim 100$ AU (see panel m).
\begin{figure*}
  \centering
 \includegraphics[width=\textwidth]{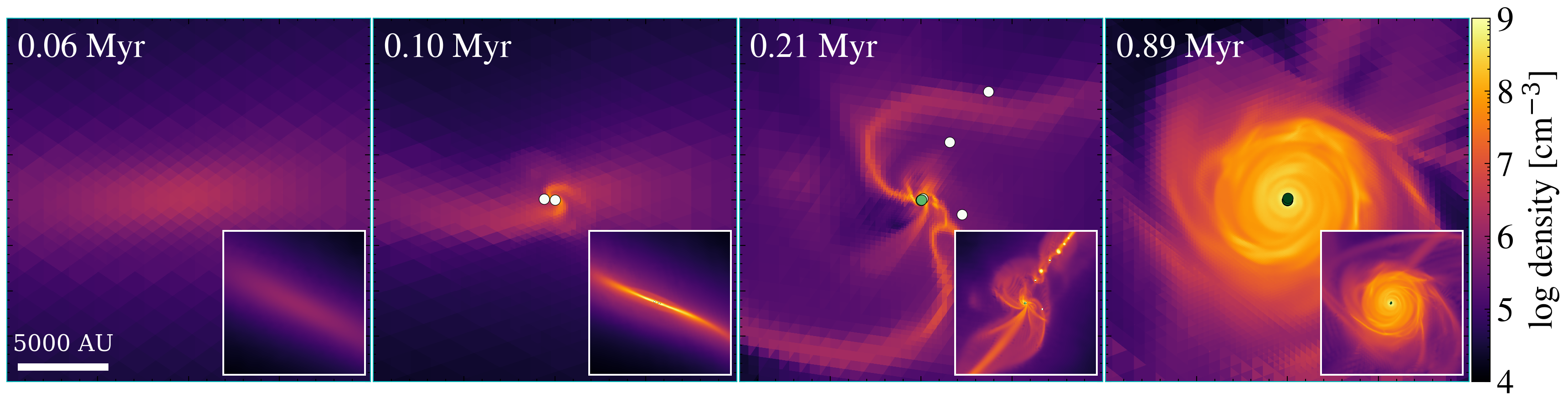}
 \includegraphics[width=\textwidth]{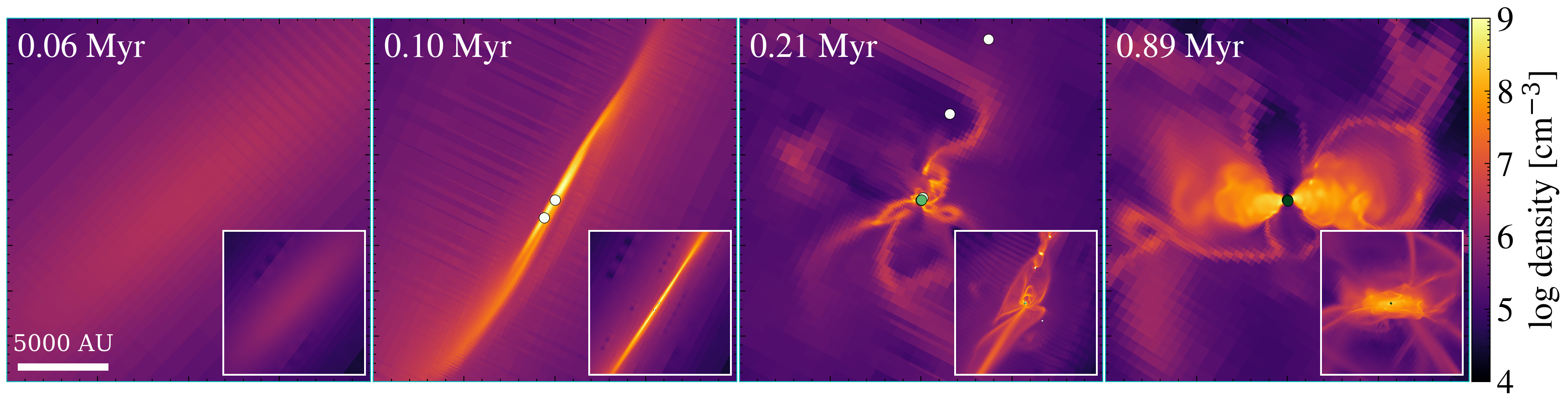}
  \caption{Same as Figure~\ref{fig:coreA} but for the very massive core (Core \B{}).
	Starting from a thin tube, the core undergoes fragmentation at the center as well as along the arms of the cylinder. With large feeding of gas along the filament, the central stars grow in mass rapidly, ending in a total mass of over $600 \ \msun{}$. The gas keeps feeding the central stars via a large, thick disk, tens of solar masses in mass.
  }
  \label{fig:coreB}
\end{figure*}

{\bf 4) Analysis of velocity gradients in comparison to observations.} 
In this part, we analyze this simulation with the objective of comparing it to observations in terms of the velocity gradients and specific angular momentum and discuss whether or not our simulations successfully reproduce some features of the observations. 

In an axis-symmetric model with initial rotation, the angular momentum measured at various scales is perfectly aligned.

We analyze the magnitude of the angular momentum in our simulated disk and its alignment at various scales.
We consider a cylinder of height $h = 500$ AU centered by the disk and aligned with the spin axis of the disk.
The average specific angular momentum are measured to be $\num{4e-3} \ {\rm km~s^{-1}~pc}$ and $\num{1.5e-2} \ {\rm km~s^{-1}~pc}$, measured within $R = 800$ AU, the radius of the disk, and $R = 5000$ AU, enclosing the envelope, respectively.
These values are consistent with observations of the protostellar regime from the CALYPSO dataset ($j \sim 10^{-3}~{\rm km~s^{-1}}$, see \citealt{Belloche2013, Gaudel2020} and the citations therein). 
The specific angular momentum as a function of distance to the center (panel o of Figure~\ref{fig:pro1}) also follows the power-law relation $j \propto r^{1.6-1.7}$, highly consistent with observations of Class 0 envelopes \citep{Goodman1993, Caselli2002}. 
The relative angle between the angular momentum measured at two scales is $12^\circ$. This misalignment is largely due to the turbulence of the initial core. This is in agreement with \cite{Verliat2020}, who find that the formation of a disk can be a result of small perturbations of the initial density field in the core in the absence of large-scale rotation.

Panel (n) in Figure~\ref{fig:pro1} displays the azimuthal velocity at various distances. 
We see a transition from the inward velocity gradient in the initial core to the outward velocity gradient of the disk and envelope, which indicates an evolution from a slowly rotating rigid body (\ie, nearly constant angular velocity) to a differentially rotating Keplerian disk.
The amplitude profile of the velocity gradient, or the angular velocity with respect to the disk center, roughly follows a power law $\Omega \sim r^{-1.4}$, close to that of a Keplerian disk with all the mass concentrated at the center, which gives $r^{-1.5}$.
These features can be tested with observations of molecular lines in nearby star-forming regions which can measure velocity gradients in the cores (at large scales) and in the disks (at smaller scales) with a precision of about $1 \ {\rm km~s^{-1}~pc^{-1}}$ \citep[\eg{}][]{Cheng2022}, which is sufficient to detect the slow rotation (at subsonic/sonic speeds) of cores out to 10,000~AU scales.

\subsection{Evolution of a 130 solar mass core}
\label{sec:diskB}

\begin{figure*}
  \centering
  \includegraphics[scale=0.8]{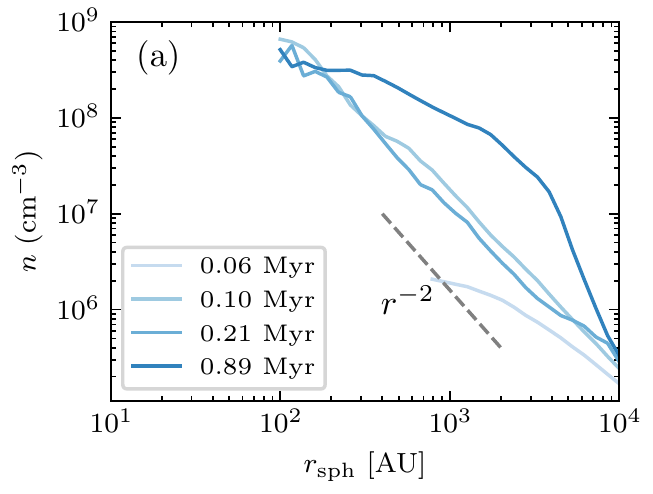}\hspace{-0.12cm}
  \includegraphics[scale=0.8]{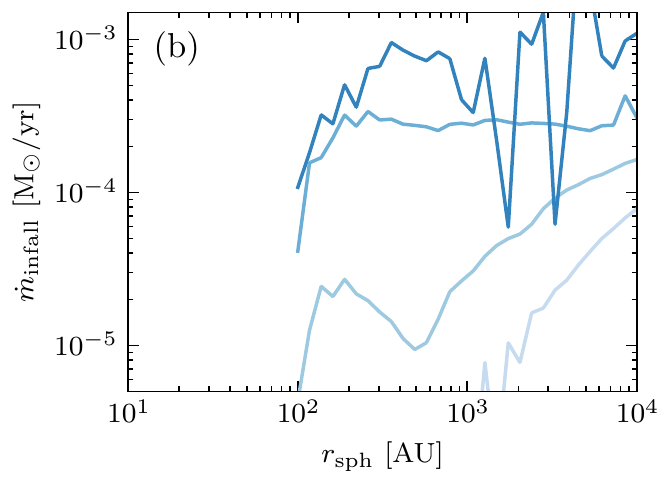}\hspace{-0.12cm}
  \includegraphics[scale=0.8]{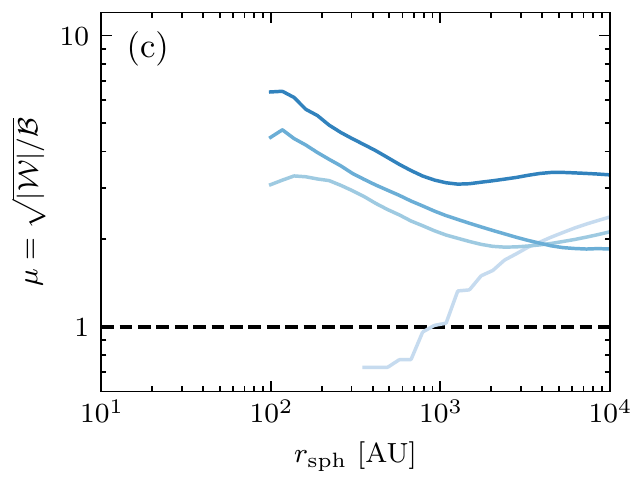}\hspace{-0.12cm}\\
  \includegraphics[scale=0.8]{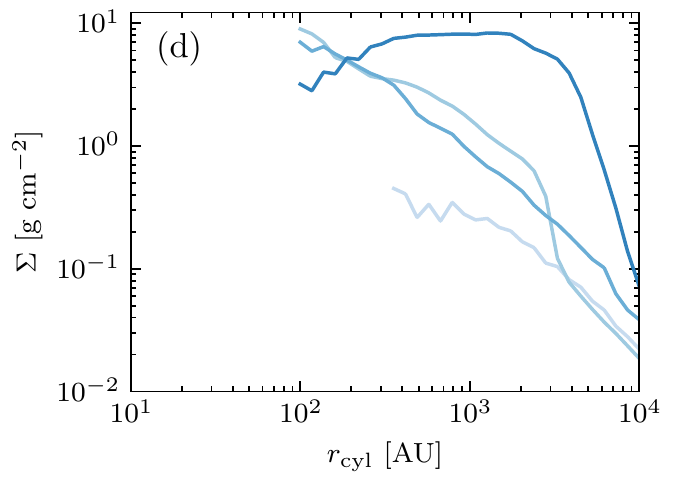}\hspace{-0.12cm}
  \includegraphics[scale=0.8]{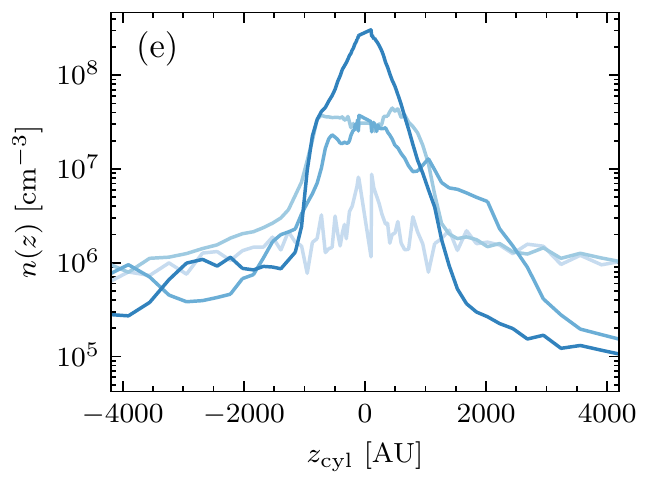}\hspace{-0.12cm}
  \includegraphics[scale=0.8]{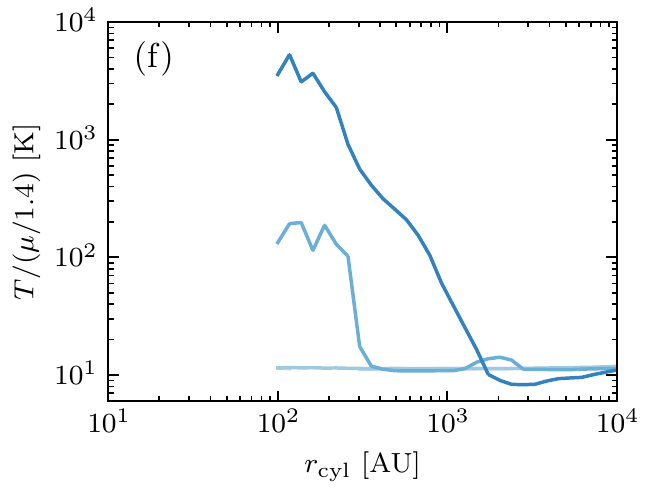}\hspace{-0.12cm}\\
  \includegraphics[scale=0.8]{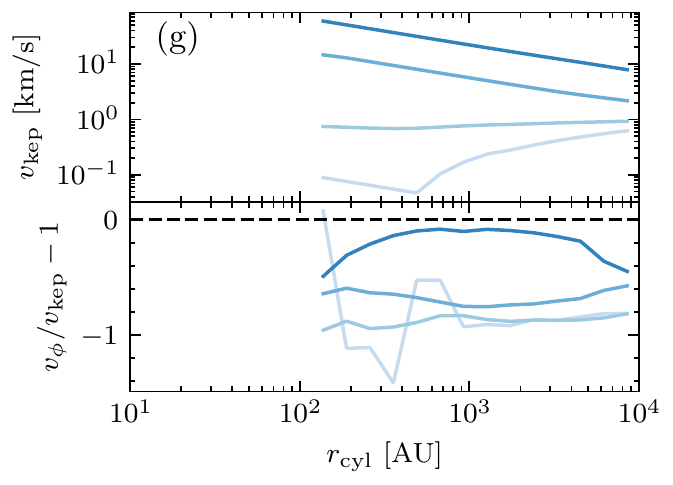}\hspace{-0.12cm}
  \includegraphics[scale=0.8]{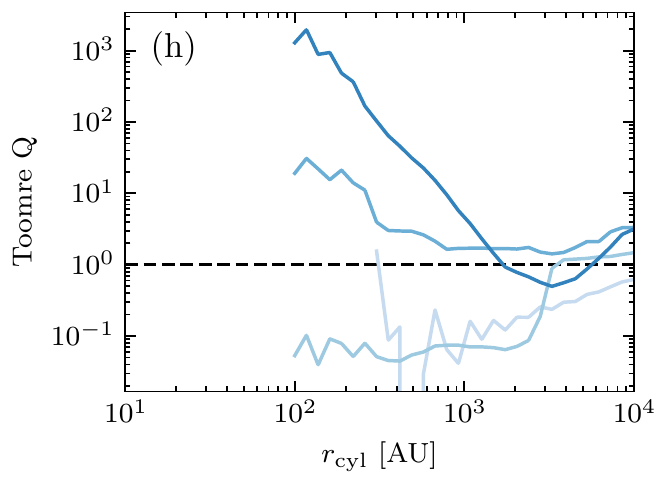}\hspace{-0.12cm}
  \includegraphics[scale=0.8]{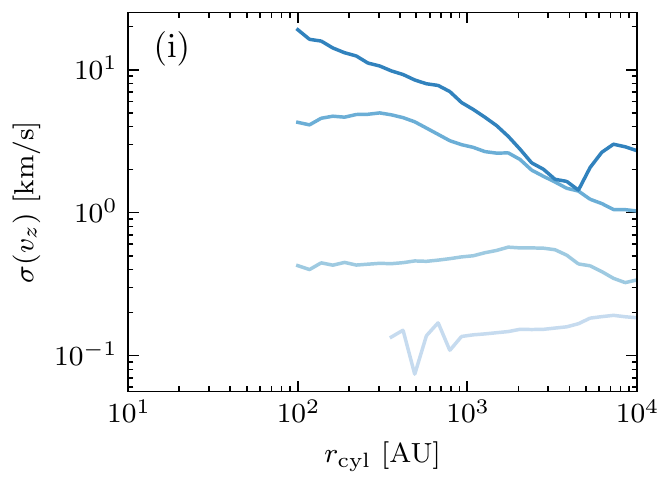}\hspace{-0.12cm}\\
  \includegraphics[scale=0.8]{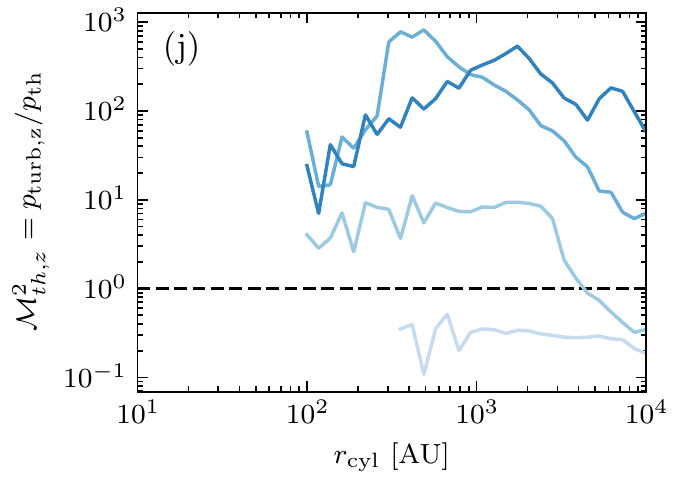}\hspace{-0.12cm}
  \includegraphics[scale=0.8]{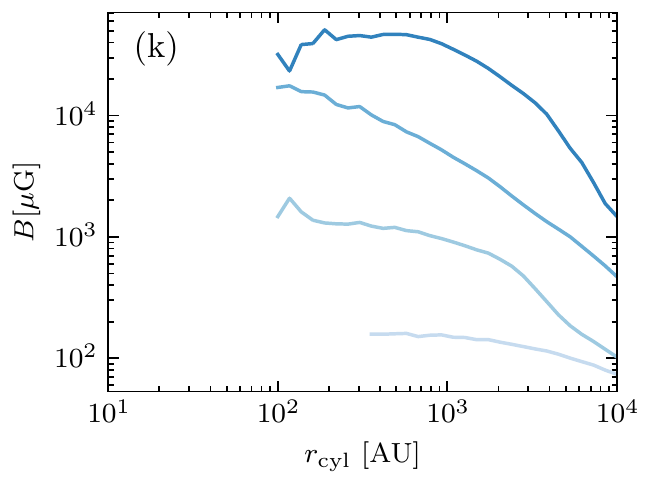}\hspace{-0.12cm}
  \includegraphics[scale=0.8]{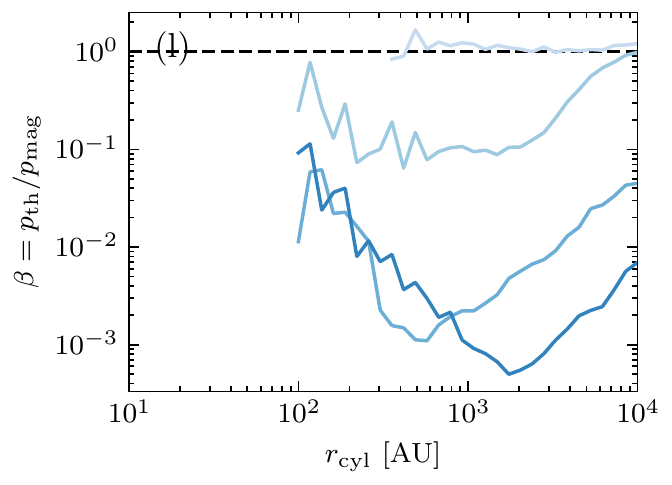}\hspace{-0.12cm}\\
  \includegraphics[scale=0.8]{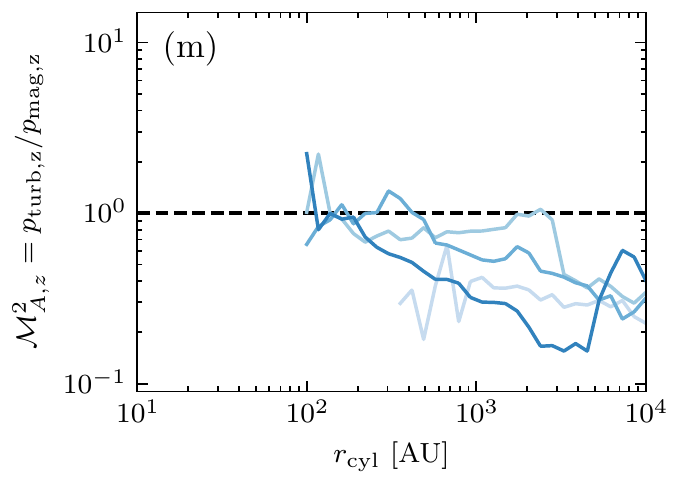}\hspace{-0.12cm}
  \includegraphics[scale=0.8]{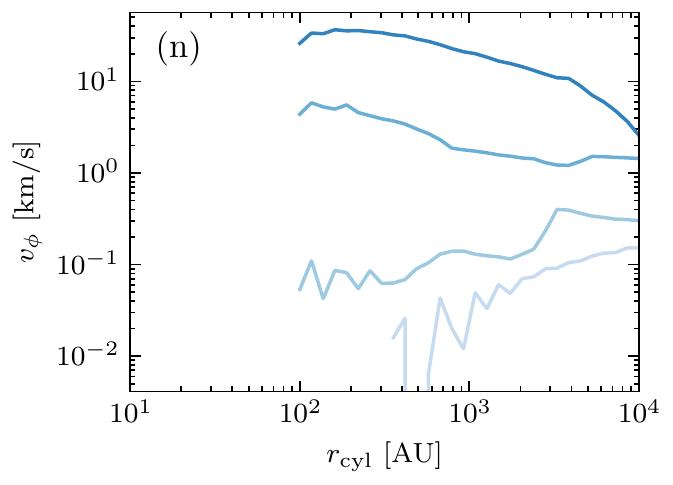}\hspace{-0.12cm}
  \includegraphics[scale=0.8]{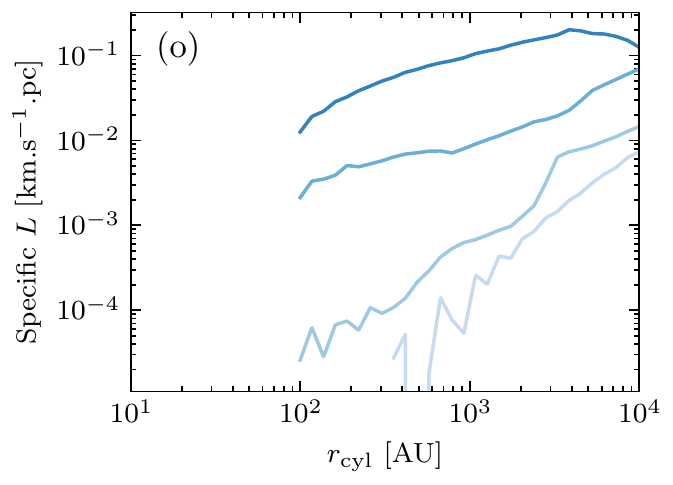} 
  \caption{Same as Figure~\ref{fig:pro1} but for the $\sim 130 \ \msun$ core \B{}.
	Refer to the texts for the implications and interpretations.
  }
  \label{fig:pro2}
\end{figure*}

The properties of core \B{}, a very massive core that grows from 130 \msun{} 
to a mass of over 600 \msun{}, are qualitatively similar to core \A{} once we 
account for the fact that it is much more massive. In this section we will 
emphasize the differences between this core and the less massive core \A{} 
discussed above. Figure~\ref{fig:coreB} shows the projected density 
distributions of the core from face-on (top) and edge-on (bottom) views, as in 
Figure~\ref{fig:coreA}.

Unlike core \A{} which collapses spherically, core \B{} starts from the collapse of a filamentary structure and fragments along the length of the filament. However, the final outcome of the collapse is a disk similar to the case of core \A{}, although much more massive.
At the time $t=0$ the initial density structure is better described by a cylindrical geometry than a spherical core (see the first and second frames in Figure~\ref{fig:coreB}). The fragmentation of the core starts at the center of the frame where the density is highest. As the density reaches $10^9 \ \pcc$, the Jeans length drops to a few hundred AU and the local free-fall time is a few kyr. The filament undergoes fragmentation along the length of the cylinder and breaks into blobs that become eddies (the third frame). Finally, these eddies migrate into the center of the system in about one dynamic time. Conservation of angular momentum turns the filamentary collapse into rotational collapse, forming a large, thick disk (the last frame).

We plot the properties of this filament/disk as a function of time in Figure~\ref{fig:pro2}, with timestamps corresponding to the four snapshots in Figure~\ref{fig:coreB}.

Even though the geometry of the core at $t=0$ is not spherical but filamentary, the spherically averaged density profile is well described by a Bonnor-Ebert density profile with a central density $\num{2e6} \ \pcc$ and core radius $\sim 2000$~AU (panel a).
Between 0.1 Myr and 0.4 Myr, the radial distribution of density displays a power-law profile with an exponent close to $-1.5$, which is flatter than an isothermal sphere, likely because the geometry is clearly filamentary at these times. After $\sim 0.4$ Myr, the core exhibit a nearly constant density profile with $n \sim 10^7$~\pcc\ within 4000~AU, indicative of a disk with nearly constant density and constant surface density as shown in panel (d). This disk is about 4 times larger than disk \A{} at the same density threshold, but disk \A{} has an increasing density and surface density toward the center. By $t=0.9$~Myr (the dark blue curve), a large, thick, near-Keplerian disk forms with a surface density of $\sim 10 \ {\rm g~cm^{-2}}$. The radius of the disk is about 4,000 AU at a cutoff density $10^7 \ \pcc$, equivalent to a column density of $3 \ {\rm g~cm^{-2}}$ (panel e).
Within this radius, the disk is near Keplerian (panel h).
The thickness of the disk is about 2,000 AU at the same cutoff density (panel f).

Unlike disk \A{} where the radial component is extremely turbulent and the mass inflow is discontinuous, disk \B{} has a steady inflow at a constant velocity of $2 - 3$~km/s, which is close to the escape velocity at the edge of the disk at 5000 AU from the center.
With a $\rho \sim r^{-2}$ relation, the inflow of mass has a constant rate $\num{3e-4} \msun/{\rm yr}$ (panel b). This inflow rate results in an accreted mass of $> 400~\msun{}$ in the accretion period of 0.5~Myr. 

Due to the large central mass, the Keplerian velocity of the disk is high, reaching $7 - 20$~km/s at 1000~AU. The turbulent velocity is also very high, between 3 and 6 km/s in the $z$ component (panel j).

Due of magnetic flux freezing, the large mass in the disk results in large magnetic strength, reaching between $10^3$ to $5 \times 10^4$ $\mu$G (panel k). Despite of the strong magnetic field, the mass-to-flux ratio $\mu$ stays high between 3 and 7 (panel c) due to the large gravitational binding energy from the large mass of the cluster at the center.
Although the inner part of the disk is heated by UV radiation to 100 - 3000 K (panel f), the disk is supported primarily by magnetic pressure and secondarily by turbulence in the axial direction (see panels j, l, and m). We will discuss the support of the disk in more detail in the next section.

\begin{figure*}
  \centering
\includegraphics[scale=1.2]{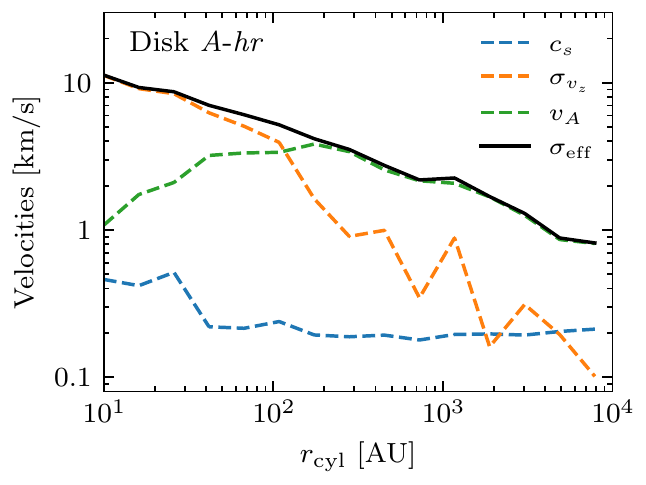}
\includegraphics[scale=1.2]{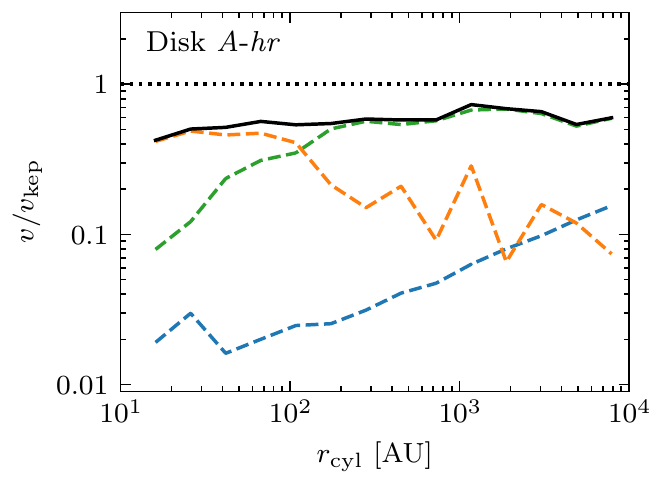}\\
\includegraphics[scale=1.2]{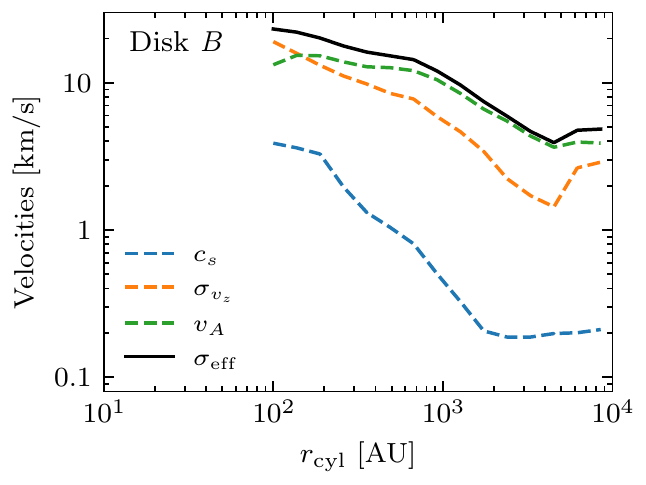} 
\includegraphics[scale=1.2]{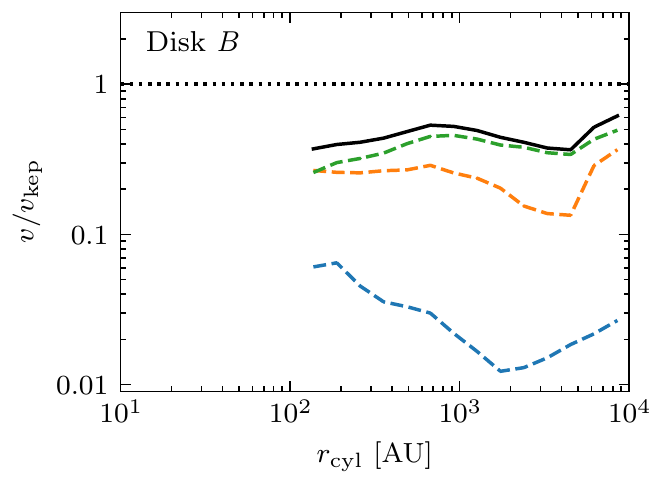}\\
\caption{\label{fig:sigmavz} ({\it Top})
  Left panel: Radial profiles of the sound speed (dot-dashed line), the Alfven speed (dashed line), and the $z$-component turbulent velocity (rms velocity in the z-direction, solid line) for our lower-mass disk (Disk \A{}). Right panel: the same quantities normalized to the Keplerian velocity $v_{\rm kep}$.  This ratio reliably predicts the scale height $H/r$ of a gas disk in hydrostatic equilibrium, where the vertical component of gravity is balanced by the pressure $\rho v^2$ gradient. The panels illustrate that the vertical support of the disk is dominated by turbulence and by the magnetic field, while thermal pressure support is negligible. The right panel also shows that $\sigma_{\rm eff}/v_{\rm kep} \sim 0.5$, and hence the disk is geometrically thick, {\it i.e.}, the disk has an aspect ratio of about 0.5. ({\it Bottom}). The same as the top panels but for our most massive disk, (Disk \B{}). We note that in both disks the turbulent support dominates from the center to 200~AU and the magnetic support from \rcyl$\sim 200$~AU to $\sim 10^4$ AU. 
}
\end{figure*}

\subsection{The disks thickness is determined by magnetic support and supersonic turbulent motions}
\label{sec:turb}

The disks in the six simulated cores measure from 200 AU to 6000 AU in radius, with Disk \B{} being the largest and most massive one. They are generally very thick as well, mostly with an aspect ratio (thickness to diameter ratio) of 0.2 to 0.5. 
Here we explore the physics behind the existence of these large disks.

For a disk of gas around a massive central object that is in hydrostatic equilibrium with the $z$ component of gravity from the center,
\begin{equation}\label{eq:dpdz}
  \frac{dP}{dz} = -\frac{G M \rho z}{(r^2 + z^2)^{3/2}} \approx - \frac{G M \rho z}{r^3}.
\end{equation}
where $r$ is the radial cylindrical coordinate for the distance from the center and $z$ is the altitude cylindrical coordinate for the distance from the disk midplane.
Assume the pressure $P$ can be expressed as $P = \rho \langle v ^2\rangle$, then
\begin{equation}
	\langle v^2 \rangle \frac{d\rho}{dz} = - \frac{GMz}{r^3} \rho,
\end{equation}
therefore
\begin{equation}
	\rho = \rho_0 \exp \left(- \frac{v_{\phi}^2 z^2}{2\langle v^2\rangle r^2} \right)
	= \rho_0 \exp \left(- \frac{z^2}{H^2}\right),
\end{equation}
where $v_{\phi} = \sqrt{GM/r}$ is the Keplerian velocity and $H$ is the disk scale height,
\begin{equation}\label{eq:HR}
H / r \approx \frac{\sqrt{2}~\langle v^2\rangle^{1/2}}{v_\phi}.
\end{equation}

The rms velocity $\sigma_{\rm eff} \equiv \langle v^2 \rangle^{1/2}$ should account for all the possible pressure supports: thermal, turbulent or magnetic. These three components can be identified as the sound speed $c_s$, the dispersion (rms) of the $z$-component velocity $\sigma_{v_z}$, or the Alfven speed $v_A$, respectively. We compare these three velocities and $\sigma_{\rm eff}=(\sigma_{v_z}^2+v_A^2+c_s^2)^{1/2}$ as a function of $r$ in Figure~\ref{fig:sigmavz}.
It is clear that turbulent pressure dominates the vertical support of the disk in the inner disk and magnetic pressure dominates in the outer disk, while thermal pressure support is negligible everywhere. For the two disks shown in Figure~\ref{fig:sigmavz}, the transition from turbulent support to magnetic support occurs at $\sim 200$~AU. However, in other disk we observed the transition radius to be closer to 1000~AU. We will further investigate the dependence of the critical radius on the physical parameters of the core in future work.

We notice that $\sigma_{v_z}$ decreases with the distance to the center with an empirical relation $\sigma_{v_z} \sim r^{-0.6}$, while $v_A$ peaks at $r_{\rm cyl} = 200$ AU and decreases slowly outward. 
The effective rms velocity $\sigma_{\rm eff}$ scales with radius as $\sigma_{\rm eff} \sim r^{-0.5}$. 
Assuming a roughly Keplerian azimuthal velocity profile  $v_{\rm kep} \sim r^{-0.5}$, we have $\sigma_{\rm eff} / v_{\rm kep} \sim r^{0}$. The predicted disk aspect ratio $H/r$ according to Equation~\ref{eq:HR} is nearly constant as a function of radius and of the order of unity, as shown in the right panels of Figure~\ref{fig:sigmavz}. Assuming an isothermal density profile $\rho \propto r^{-2}$ as for disk \A{}, the disk surface density is $\Sigma \propto r^{-1}$, while  disk \B{} has $\rho \propto r^{-1}$ and a nearly constant surface density profile, $\Sigma \sim {\rm const}(r)$. Indeed the disks have nearly constant opening angles and the surface density profiles we derived above are consistent with the actually geometrical properties of the disks shown in the fourth column of Figures \ref{fig:coreA} and \ref{fig:coreB}.

The magnetic field plays a more significant role at determining the structure of the outer disk. At large radii ($\gtrsim 200$ AU in both disk \A{} and disk \B{}) the magnetic pressure dominate over turbulent pressure.
Even though the initial strength of the magnetic field ($10 - 25 \ \mu {\rm G}$) of the cloud at density $10^3 \ \pcc$ is in the typical range of what is observed in present-day molecular clouds and the initial velocity dispersion is 5 times higher than $v_A$, the strengths inside both disks are above 1000 $\mu$G and even stronger at the very center.
The Alfven velocity is between 1 and 4~km/s in Disk \A{} and between 4~km/s and 15~km/s in Disk \B{}, many times higher than the thermal sound speed.
Therefore, we expect that magnetic field could become dominant dynamically over turbulence even in the inner parts of the disk in simulations with mildly stronger initial values of the magnetic field. 
We will explore this possibility in a followup work. 

In our simulations, the gas temperature is underestimated near sources of radiation due to neglecting heating processes important at high densities (\ie, gas heating from stellar radiation absorbed by dust grains). However, this missing physics should have a negligible impact on the simulation results since the gas thermal pressure does not play an important role in determining the structure of the disk and the protostars.

\section{Results. II. Low-mass stars form from the fragmentation of massive pre-stellar cores}
\label{sec:3}

\begin{figure}
  \centering
  \includegraphics[width=3in]{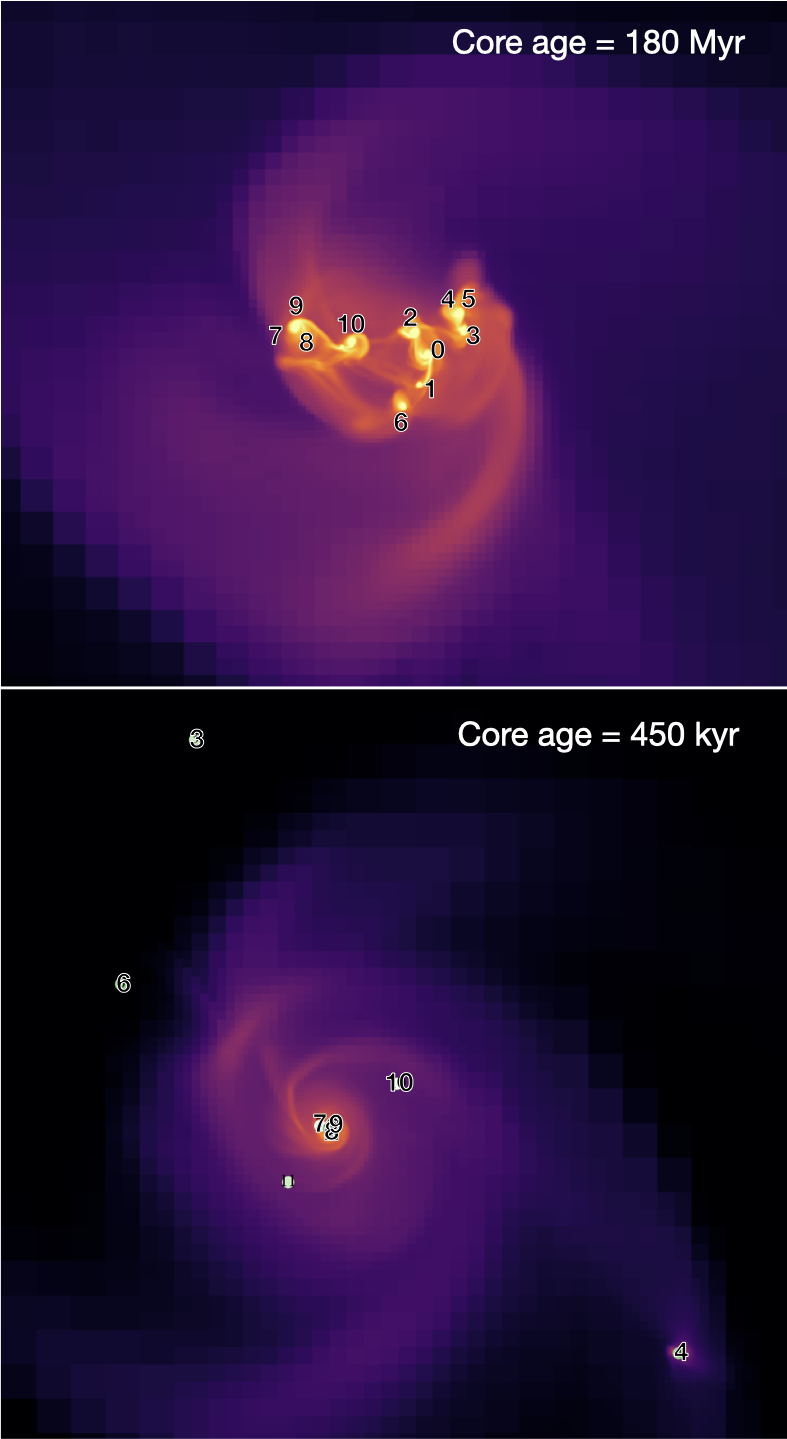}
  \caption{
  Snapshots of projected density of core \A{} after 180 kyr ({\it top}) and 450 kyr ({\it bottom}) from the isothermal core formation (defined as when the core density reaches $10^6 \ \pcc$). In the top panel, we identify the positions of the prestellar cores where sink particles form at later times, and show the IDs of the sink particles that form in them. In the bottom panel, the circles and numbers indicate the locations and IDs of the sink particles existing at the time shown in the snapshot. The figure shows that most sinks form from the fragmentation of a turbulent disk at an early time, but the sinks form with a time delay, and they either spiral in toward the center of the disk or are ejected (or in the process of being ejected). The image measures $2 \times 10^4$ AU on a side. .
  \label{fig:blobs}}
\end{figure}

\begin{figure*}
    \centering
    \begin{tabular}{ccc}
    Core \A{} & Core {\it A} & Core \B{}  \\
    \includegraphics[width=2in]{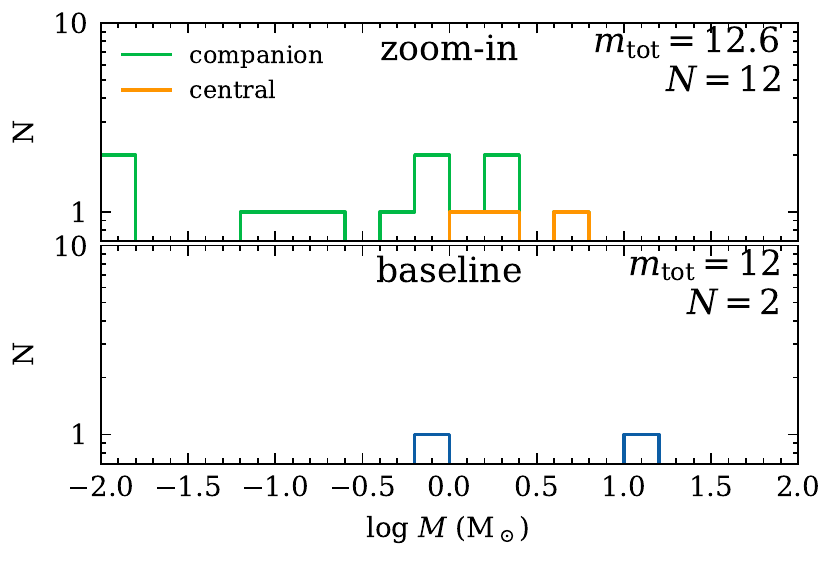} &
    \includegraphics[width=2in]{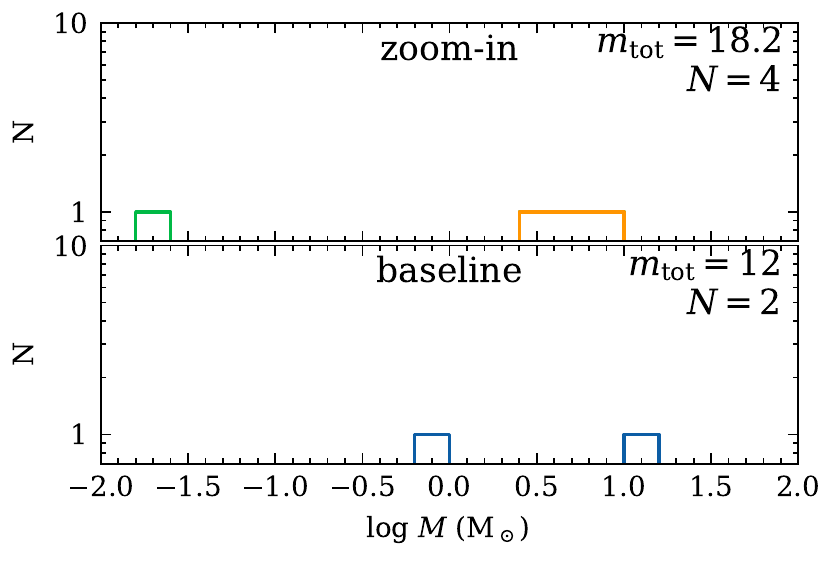} &
    \includegraphics[width=2in]{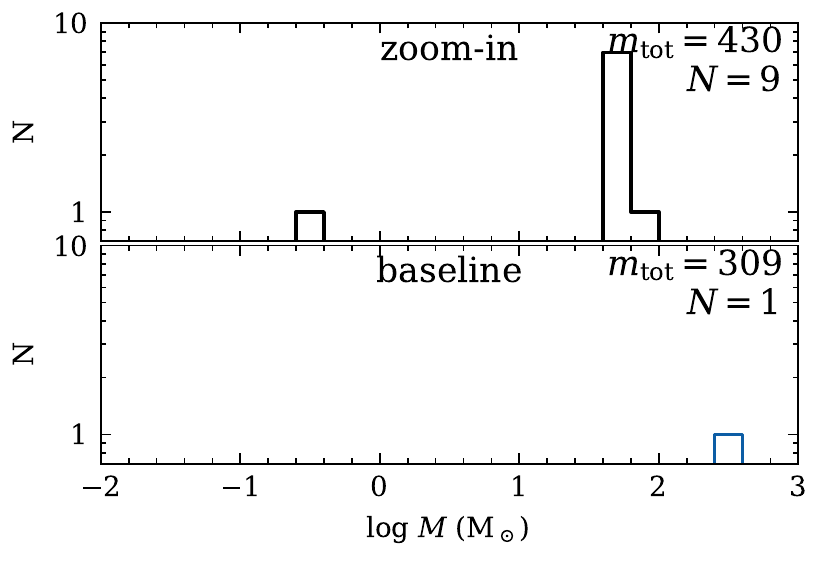} \\
    Core {\it C} & Core {\it D} + {\it E} & Core {\it F} \\
    \includegraphics[width=2in]{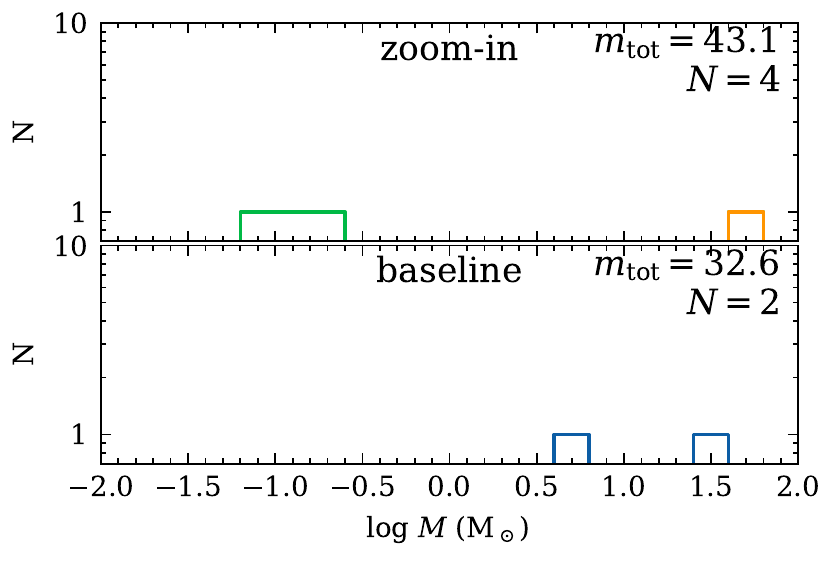} &
    \includegraphics[width=2in]{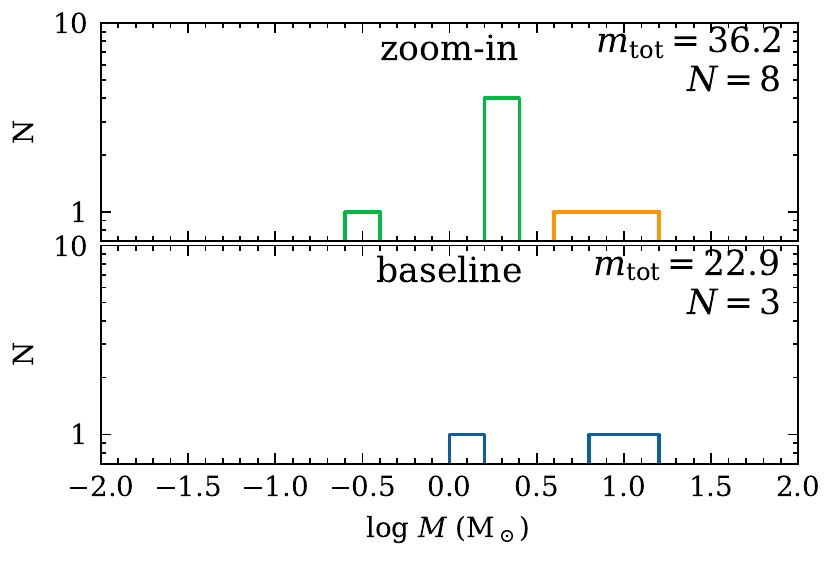} &
    \includegraphics[width=2in]{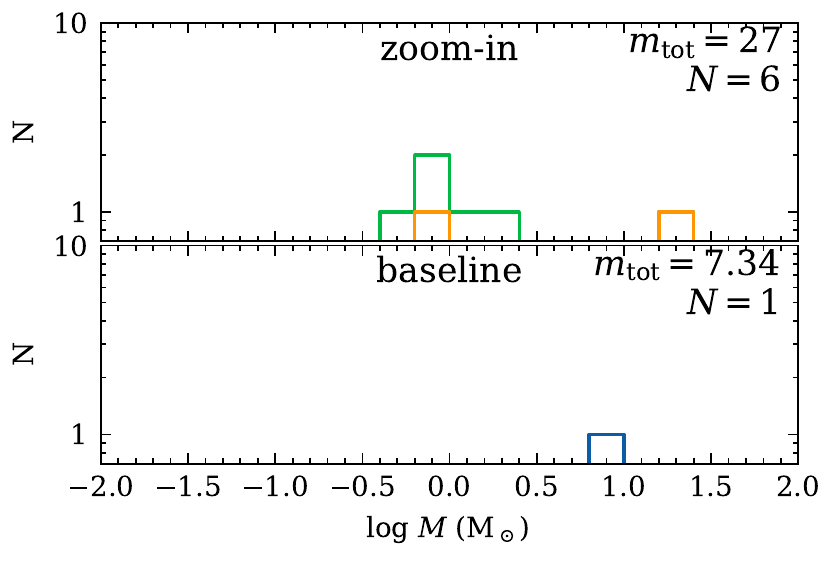} 
    \end{tabular}
    \caption{The mass distribution of the stars forming in each core in our `zoom-in' simulations, labeled `zoom-in', compared to the stars that form from the same core in the baseline run, labeled `baseline'. The mass functions of the stars in the `zoom-in' simulations are plotted in two colors to distinguish the central stars (orange) and companion stars (green), or in black when these two groups of stars are indistinguishable. In the case of cores {\it D} and {\it E}, because they are in close proximity to each other, their central stars, as well as their companion stars, are grouped together. Overall, the number of stars that form from the fragments of a core ranges from 1 to 12.
 }
    \label{fig:MF}
\end{figure*}

\subsection{Fragmentation into low-mass stars before the formation of a steady disk structure}
\label{sec:frag}

In the classical picture of prestellar disk formation, an unstable disk undergoes fragmentation and stars likely form from the disk fragments \citep{Agertz2007, Kratter2016}. Simulations of Pop~III star formation clearly show the behaviors for the formation of high-mass stars, which form near the center of the disk that is most unstable, and migrate outward \citep{Machida2008, Stacy2010,park2021a,park2021b}. A disk that accretes gas from a surrounding envelope tends to fragment by gravitational instability, producing self-gravitating clumps.
However, although we recognize a few low-mass disk fragments in our simulations, in our simulations steady quasi-Keplerian disks form only after the initial collapse and fragmentation phase of the core: when clear disk morphologies can be identified, these disks are relatively Toomre-stable.

\cite{AndreOliva2020} characterize the temporal evolution of protostellar disks starting from idealized initial conditions into four epochs: the initial setup, the disk formation epoch, the fragmentation epoch, and the quiescent epoch.  
In our simulations, we also identify four phases of the core evolution: the quasi-hydrostatic phase, the core fragmentation phase, the disk formation phase, and the steady-state (quiescent) phase.

In core \A{}, almost all of the self-gravitating fragments in which a star or a multiple stellar system forms, originate from high-density perturbations that appear and grow during the quasi-spherical collapse phase of the core, well before the formation of the pseudo-disk, as shown in Figure~\ref{fig:blobs}. %
Keeping in mind the caveat that our disks are generally more massive than those in \cite{AndreOliva2020}, the main qualitative difference between these two sets of simulations is that in the present work the fragmentation epoch precedes the disk formation phase. In other words, the formation of low-mass (disk) stars is initiated during the fragmentation of the core and before the disk formation phase.
These phases are discussed in \S{}~\ref{sec:2}, and illustrated in Figure~\ref{fig:schematic}. 

A more realistic model of star formation from core fragmentation should take into account accurate modeling of stellar evolution. In our simulations, the formation time of protostar particles is instantaneous when a clump center exceeds a given density threshold (see \S~\ref{sec:2} for details on the sink formation criteria in \ramses{} \citep{Bleuler2014}). More realistically, the formation of protostars follows a thermal timescale, or the Kelvin-Helmholtz (KH) timescale. This time is extremely short for massive stars. For a low-mass star, say a solar-mass star, even though the KH timescale is about 30 Myr long, the protostar shrinks to 100 solar radii, or about 0.5 AU, by 1/100 of the KH time, or 0.3 Myr, assuming a constant rate of radiating thermal energy. This means that in a very short time after sink particle formation, a star becomes a subgrid particle in the simulation, well tracked by a point-source sink particle. 

The N-body integrator in the code uses a softening length of $2 \Delta x_{\rm min}$, which is 14~AU for Core \A{} and 120~AU for Core \B{}. As a consequence, the formation of hard binaries is not captured and their dynamical evolution is not accurately resolved. We leave a robust treatment of the dynamics of these multiple systems for future work. 

\subsection{Star formation efficiency in cores and multiplicity}

The observed core mass function (CMF) closely resembles the stellar IMF but is shifted to the higher-mass end by a factor of $\sim 3$ \citep[e.g.][]{Alves2007}. This similarity seems to suggest the idea that the efficiency of star formation in dense cores ($n > 10^4 \ \pcc$) is of the order of 30 percent.
Previous studies tend to explain this offset invoking feedback, namely, protostellar outflows acting on core scales, entraining and expelling a large fraction of the core \citep[][]{Hansen2012, Kuiper2016}.  
Numerical simulations \citep{Kuiper2010, Offner2017} of isolated core collapse suggest outflows have a mass-loading factor of $\sim 3$. In \cite{He2019} we instead argue that cores have close to 100 percent star formation efficiency but the cores fragment into several smaller mass stars with a relatively flat IMF. We show that such a model can reproduce both the shape and normalization of the IMF in our MC scale simulations, which we refer to here as "baseline" simulations. One of the main motivations of this present work is to test this hypothesis by zooming on a few selected cores with high resolution. 

Figure~\ref{fig:MF} shows the mass function of the stars forming in each core in our simulations. The mass functions are plotted in two colors to distinguish the central stars (orange) from the companion stars (green). Black histograms are used when these two groups of stars are indistinguishable.
We can see that each core fragments into a mini cluster consisting of 1 to 12 stars, as found in previous numerical studies \citep{Bate1997, Goodwin2004a}.
In each panel, we compare the stellar masses in our zoom simulations to the sink masses of the ``baseline'' lower-resolution simulation of the same core, shown in the lower half of each panel. In the zoom-in simulations the spatial resolution increases by a factor $\sim 20$ ($\sim 60$ for core \A{}), reaching densities 3 orders of magnitude larger with respect to the baseline. 

The labels in each panel of Figure~\ref{fig:MF} compare the total stellar masses in the zoom-in simulations to the corresponding sink mass (representing a prestellar core) in baseline simulations. We find that the total masses of stars in the zoom-in simulations are either equal to or higher than the sink mass in the baseline, indicating that star formation efficiencies in cores are close to 100 percent, independent of the core mass. In addition, we find that the cores in the zoom-in simulations have nearly 100 percent efficiency of conversion of gas into stars: \ie, the final mass in star is between 50 to 100 percent the initial core mass, and for core \B{} the mass is stars is higher than the initial core mass. However, the core forms multiple stars with the masses of the highest mass stars reduced by approximately $1/3$. Hence, we argue that the CMF/IMF scaling parameter is due to the fragmentation of the core into multiple smaller mass stars rather than the inefficient conversion of gas into stars due to feedback effects. However, most of the mass in stars is locked in a few (2 to 3) relatively massive stars, while low mass stars -- that can be numerous, \eg, core \A{} forms a total of 12 stars -- account for a minority of the total mass of the core. Due to the small-number statistics, we cannot infer a shape for the mass function of stars in cores, but it appears to be flat.

In most of our zoom-in simulations where the resolution is lower and the peak density is below $10^9 \pcc$, the total number of stars is $<9$, the total mass in stars exceeds by roughly 30 percent the sink mass in the corresponding baseline simulation and the most massive star have sometimes a higher mass than $1/3$ of the sink mass. We partially attribute this to non-convergent results due to the limited resolution. Indeed in core \A{}, that is our highest resolution zoom-in simulation, where the sink formation density reaches $\sim 10^{10} \ \pcc$ (close to the density where the gas becomes adiabatic), 12 stars are formed, we reproduce very closely the stellar mass of the sink in the baseline simulation, our highest mass star is $<1/3$ of the baseline sink mass and we resolve low mass stars down to $0.01$~M$_\odot$.

It has been argued that the formation of low-mass stars from the fragmentation of massive cores contradicts the observations of a high binary fraction of low-mass stars \citep{Goodwin2007}. However, if the low-mass stars form in a hierarchical system as in our simulations, when they are ejected from the systems they can retain their binary companion. Indeed in core \A{} we observe binaries in disk stars and ejected binary systems, even though the limited resolution prevents us from resolving close binaries with separations $<14$~AU. We will study stellar dynamics in more detail in future work.

\section{Discussion}
\label{sec:dis}

\subsection{Formation of ultra-high-mass stars -- a competitive accretion scenario}

Recent advancement of radio/mm and optical/IR interferometers has enabled important progress in the field of disks around high-mass (early-B to late-O type) YSOs \citep[see][]{Davies2011, Mottram2011}.
The current unambiguous evidence for circumstellar disks around high-mass young stars is limited to objects with masses up to 30 \msun{} (late-O type) \citep[][]{Beltran2016}. 
Stars with these spectral types or brighter have strong UV emissions that can heat and disperse the gas. 
Typical disk radii of these sources are a few thousand of AU with rare exceptions of radii as small as 300-400 AU. 
Most of the protostars in our simulations (Cores \A{}, {\it D}, {\it E}, {\it F}) fall in the range of high-mass stars, with Cores {\it B} and {\it C} forming stars that are over 40 \msun{}. 
The range of the radii of the simulated disks in this work (see Table~\ref{tab:init}) agrees well with the observations. 

In the observations mentioned above, the circumstellar disks have typical gas masses ranging from 4 to a few $\times 10$ \msun{}. In our simulations the masses of disks,  defined as disk gas above a density threshold of $\sim 10^5~\pcc{}$, ranges between 3 \msun{} to 50 \msun{}, in agreement with the observed range. The disk mass remains relatively low ($\sim 40$~\msun) even in the very massive core \B{} where the central star cluster mass is above 600 \msun{}. This is due to the surface density being roughly constant at $ 8$~g/cm$^2$ up to a disk radius of $\sim 6000$~AU, where the azimuthal velocity becomes comparable to the gas velocity dispersion, typically 1 to 3~km/s. Higher mass (up to 200 \msun{}) disks are only reported in a few cases where the angular resolution is not enough to properly separate the envelope from the disk \citep{Beltran2004}. 

At the lower bound of the high-mass range where there are enough nearby sources, observers are able to estimate the hydrostatic scale height of these structures from (sub)millimeter observations of the line width. The typical line width is found to be $\sim 2 \ {\rm km~s^{-1}}$ and the estimated scale height is in most cases >30-40 percent of the disk radius, indicating that the disks of embedded protostars are likely geometrically thick \citep[e.g.,][]{Beltran2006}. 

Both the velocity field probed via molecular lines at high angular resolution ($\le 0.5^{''}$) and the CO bandhead proﬁle suggest that the rotation of the majority of the disks is consistent with Keplerian or quasi-Keplerian rotation \citep[e.g.][]{Wang2012, Cesaroni2014, Beltran2014}. 
For some sources, observations reveal both sub-Keplerian \citep[e.g.][]{Cesaroni2005, Wang2012, Beltran2014} and super-Keplerian \citep{Beuther2008}. \cite{Wang2012} argues that sub-Keplerian motions suggest a role for magnetic fields that could slow down the rotation below pure Keplerian by, i.e., magnetic braking. 
In our simulations, the disks are generally sub-Keplerian, with the non-Keplerianity parameter $\beta_{\rm kep} $ being in the range of -0.5 to -0.1. The low values of $\beta_{\rm kep}$ indicate a relatively large accretion rate from the large-scale envelope.

Another feature of protostellar disks of high-mass stars is the presence of asymmetries and inhomogeneities \citep[e.g.][]{Cesaroni2014}. The authors claim that these asymmetries could be caused by the presence of spiral arms or infalling filaments accreting material onto the disk or by interacting with companions nearby. 
Disks that have been observed with an extremely high angular resolution where a clear disk structure could be resolved, appear to be slightly elongated and oriented perpendicular to molecular outflows \citep{Wang2012, Beltran2014}. 
This structure is similar to the early phase of collapse of our simulated Case \B{} core.

\begin{figure}
    \centering
    \includegraphics[scale=0.8]{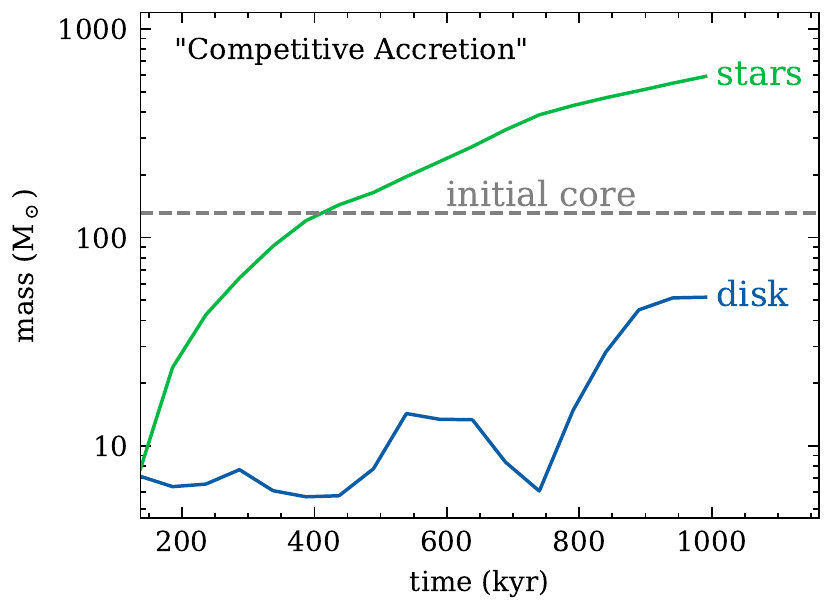}
    \includegraphics[scale=0.8]{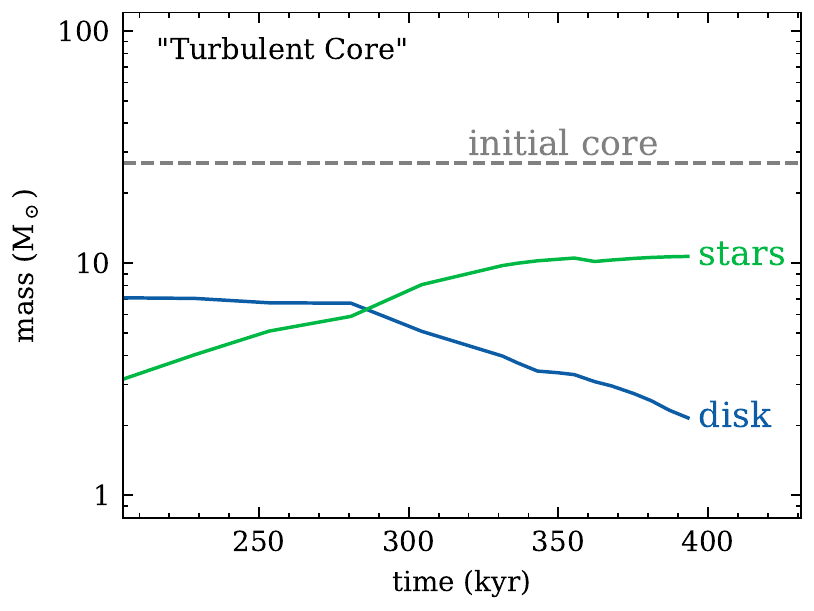}
    \caption{{\it Top}: The growth of Core \B{} from a 130 \msun{} core into a mini-cluster of stars with a total mass of $\sim  600$ \msun{} and average mass of 70 \msun{} via filamentary accretion from the background, demonstrating a competitive accretion scenario for ultra-high-mass star formation. {\it Bottom:} The collapse of 27 \msun{} Core \A{} into a central binary of $\sim 10$ \msun{} plus low-mass companions, demonstrating a ``turbulent core'' scenario for intemediate- to high-mass star formation. 
    The color curves show the evolution over time of the total mass in stars, labeled as `stars', or the mass of the disk, labeled as `disk'.
    }
    \label{fig:accretion}
\end{figure}

During the initial phase of collapse, our simulated core \B{} has a spherical shape with a radius of $\sim 0.5$ parsec within a density threshold of  $3000~\pcc{}$, enclosing a total mass of $131~\msun{}$. The core has a density profile that is shallower than that of a Bonnor-Ebert sphere, with both magnetic and turbulent pressure in the envelope being 10 times stronger than thermal pressure (Figure~\ref{fig:pro1}). However, the mass accreted in the central (proto)star cluster is $> 600~\msun{}$ over a timescale of about 0.9~Myr, more than four times the initial gas reservoir in the core (Figure~\ref{fig:accretion}). This is due to the sustained high accretion rate ($10^{-4} \text{--} 10^{-3} \ \msun{}/{\rm yr}$) over 0.9~Myr at a inflow velocity of 1 to 3~km/s. We, therefore, argue that the large masses (50 -- 100 \msun{}) of the YSOs in core \B{} can be described, at least partially, in the context of the competitive accretion scenario, in which the gas is collected over time from scales beyond the initial core radius.
This agrees with \cite{Gong2015} who, through a set of simulations of turbulent, unmagnetized GMCs, find that sink particles accrete at a nearly constant rate even after the initial mass reservoir is depleted. However, this kind of ``competitive accretion'' is only seen in the most massive (> 50 \msun{}) core in our simulations.

Observational evidence of circumstellar structures in the most massive protostars, \ie, early-O type, is very limited. Huge, dense, massive, rotating cores have been detected around early-O-type protostars in studies performed at moderate spatial resolutions.
These objects are in all likelihood non-equilibrium structures surrounding clusters of young protostars and not merely individual massive stars \citep[see][and references therein]{Beltran2011}. These structures are characterized by a much higher mass and larger size than the rotationally supported disks around lower-mass protostars discussed above. \cite{Beltran2005} referred to these massive structures as ``toroids'' to make a distinction. The reported toroids have radii of a few $\times 1000$ AU to up to $10^4$ AU \citep[e.g.][]{Zapata2008}. The hydrostatic scale height of these toroids, estimated by assuming hydrostatic equilibrium, is $> 50 \%$ of the radius -- these structures are extremely thick.

We find that the most massive stars in our star cluster formation simulations form as clusters inside large and dynamically stable toroids with significant mass infall. Our simulated Core \B{} matches the properties of the toroids discussed above. The structure that enshrouds the central protostar cluster forms a toroid that is both large (4000--8000 AU in radius) and thick (3000--8000 AU in thickness) and is largely sub-Keplerian.
The high infall rate, of the order of $10^{-3} \ \msun{}/{\rm yr}$, could be high enough to quench the formation of an HII region or to slow down its expansion (\citealt{Yorke1986}, see also \S{}~\ref{sec:HII}). In the next section, we will show that core \B{} produces a bipolar HII region and an outflow. However,
in general, the question of whether or not the launching of outflows could be quenched initially by the massive envelope requires further studies that take into account jet-driven outflow and perhaps stellar winds.

\subsection{UV radiation trapping}
\label{sec:HII}

\begin{figure*}
    \centering
	\includegraphics[width=\textwidth]{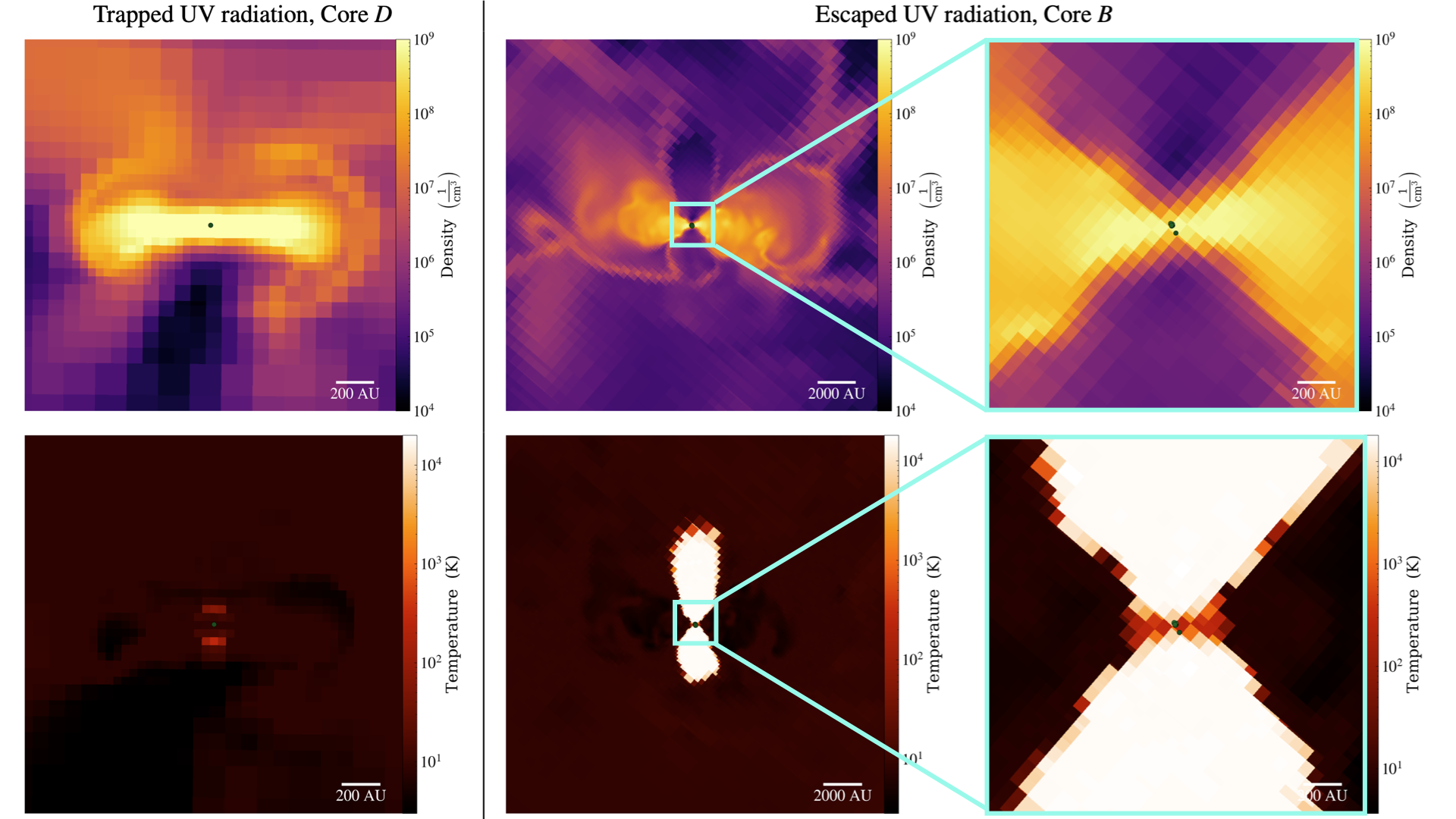}
    \caption{The trapping (left) and escaping (right) of HII regions.
	  {\it Left}: Density (top) and temperature (bottom) slices of core {\it D} showing the UV radiation from a $\sim 10 \msun{}$ star is trapped inside an ultra-compact region at the center of a disk.
	  {\it Right}: Similar to {\it Left} but for the Core \B{} showing the escaping of an ultracompact HII region. The dynamical motions of the multi-star system create a channel for radiation to escape.
	}
    \label{fig:trap2}
\end{figure*}

\begin{figure*}
  \centering
  \includegraphics[width=1.6in]{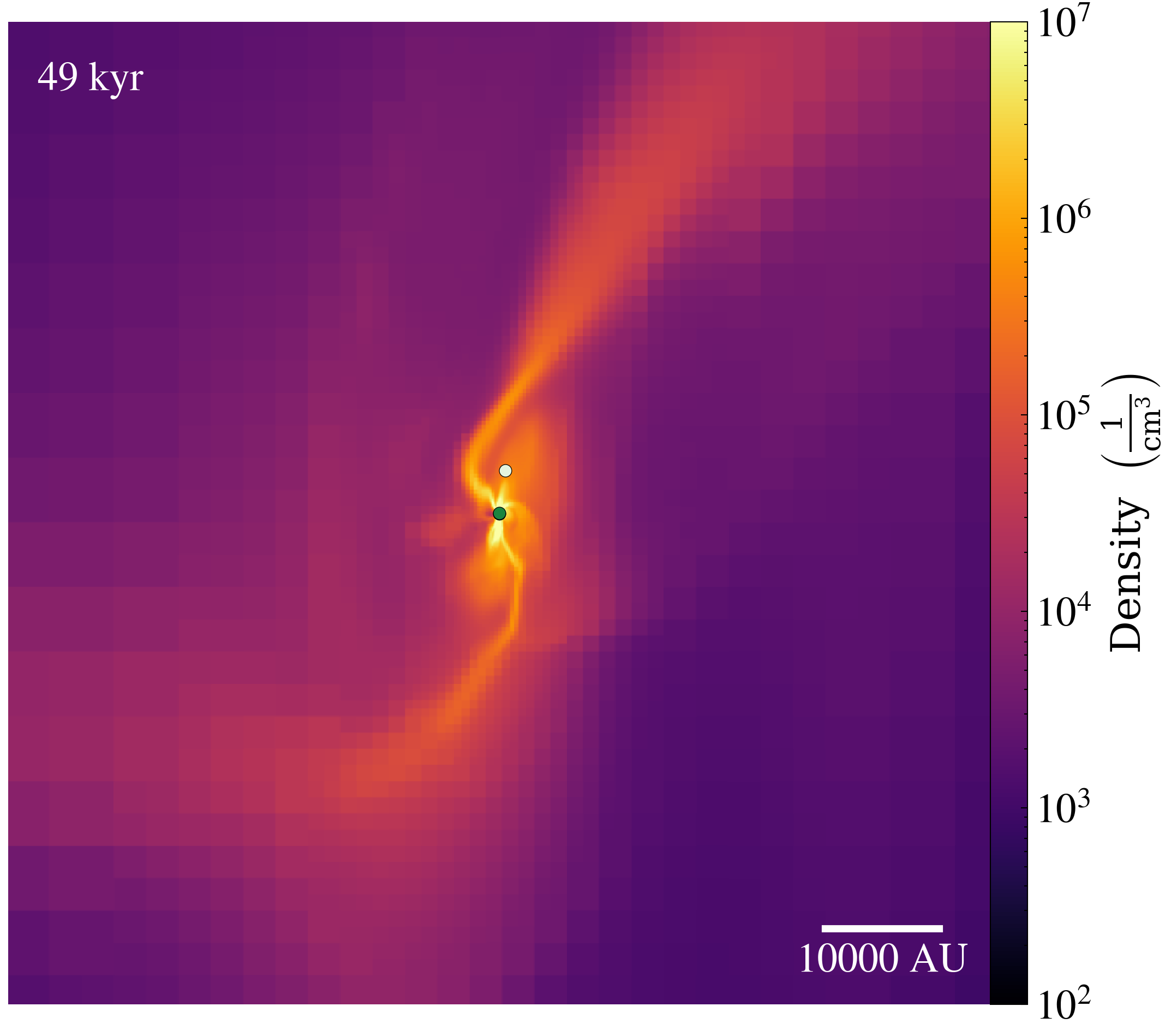}\hspace{-0.1in}
  \includegraphics[width=1.6in]{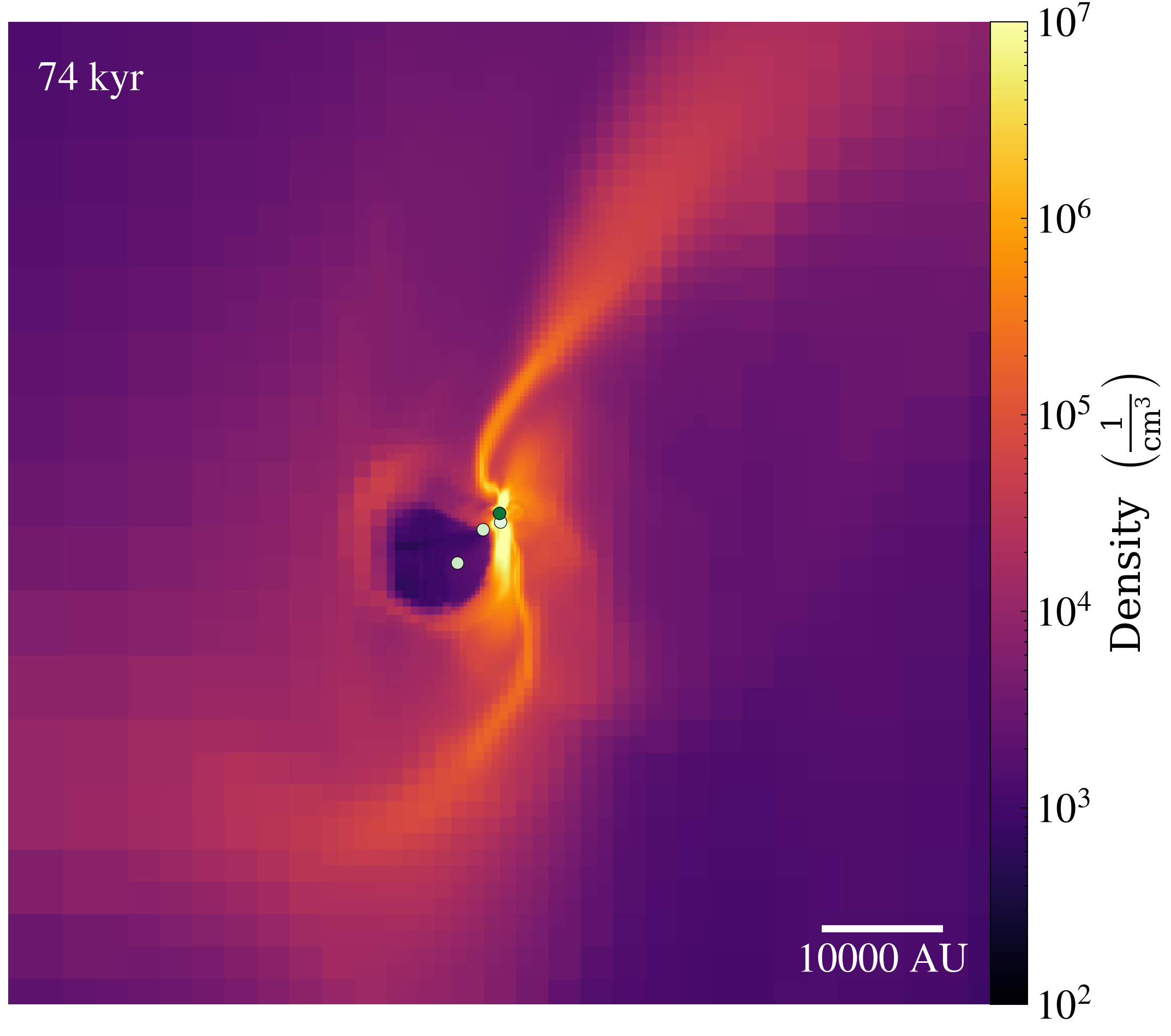}
  \includegraphics[width=1.6in]{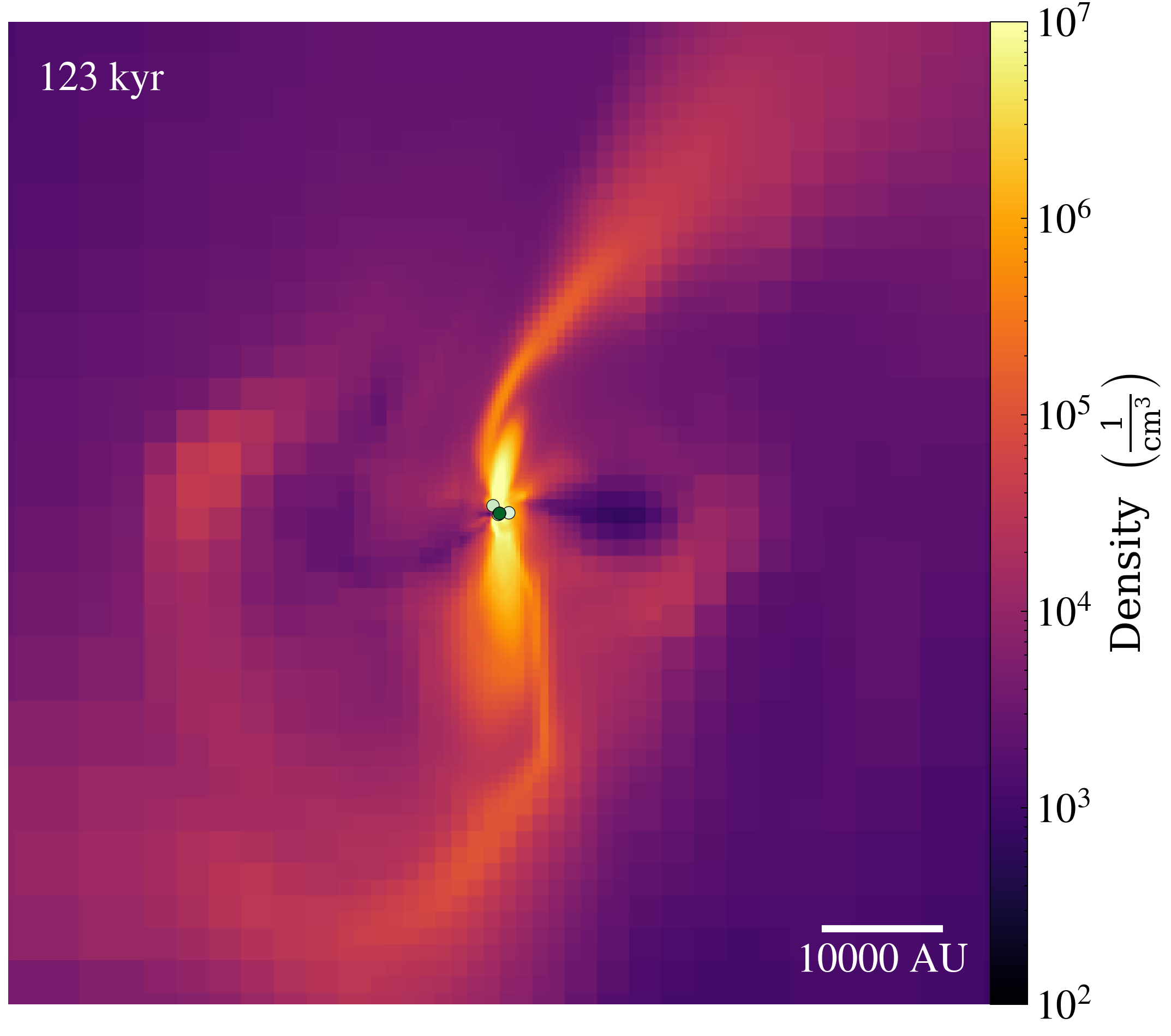}
  \includegraphics[width=1.6in]{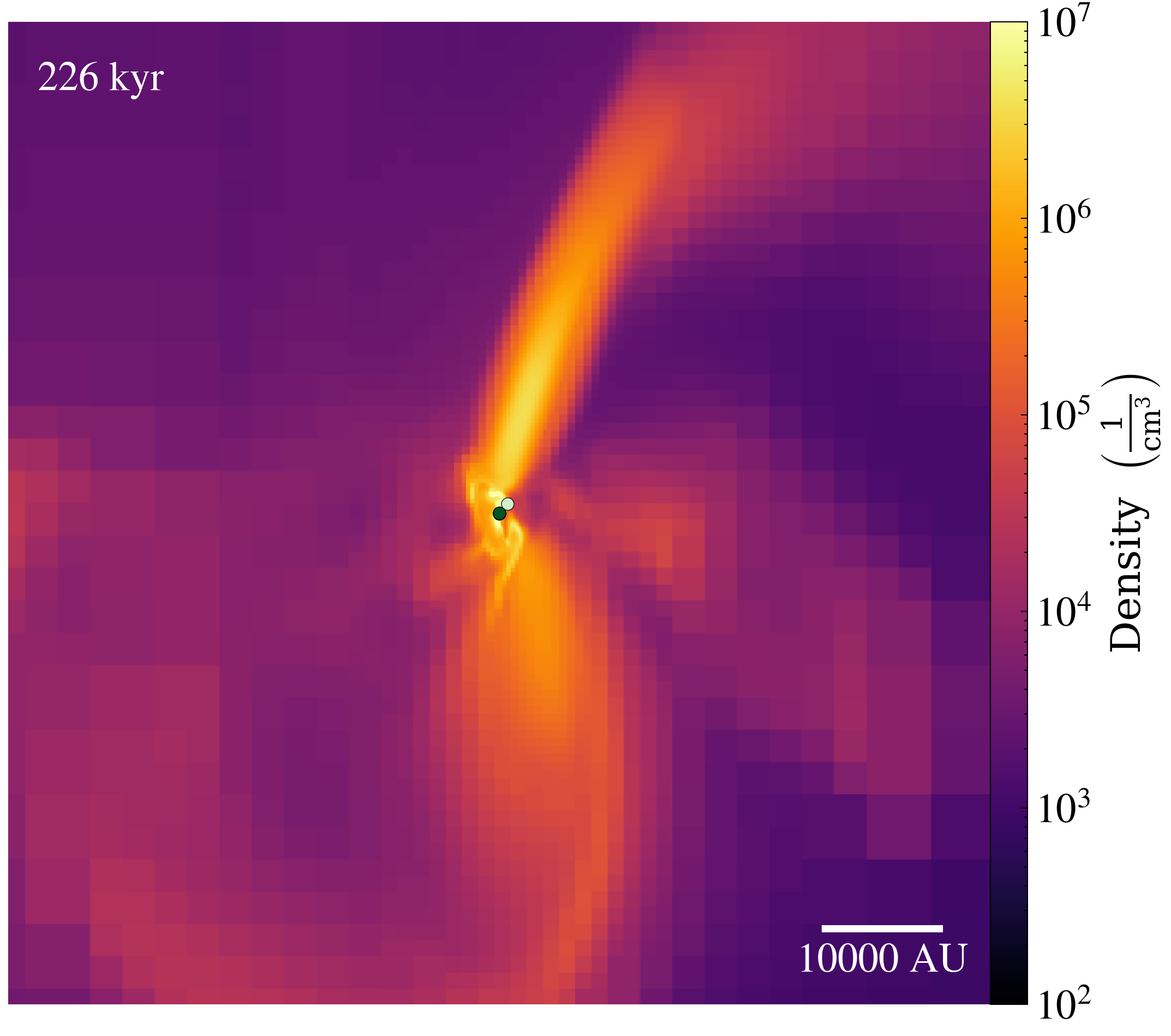}\\
  \includegraphics[width=1.6in]{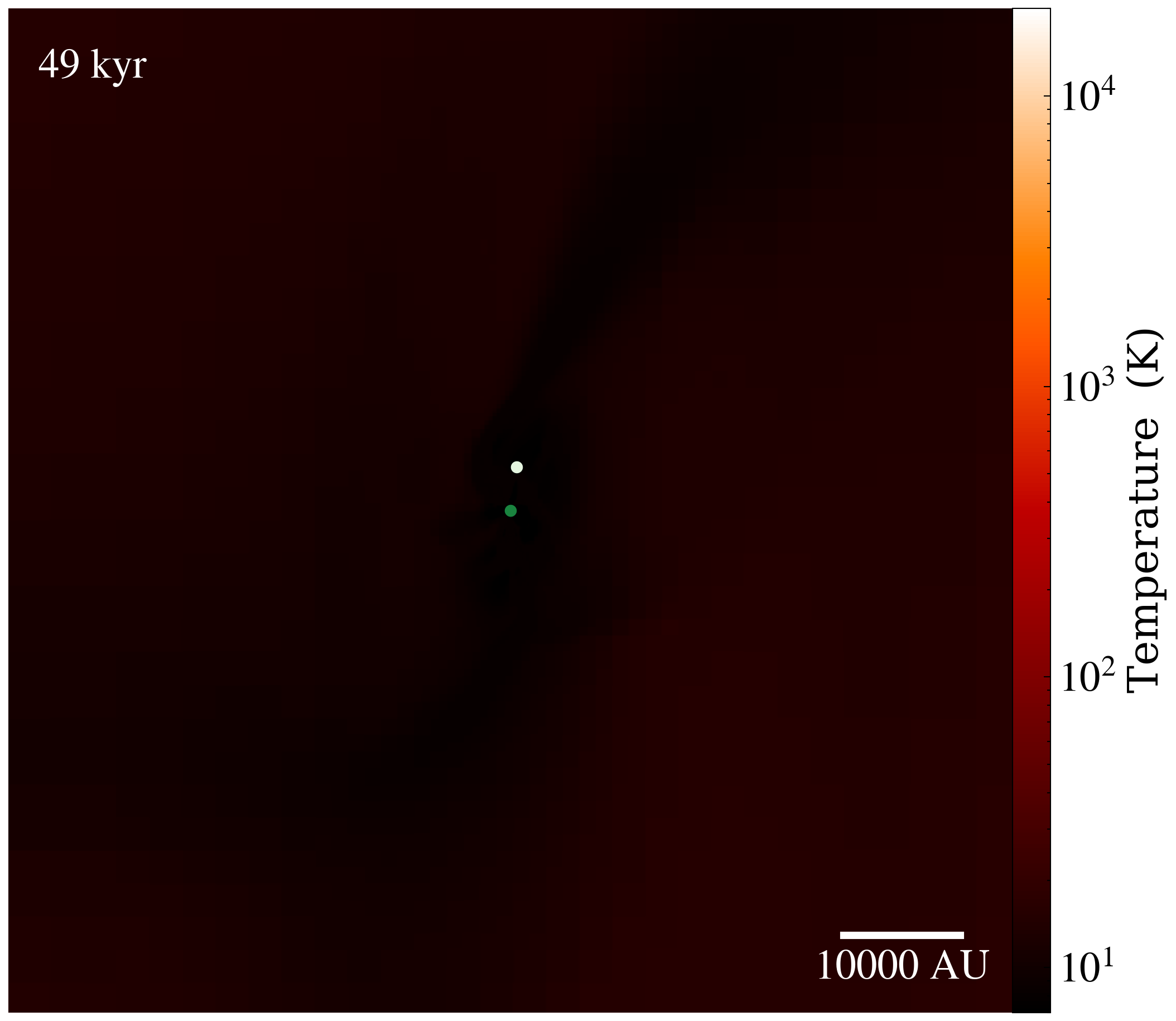}
  \includegraphics[width=1.6in]{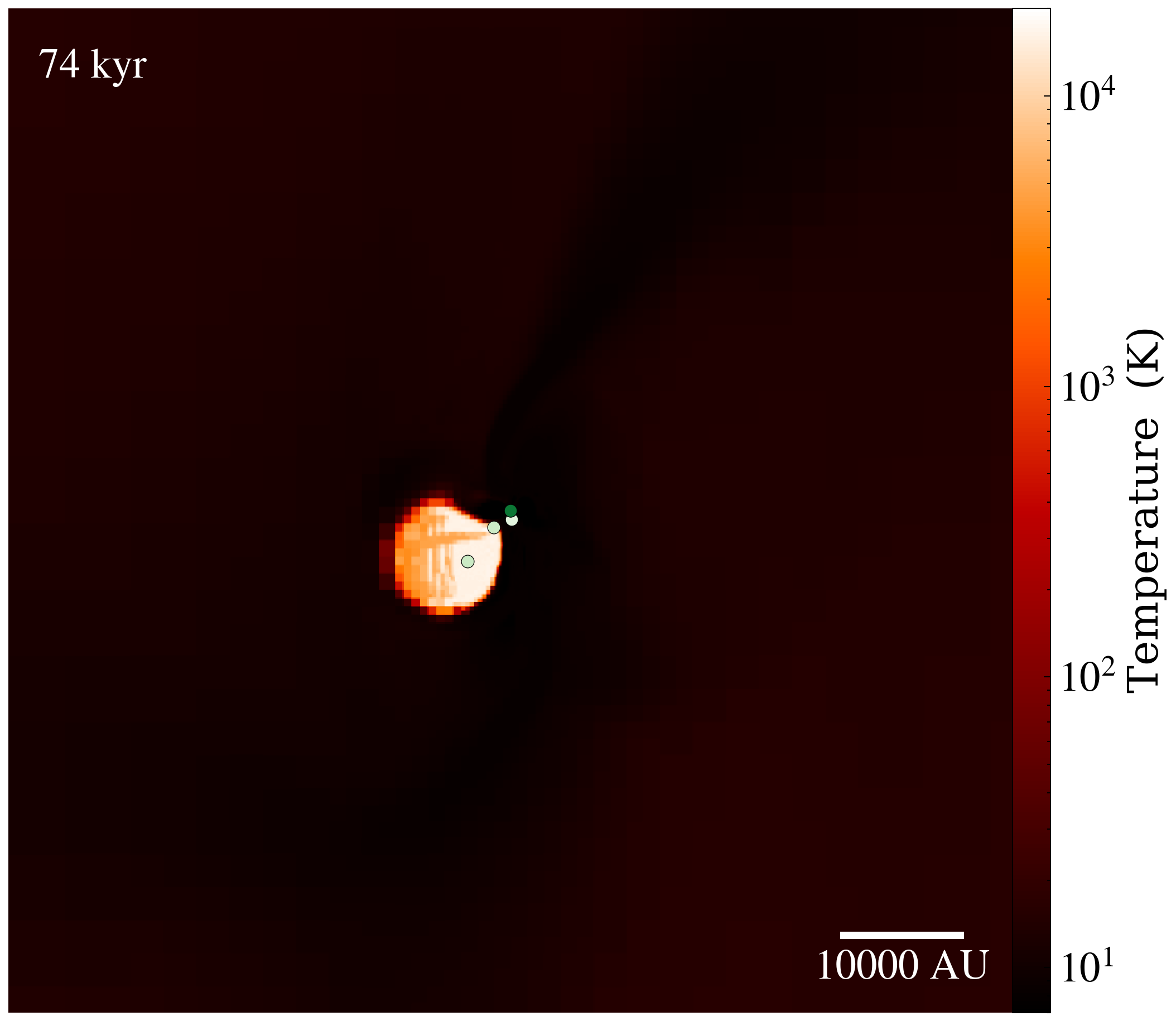}
  \includegraphics[width=1.6in]{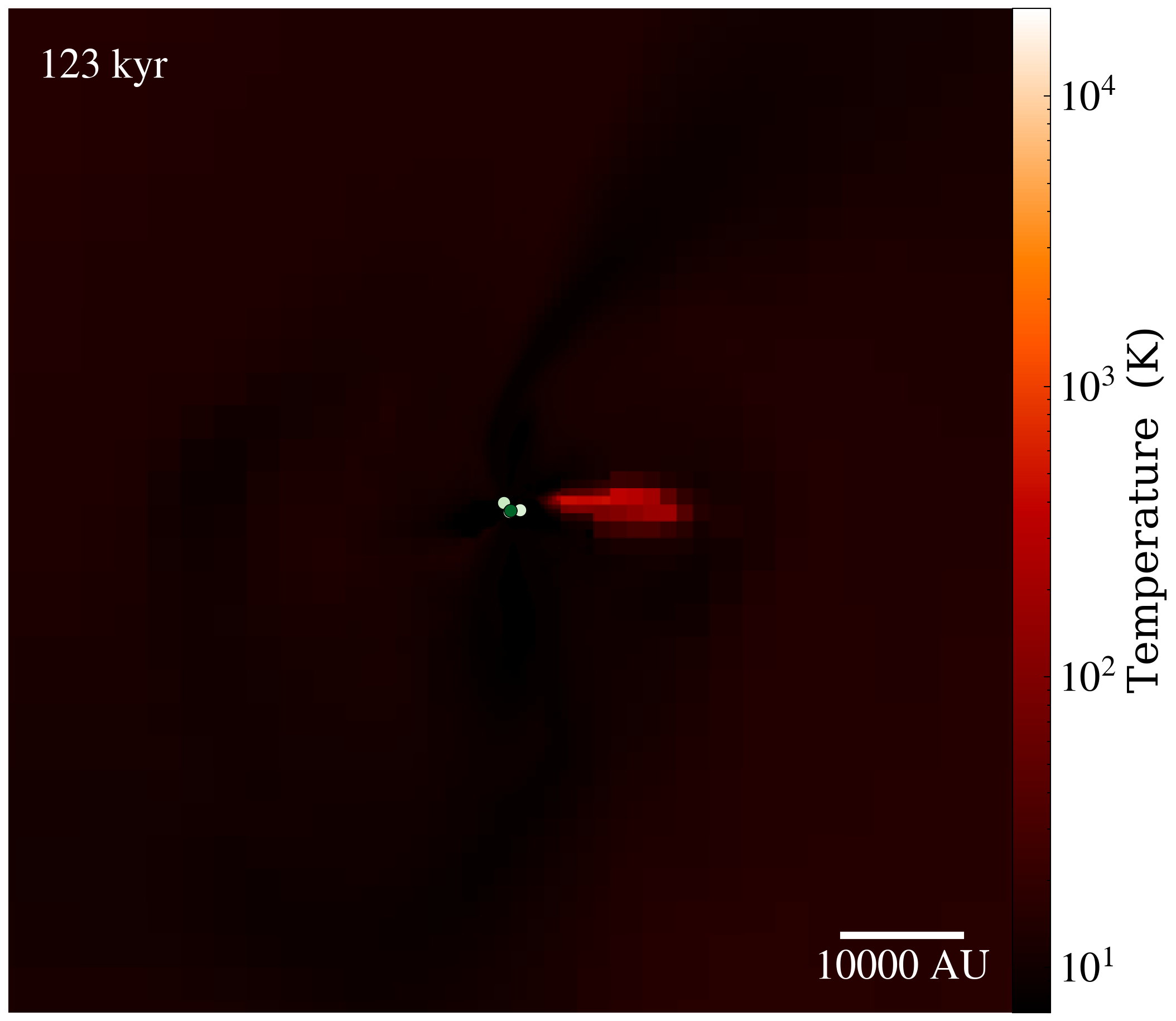}
  \includegraphics[width=1.6in]{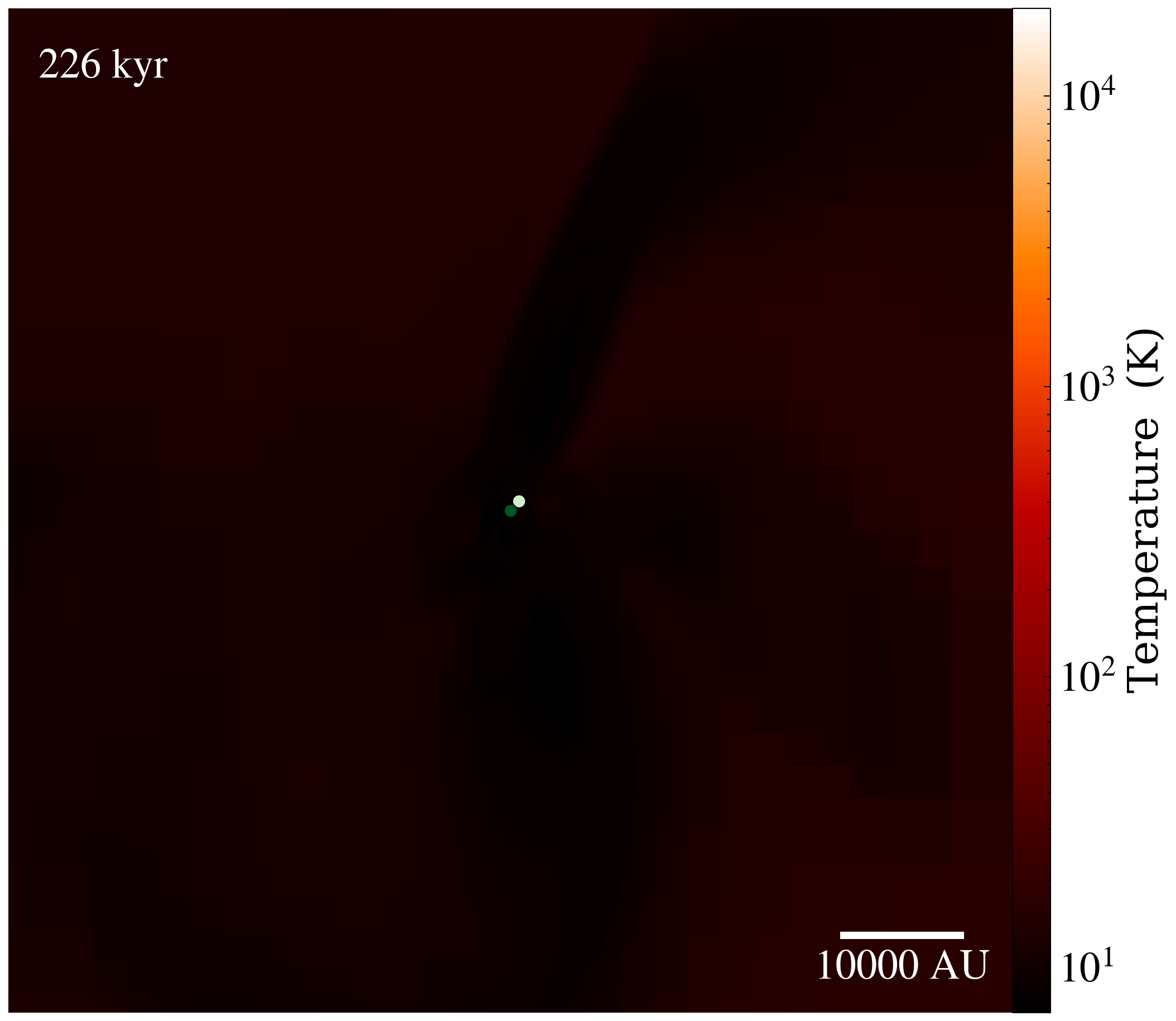}
  \caption{
	The escaping of UV radiation from a dense filament due to dynamical motion. Here it shows a time sequence of the density (top) and temperature (bottom) slice. The dancing of the stars creates a channel for UV photons to escape from the dense region when 
  \label{fig:trap}}
\end{figure*}

High-mass stars are often observed to be deeply embedded in dense gas and their HII regions in the gas above a density of $\sim 10^4 \pcc$ can remain trapped forming ultracompact HII regions \citep{Churchwell2002}. 
In our simulated prestellar cores, we notice two main distinct scenarios: A) when a single massive star is deeply embedded in a thick disk, where the density reaches $\sim 10^8~\pcc$, the HII regions remains trapped at least during the first 0.5~Myr while the disk is still relatively massive. B) Often massive stars orbit a common center of mass in binary or multiple star systems: in these cases the stellar orbit may displace the massive star from the densest regions in the disk or in a filament, allowing the HII regions to break out the dense disk or filament.
Case A) can be observed in the left panels of Figure~\ref{fig:trap2}), showing core {\it D}, where a single massive star forms from the collapse of the small core. The UV radiation of the protostar is trapped inside the dense neutral gas The HII region remains trapped for a few hundred kiloyears. 
In Core \A{}, however, multiple stars form at or migrate into the disk center (see right panels of Figure~\ref{fig:trap2}). The dynamic of the few-body star system displaces the stars by about the virial radius of the star system, which is tens of AU, close to or higher than the disk thickness. In this case, radiation can escape when one of the massive stars approaches the edges of the disk. Once a channel of lower-density ionized gas is created a long-lasting bipolar HII region and outflow is created. 

Finally Figure~\ref{fig:trap}) shows another interesting mechanism that allows UV ionizing radiation to break out of a dense filament. The figure shows the density (top panels) and the temperature (bottom panels) in a set of snapshots showing a disk embedded in a filament in which a multiple systems, including a massive star, forms. The stars are shown as circles color-coded according to their masses from white (1 \msun{}) to dark green (10 \msun{}). From the time sequence, it is clear that HII regions are created intermittently on either side of the filament when the massive star in its orbit is further from the densest part of the filament, allowing the HII region to break out of the filament. This is more evident in the animation that we make available in the electronic version of this paper.

\subsection{Influence of metallicity on disk stability}

Our simulations are conducted at solar metallicity and the cooling from hydrogen, helium, carbon, oxygen, and dust grains. Lower metallicity could make a big difference in how the disk fragments.
Recent study \citep{Matsukoba2022} of the metallicity dependence of protostellar-disk fragmentation has shown that fragmentation of spiral arms is more common in lower metallicities where the dust cooling is effective. At high metallicity, the disk is stabilized by stellar irradiation.
We find that despite being cold (close to 10 K) due to the lack of effective heating from the stars, the disks are stable and do not undergo fragmentation, as discussed in \S{}~\ref{sec:frag}. 

\section{Summary}

We have simulated the formation and collapse of prestellar cores in a set of grid-based radiation-MHD simulations, resolving from GMC scales, tens-of-parsec in size, down to disk scales, with resolutions up to 7 astronomical units in our highest resolution simulation.

We studied a set of 6 massive ($\sim 10$~M$_\odot$) or very massive ($\sim 100$~M$_\odot$) cores in two GMCs, following their collapse, fragmentation, and the formation of (proto)stars embedded in circumstellar disks with sizes ranging from 200~AU to 6000~AU. The properties of the simulated cores, and the (proto)stars and disks that form therein are listed in Table~\ref{tab:init}.

The following is a list of the main results:
\begin{enumerate}[label=\arabic*.]
    \item The disks are generally large ($R = 200 - 6000$ AU), thick (aspect ratio $H/R = 0.2 - 1.3$), and massive with masses spanning from a few to 40 \msun{} (Table~\ref{tab:init} and Figure~\ref{fig:gallery}). These disks or toroids sit in the range of observed disk properties around high-mass YSOs.

    \item Each core undergoes fragmentation in the early collapsing phase with geometries that can be separated into two main distinct modes: ``quasi-spherical'' collapse and ``filamentary'' collapse (see Figure~\ref{fig:schematic}). However, in both modes of collapse, the fragments eventually become embedded in a quasi-steady accretion disk or toroid.
    
    \item We observe the formation of multiple massive stars  at the center of the disk, but also lower-mass stars apparently forming in outer parts of the disk. However, the disk is on average Toomre-stable. We find that ``disk stars'' form from pre-existing self-gravitating fragments created before the formation of the gravitationally stable disks and are accreted into the disk as mentioned in the point above (see Figures~\ref{fig:schematic} and \ref{fig:blobs}). We, therefore, conclude that in order to realistically simulate the formation and evolution of massive stars and their circumstellar disks it is crucial to capture the environment and initial conditions of the protostellar core. Idealized initial conditions starting from smooth disk structures, often used to model circumstellar disks around solar-mass stars, will likely not capture the full picture of fragmentation of the disk for the high-mass case. 
    
    \item Large and massive disks around high-mass stars are supported by both magnetic and turbulent pressure. This is in contrast to the case of disks around lower-mass stars, supported instead by thermal pressure. Regardless of the core mass/size, the magnetic pressure dominates in the envelope as well as the outer disk at radii $\gtrsim 200-1000$~AU, while turbulent pressure dominates in the inner disk at $< 200-1000$~AU (see Figure~\ref{fig:sigmavz}). The opening angle $H/R$ is nearly constant as a function of radius.

    \item The final number of (proto)stars that form in a core is between 1 and 12 (Figure~\ref{fig:MF}). Most of the accreted mass is distributed among 1 to 3 stars of similar mass (up to a mini cluster of 8 stars for our most massive core) that form near the center of the disk/toroid. The disk stars account for a small or negligible fraction of the mass of stars.
  
    \item 
    In our highest resolution simulation ($\Delta x_{\rm min} =$ 7~AU) where the sink formation density is above $10^{10} \ \pcc{}$, close to the density where the gas becomes adiabatic, the total mass in stars ($12.6 \ \msun{}$) is approximately equal to the mass of the sink particle in the baseline run, which is believed to represent a prestellar core. This suggests a nearly 100 percent star formation efficiency in high-mass cores.
  
    \item In the most massive core we simulated, the core evolves from a spherical shape with a radius of 0.5 parsec and a mass of 131 \msun{} into a (proto)star cluster that is over 600 \msun{} enshrouded by a massive toroid over a timescale of about 0.9 Myr (Figure~\ref{fig:accretion}). We account for the large masses in these ultra-high-mass (proto)stars to the competitive accretion scenario in which gas is continuously supplied from larger scales beyond the mass reservoir of the core.
  
    \item O/B stars that form as a single star typically produce an ultracompact HII region that remains trapped in the dense and thick circumstellar disk for an extended period of time ($\sim 500$~kyrs). However, when high-mass stars form as wide binaries or in multiple systems, the dynamic motion of the system displaces the stars periodical from the densest parts in the disk plane or filament where the density is lower allowing UV radiation to escape and creating a long-lasting or periodic bipolar HII regions (see Figures~\ref{fig:trap2}-\ref{fig:trap}).
\end{enumerate}

In a companion paper, we will further study the properties and growth of the magnetic field in magnetically critical and sub-critical cores, where we will also address the origin of the density-$B$ relationship and the magnetic breaking problem.

\section*{Acknowledgements}

CCH acknowledges the support by the NASA FINESST grant 80NSSC21K1850. The authors acknowledge the support of the NASA grant 80NSSC18K0527. The authors acknowledge the University of Maryland supercomputing resources (http://hpcc.umd.edu) made available for conducting the research reported in this paper.

\section*{Data Availability}

The data underlying this article were accessed from the University of Maryland supercomputing resources (http://hpcc.umd.edu). The derived data generated in this research will be shared on reasonable request to the corresponding author. The software used to do the analysis in this paper is \textsc{ramtools}, a toolkit to postprocess \textsc{ramses} simulations that is based on the \textsc{yt} toolkit (https://yt-project.org/doc/index.html) and is available to download from https://chongchonghe.github.io/ramtools-pages/.

\bibliographystyle{mnras}
\bibliography{BIB_HE,disk}

\listoffixmes

\bsp	%
\label{lastpage}
\end{document}